\documentclass[iop]{emulateapj}

\usepackage{txfonts}
\usepackage{graphicx}
\usepackage{color}
\usepackage{natbib}
\usepackage{amssymb}

\newcommand{\wi}{5.0cm}
\newcommand{\wib}{8.0cm}

\shorttitle{Effect of a DM halo on the determination of dynamical black hole masses}
\shortauthors{Schulze \& Gebhardt}

\begin{document}

\title{Effect of a dark matter halo on the determination of black hole masses}

\author{Andreas Schulze}
\affil{Astrophysikalisches Institut Potsdam, An der Sternwarte 16, 14482 Potsdam, Germany}
\email{aschulze@aip.de}
\and
\author{Karl Gebhardt}
\affil{Department of Astronomy, The University of Texas at Austin, 1 University Station, C1400, Austin, TX 78712, USA}
\email{gebhardt@astro.as.utexas.edu}

\begin{abstract}
Stellar dynamical modeling is a powerful method to determine the mass of black holes in quiescent galaxies. However, in previous work the presence of a dark matter halo has been ignored in the modeling. Gebhardt \& Thomas (2009) showed that accounting for a dark matter halo increased the black hole mass of the massive galaxy M87 by a factor of two. We used a sample of 12 galaxies to investigate the effect of accounting for a dark matter halo in the dynamical modeling in more detail, and also updated the masses using improved modeling. The sample of galaxies possesses \textit{Hubble Space Telescope} and ground-based observations of stellar kinematics. Their black hole masses have been presented before, but without including a dark matter halo in the models.
Without a dark halo, we find a mean increase in the estimated mass of 1.5 for the whole sample compared to previous results. We attribute this change to using a more complete orbit library. When we include a dark matter halo, along with the updated models, we find an additional increase in black hole mass by a factor of 1.2 in the mean, much less than for M87. We attribute the smaller discrepancy in black hole mass to using data that better resolve the black hole's sphere of influence.
We redetermined the $M_\bullet-\sigma_\ast$ and $M_\bullet-L_V$ relationships using our updated black hole masses and found a slight increase in both normalization and intrinsic scatter.
\end{abstract}


\section{Introduction}  \label{sec:intro}
It is now well established that almost every massive galaxy harbors a supermassive black hole in its center. Furthermore, close relations between the mass of this supermassive black hole and the properties of the galaxy's spheroid component have been found, namely with the mass \citep{Magorrian:1998, Haering:2004}, luminosity \citep{Kormendy:1995,Kormendy:2001,Marconi:2003,Gultekin:2009}, and with the velocity dispersion \citep{Gebhardt:2000,Ferrarese:2000, Tremaine:2002,Gultekin:2009}. These relations imply a link between the growth of black holes and galaxy evolution, usually attributed to active galactic nucleus (AGN) feedback \citep[e.g.,][]{Silk:1998,DiMatteo:2005,Springel:2005,Ciotti:2007}, but at least to some degree they are a natural result within a merger-driven galaxy evolution framework \citep{Peng:2007, Hirschmann:2010, Jahnke:2010}. In general, the black hole-bulge relations and especially their evolution with cosmic time are able to provide deep insight into galaxy formation and black hole growth. Therefore, it is essential to properly establish the local relationships as precisely as possible.

The black hole-bulge relations are based on a sample of $\sim50$ quiescent black holes, whose masses have been determined based on maser emission \citep[e.g.,][]{Greenhill:2003,Herrnstein:2005,Kuo:2010}, gas kinematics \citep[e.g.,][]{Ferrarese:1996,Marconi:2001,DallaBonta:2009}, and stellar kinematics \citep[e.g.,][]{vanderMarel:1998,Gebhardt:2000b,Shapiro:2006,Gebhardt:2007,Gultekin:2009b}. 
In particular, stellar dynamical modeling using orbit superposition is a powerful method to estimate black hole masses in quiescent galaxies, also recovering the orbital structure within the galaxy. Usually, axisymmetry is assumed in these models. However, there are still uncertainties and possibly systematic biases within these methods. Uncertainties may arise from the deprojection of the observables onto three dimensions as the true inclination of the galaxy often is not well known, the presence of dust, some triaxiality that cannot be modeled properly with axisymmetric models \citep{vandenBosch:2010}, or the presence of an AGN at the center. So far, in most models the contribution of the galaxy's dark matter (DM) halo has been neglected. \citet{Gebhardt:2009} recently showed that the black hole mass can be underestimated in this case. For the massive galaxy M87 they found an increase of more than a factor of two in the measured black hole mass, just by including a DM halo in the modeling. The reason is that without a DM halo the mass-to-light ratio is overestimated in order to account for the mass in the outer parts of the galaxy. Under the usually applied assumption of a constant mass-to-light ratio for the whole galaxy, this will propagate inward and lead to an underestimation of the black hole mass at the center due to overestimation of the stellar contribution. 

A similar result has been obtained by \citet{McConnell:2010}. They measured the black hole mass in the brightest cluster galaxy NGC~6086. They report a factor of six difference between the black hole mass obtained from models without a DM halo and their most massive DM halo models. However, the black hole's sphere of influence is barely resolved in their work.

\citet{Shen:2010} found for NGC~4649, also a massive galaxy, no change in the black hole mass by including a DM halo. In this case the sphere of influence is well resolved by the data. A larger sample, especially spanning a larger range in mass, is clearly required. 

\citet{Gebhardt:2003} (hereafter G03) studied a sample of 12 galaxies with kinematics derived from \textit{Hubble Space Telescope} (\textit{HST}) and ground-based observations, using axisymmetric orbit superposition models. They do not include a DM halo in their modeling. Since then the orbit superposition code used by our group has been improved \citep{Thomas:2004, Thomas:2005, Siopis:2009}, by including a more complete orbit sampling.

The aim of this paper is to reanalyze the data set presented by G03, using the most up-to-date dynamical modeling code and investigating the effect of accounting for the dark matter contribution on the derived black hole masses. One of the galaxies in the G03 sample, NGC~4649, was recently analyzed by \citet{Shen:2010}, including a DM halo.
We have reanalyzed this galaxy for consistency with the remaining sample, but find consistent results with this previous investigation. 

\section{Data} \label{sec:data}
The data set used in this work is identical to those used in the work of G03. Thus, we will only give a brief summary and refer to G03 for more detail. The data consist of three sets of observations for each galaxy: imaging, \textit{HST} stellar kinematics, and ground-based stellar kinematics.

High-resolution imaging is required to obtain the stellar surface brightness profile for each galaxy. This imaging has been obtained in the V band with the \textit{HST} WFPC2 \citep{Lauer:2005}, except for NGC~4697, which was observed with the \textit{HST} WFPC1 \citep{Lauer:1995}. Surface brightness profiles were measured from the pont-spread function deconvolved images and were augmented with ground-based imaging at the outer parts, not covered by \textit{HST}. For the deprojection of the surface brightness profile to a luminosity density profile, we assume axisymmetry, an inclination angle of $90^\circ$, which we refer to as edge-on, and used the technique outlined in \citet{Gebhardt:1996}.

The \textit{HST} observations and kinematics are presented by \citet{Pinkney:2003} and G03. They consist of Space Telescope Imaging Spectrograph (STIS) long-slit spectra along the major axes, except for NGC~3377 and NGC~5845, which have Faint Object Spectrograph (FOS) aperture spectra. The spectra cover the \ion{Ca}{2} triplet around 8500~\AA{}. For the dynamical modeling, the line-of-sight velocity distributions (LOSVDs), extracted from the spectra, are used. The LOSVDs are given in a non-parametric form, binned into 15 equidistant bins, compared to 13 bins in G03. 

The ground-based kinematics are presented by \citet{Pinkney:2003} and G03 as well. They consist of long-slit spectra along several position angles, mainly obtained at the MDM observatory. They also cover the \ion{Ca}{2} triplet, or alternatively the Mg$b$ absorption at 5175 \AA{}. The individual LOSVDs are binned into 15 points as well.

\begin{table*}
\caption{Results for the Galaxy Sample}
\label{tab:res}
\centering
\begin{tabular}{lcccccccccc}
\hline \hline \noalign{\smallskip}
Galaxy & $D$ (Mpc) & $M_{\bullet,\mathrm{G03}} \, (M_\odot)$ & $M/L_\mathrm{G03}$ & $M_{\bullet,\mathrm{noDM}} \, (M_\odot)$ & $M/L_\mathrm{noDM}$ & $M_{\bullet,\mathrm{DM}} \, [M_\odot] $ & $M/L_\mathrm{DM}$ & $V_c$~(km s$^{-1}$) & $r_c$~(kpc) & $R_\mathrm{inf}$ ($\arcsec$) \\ 
(1) & (2) & (3) & (4) & (5) & (6) & (7) & (8) & (9) & (10) & (11)\\
\hline \noalign{\smallskip}
NGC 821 & $25.5$ & $9.9 \pm 4.1 \times 10^{7}$ & $ 6.8$ & $1.1 \pm 0.4 \times 10^{8}$ & $  7.7 \pm   0.5$ & $1.8 \pm 0.8 \times 10^{8}$  & $  6.2 \pm   0.7$ & 450 & 14.0 & 0.14 \\ 
NGC 2778 & $24.2$ & $1.6 \pm 1.0 \times 10^{7}$ & $ 7.2$ & $1.4 \pm 1.1 \times 10^{7}$ & $ 11.9 \pm   1.1$ & $1.5 \pm 1.5 \times 10^{7}$  & $ 11.8 \pm   1.2$ & 300 &  5.0 & 0.02 \\ 
NGC 3377 & $11.7$ & $1.1 \pm 0.6 \times 10^{8}$ & $ 2.7$ & $1.6 \pm 1.0 \times 10^{8}$ & $  2.6 \pm   0.5$ & $1.9 \pm 1.0 \times 10^{8}$  & $  2.3 \pm   0.4$ & 350 &  6.0 & 0.69 \\ 
NGC 3384 & $11.7$ & $1.8 \pm 0.2 \times 10^{7}$ & $ 2.5$ & $8.0 \pm 4.2 \times 10^{6}$ & $  2.4 \pm   0.1$ & $1.1 \pm 0.5 \times 10^{7}$  & $  2.2 \pm   0.1$ & 400 &  8.0 & 0.04 \\ 
NGC 3608 & $23.0$ & $1.9 \pm 0.9 \times 10^{8}$ & $ 3.7$ & $4.6 \pm 0.9 \times 10^{8}$ & $  3.5 \pm   0.3$ & $4.7 \pm 1.0 \times 10^{8}$  & $  3.3 \pm   0.3$ & 400 & 10.0 & 0.55 \\ 
NGC 4291 & $25.0$ & $3.2 \pm 1.6 \times 10^{8}$ & $ 6.0$ & $9.7 \pm 2.0 \times 10^{8}$ & $  6.0 \pm   0.5$ & $9.2 \pm 2.9 \times 10^{8}$  & $  6.0 \pm   0.8$ & 400 &  8.5 & 0.56 \\ 
NGC 4473 & $17.0$ & $1.3 \pm 0.7 \times 10^{8}$ & $ 5.1$ & $5.9 \pm 5.0 \times 10^{7}$ & $  7.4 \pm   0.2$ & $1.0 \pm 0.5 \times 10^{8}$  & $  6.8 \pm   0.3$ & 400 & 10.0 & 0.15 \\ 
NGC 4564 & $17.0$ & $6.9 \pm 0.7 \times 10^{7}$ & $ 1.6$ & $9.8 \pm 2.3 \times 10^{7}$ & $  1.6 \pm   0.1$ & $9.4 \pm 2.6 \times 10^{7}$  & $  1.5 \pm   0.1$ & 350 &  7.0 & 0.19 \\ 
NGC 4649 & $16.5$ & $2.1 \pm 0.6 \times 10^{9}$ & $ 8.8$ & $3.9 \pm 1.0 \times 10^{9}$ & $  8.6 \pm   0.6$ & $4.2 \pm 1.0 \times 10^{9}$  & $  8.0 \pm   0.7$ & 500 & 20.0 & 1.51 \\ 
NGC 4697 & $12.4$ & $2.0 \pm 0.2 \times 10^{8}$ & $ 4.2$ & $2.2 \pm 0.3 \times 10^{8}$ & $  4.5 \pm   0.3$ & $2.0 \pm 0.5 \times 10^{8}$  & $  4.3 \pm   0.3$ & 450 & 12.0 & 0.45 \\ 
NGC 5845 & $28.7$ & $2.9 \pm 1.1 \times 10^{8}$ & $ 4.5$ & $4.5 \pm 1.2 \times 10^{8}$ & $  5.4 \pm   0.2$ & $5.4 \pm 1.7 \times 10^{8}$  & $  5.1 \pm   0.2$ & 300 &  5.0 & 0.30 \\ 
NGC 7457 & $14.0$ & $4.1 \pm 1.4 \times 10^{6}$ & $ 2.8$ & $7.4 \pm 4.2 \times 10^{6}$ & $  2.7 \pm   0.5$ & $1.0 \pm 0.6 \times 10^{7}$  & $  2.6 \pm   0.5$ & 300 &  5.0 & 0.14 \\ 
\noalign{\smallskip} \hline
\end{tabular}
\tablecomments{Column 1: name. Column 2: distance. Columns 3-4: black hole mass and mass-to-light ratio from study of \citet{Gebhardt:2003}. Columns 5-6: black hole mass and mass-to-light ratio using updated code but without including a DM halo in the dynamical modeling. Columns 7-8: black hole mass and mass-to-light ratio when a DM halo is included. Column 9-10: parameters of the circular velocity and the core radius for the logarithmic DM density profile used. Column~11: radius of the black hole's sphere of influence, based on the black hole mass including the contribution of DM.}
\end{table*}

\section{Dynamical Models} \label{sec:model}
The dynamical modeling uses the orbit superposition method, first proposed by \citet{Schwarzschild:1979}. This general method has been widely used by various groups \citep{Rix:1997,vanderMarel:1998,Cretton:1999,Valluri:2004}.
Our technique is described in detail in G03, \citet{Thomas:2004,Thomas:2005} and \citet{Siopis:2009}. We will give a brief summary here and especially point out the differences compared to the work of G03. The basic approach consists of the following steps: (1) deprojection of the surface brightness profile to a three-dimensional luminosity distribution, (2) computation of the specified gravitational potential, (3) generation of a representative orbit library in this potential, (4) fitting the orbit library to the observed light distribution and kinematics, and (5) modifying the input potential to find the best match to the data, based on a $\chi^2$ analysis.

As described in Section~\ref{sec:data}, we deproject the surface brightness profile following \citet{Gebhardt:1996} and assume an edge-on configuration, as used by G03. The only exception is NCG~4473, where we assume an inclination of $72^{\circ}$, as has been done in G03. To determine the potential, we assume a constant mass-to-light ratio throughout the galaxy, a specific black hole mass and optionally also include a DM halo. The mass distribution is then given by
\begin{equation}
 \rho(r,\theta) = M_\bullet \delta(r) + \Upsilon \nu(r,\theta) + \rho_\mathrm{DM}(r)   \label{eq:rho}
\end{equation} 
where $M_\bullet$ is the black hole mass, $\Upsilon$ is the mass-to-light ratio, $\nu$ is the stellar luminosity distribution, and $\rho_\mathrm{DM}$ is the DM density profile. The potential $\Phi(r,\theta)$ is derived by integrating Poisson's equation.

In this potential, a representative orbit library is constructed that samples the phase space systematically. The generation of the orbit library is described in detail in \citet{Thomas:2004} and \citet{Siopis:2009}. 
For the comparison with the data, we use a spatial grid of $N_r=20$ radial bins and $N_\theta=5$ angular bins and use $N_v=15$ velocity bins for the LOSVD at each spatial gridpoint. The galaxy potential and the forces are evaluated on a grid with 16 times finer resolution. 
For our axisymmetric code, there are three integrals of motion that sample the phase space accordingly: the energy $E$, the angular momentum $L_z$, and a non-classical third integral $I_3$. The ($E$, $L_z$)-plane is sampled based on the spatial binning \citep{Richstone:1988}. We choose $E$ and $L_z$ such that the respective orbits have their pericenter and apocenter in every pair of the radial grid bins.
The third integral $I_3$ is sampled as outlined by \citet{Thomas:2004}. First, orbits are dropped from the zero-velocity curve (defined by $E=L_z^2/(2r^2\cos^2\theta) + \Phi(r,\theta)$), as in G03. This is done by using the intersections of the angular rays of the spatial grid with the zero-velocity curve as starting points for the integration of the orbit's motion.
However, this does not ensure a representative sampling of orbits. Such a sampling is indicated by a homogeneous coverage of the surface of section, i.e., the position of radii and radial velocities of orbits during their upward crossing of the equatorial plane. Therefore, additional orbits are launched to give such a homogeneous coverage. This method provides a complete sampling of the surface of section for given $E$ and $L_z$ and thus a proper coverage of phase space. We typically have 13,000$-$16,000 orbits in our library. The allocation of the individual orbits to the spatial grid points is based on the time they spend there.

Given this orbit library, the orbit weights are chosen by matching the orbit superposition to the observed light distribution and the LOSVDs of the galaxy on the spatial grid. To fit the orbit library to the data, we use the maximum entropy technique of \citet{Richstone:1988}. This method maximizes the function
\begin{equation}
 \hat{S} = S - \alpha \chi^2 \label{eq:maxentropy} \ ,
\end{equation}  
where $\chi^2$ is the sum of the squared residuals over all spatial and velocity bins, e.g.,
\begin{equation}
 \chi^2 = \sum_{k=1}^{N_d} \frac{(l_{\mathrm{mod},k} -l_{\mathrm{dat},k})^2}{\sigma_k^2} \ ,
\end{equation} 
where $l_k$ is the light in the $k$th bin, with the bins composed of the spatial position on the sky and the line-of-sight velocity, thus the bin in the LOSVD, at that position. Hence, $k$ is varying from $1$ to $N_d =N_r N_\theta N_v$. While $l_{\mathrm{dat},k}$ refers to the measured light at that position, $l_{\mathrm{mod},k}$ is given by the 
weighted sum of the contribution of all orbits to the $k$th bin.

$S$ is the Boltzmann entropy
\begin{equation}
 S = \sum_{i=1}^{N_\mathrm{orb}} w_i \log \left( \frac{w_i}{V_i} \right)  \ ,
\end{equation} 
where $w_i$ is the weight of the individual orbit and $V_i$ is the phase space volume of this orbit, i.e., the volume of the region in phase space that is represented by this orbit $i$, given by
\begin{equation}
 V = \Delta E \Delta L_z \int T(r,v_r) dr dv_r \ ,
\end{equation}
where $T(r,v_r)$ is the time between two successive crossings of the equatorial plane and $\Delta E$ and  $\Delta L_z$ are the ranges in energies and angular momenta of the respective orbits \citep{Binney:1985,Thomas:2004}.

The parameter $\alpha$ in Equation~(\ref{eq:maxentropy}) controls the relative weight of entropy and $\chi^2$ for the maximization. We start with a small $\alpha$, being dominated by the entropy maximization in the fit, and then iteratively increase it until there is no longer an improvement in the $\chi^2$.

This procedure provides a value for the $\chi^2$ for one combination of $M_\bullet$, $\Upsilon$ and DM halo.
The best fit is found by the global minimum of $\chi^2$ for the variation of these parameters. For the estimation of the parameter uncertainties we adopt the usual $\Delta \chi^2=1$ criterion \citep{Press:1992} to obtain the 68\% confidence intervals for one degree of freedom, thus when marginalizing over the other free parameters.

\section{Results} \label{sec:res}
\subsection{Models without a DM halo} \label{sec:nodm}
We first ran a set of models without including the contribution of a DM halo; thus, we set $\rho_\mathrm{DM}=0$ in Equation~(\ref{eq:rho}). This assumption is consistent with most previous studies on black hole masses using dynamical models as well as with G03. As we are using the same data asthose of G03, the main difference is the improved modeling code. Thus, we would expect to recover similar black hole masses as in G03. We also use slightly different distances to the galaxies, as given by \citet{Gultekin:2009}.

For each galaxy we ran models on a fine grid in $M_\bullet$ and mass-to-light ratio ($M/L$). Each model gives a best-fitting orbit superposition, and thus orbital structure for the galaxy, with a corresponding value of $\chi^2$. The best fit $M_\bullet$ and $M/L$ are determined by the global minimum of the $\chi^2$ distribution.

\begin{figure*}
\centering
\setlength{\unitlength}{1mm}
\begin{picture}(156,210)
\put(0,156){\includegraphics[width=\wi]{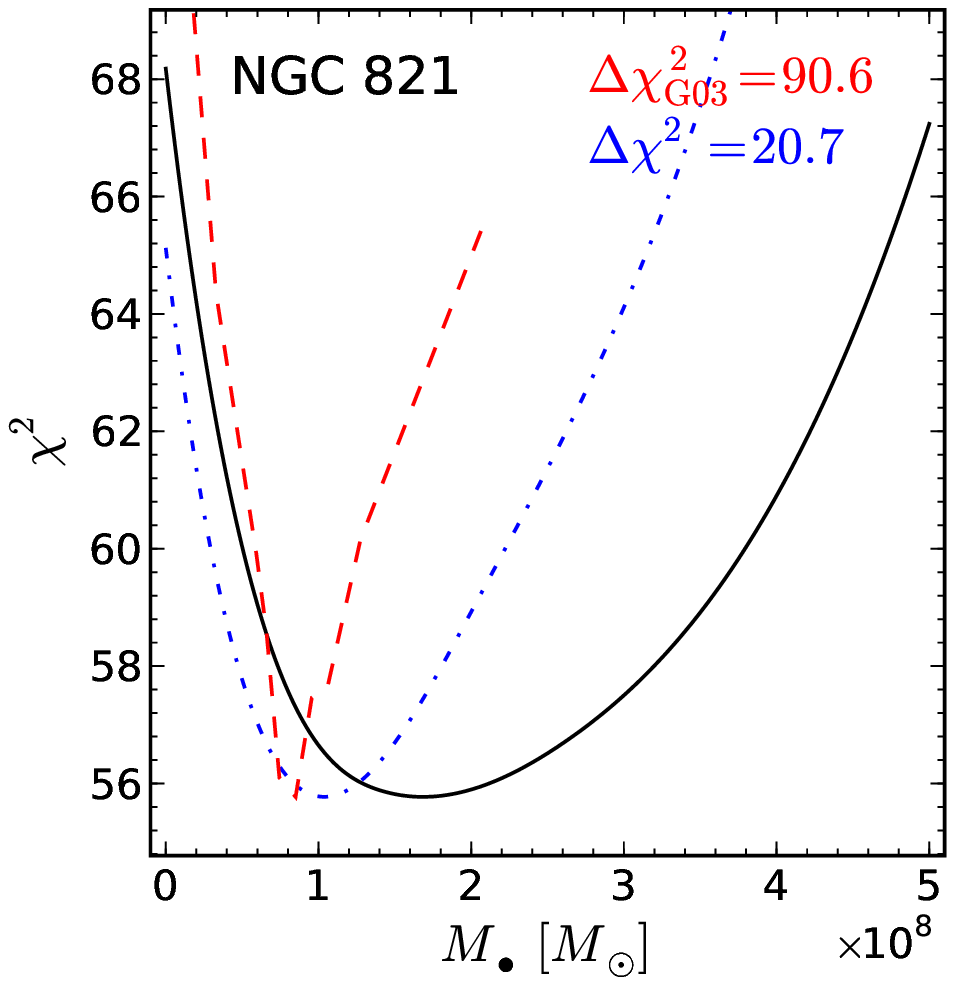}}
\put(52,156){\includegraphics[width=\wi]{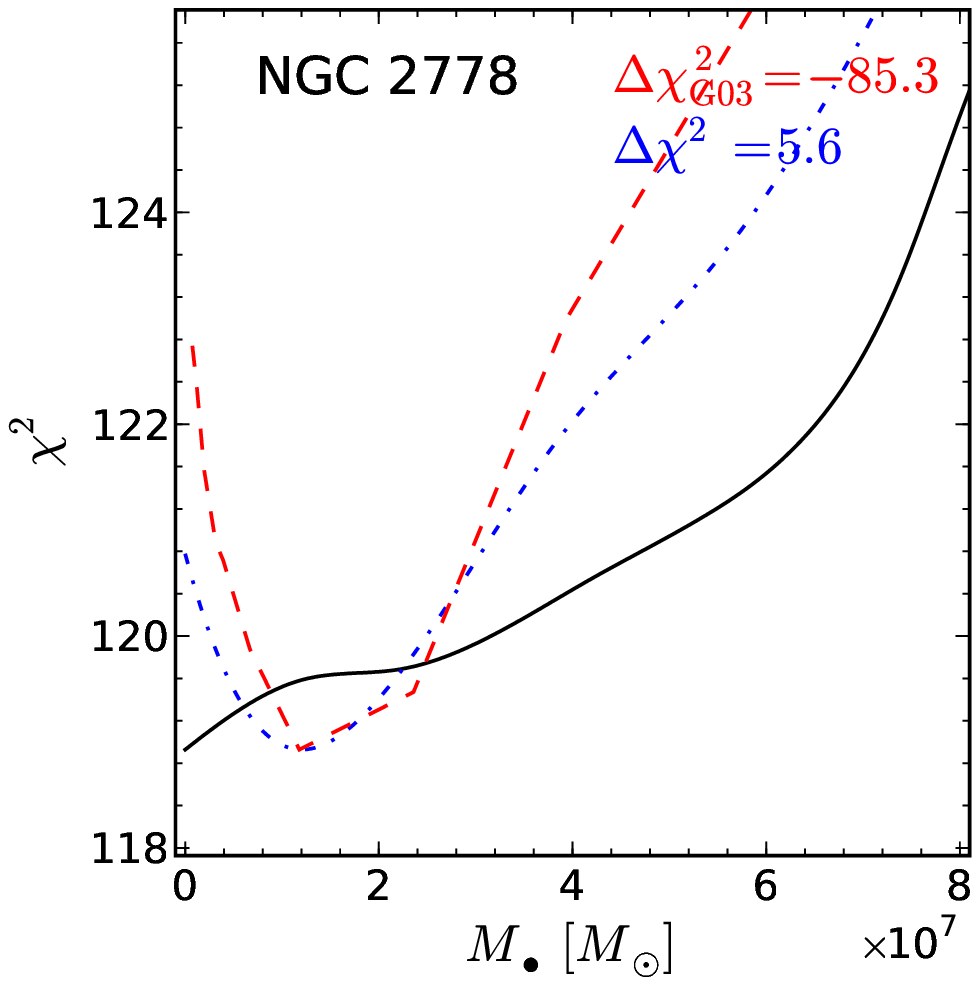}}
\put(104,156){\includegraphics[width=\wi]{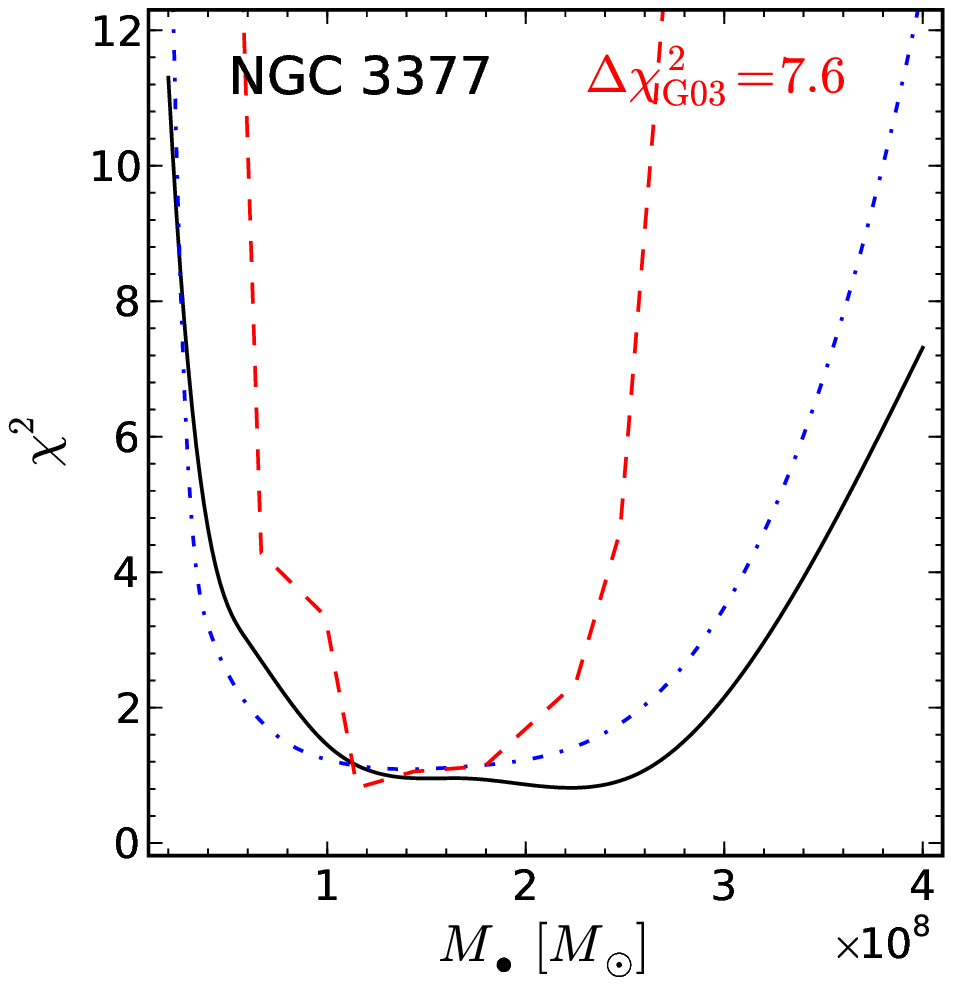}}

\put(0,104){\includegraphics[width=\wi]{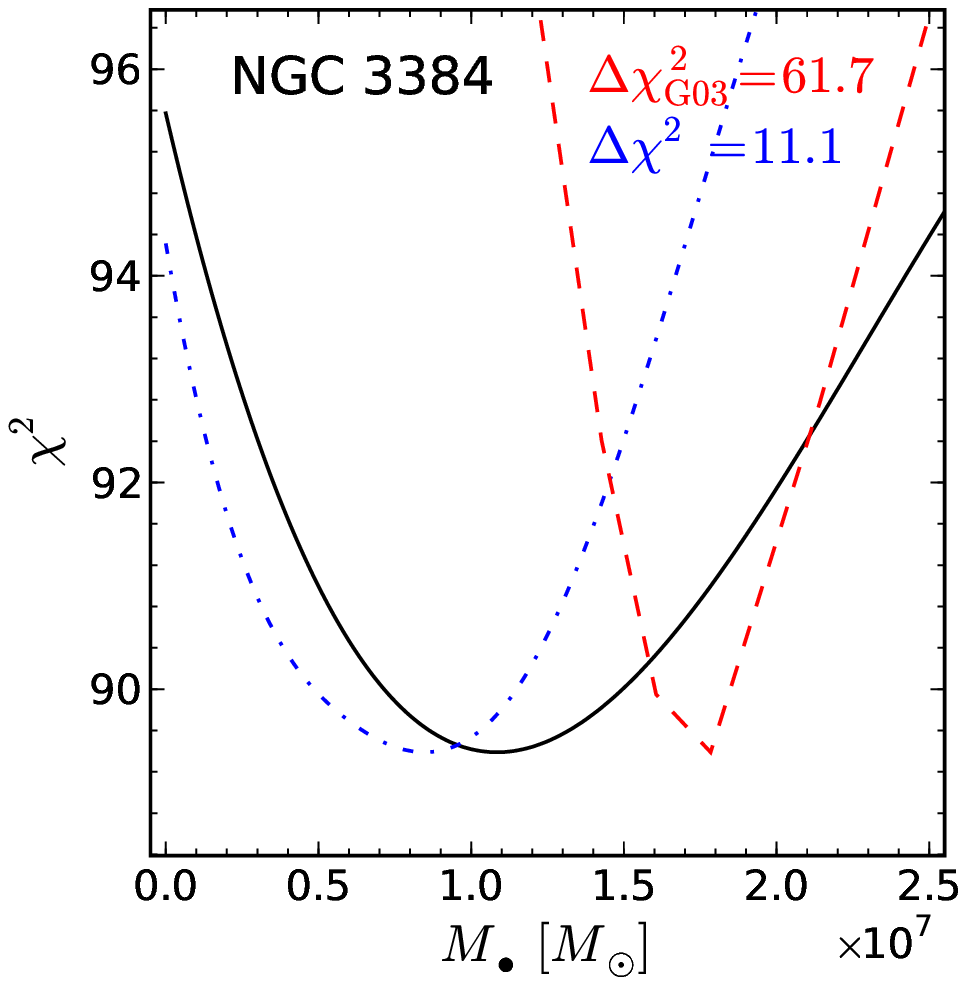}}
\put(52,104){\includegraphics[width=\wi]{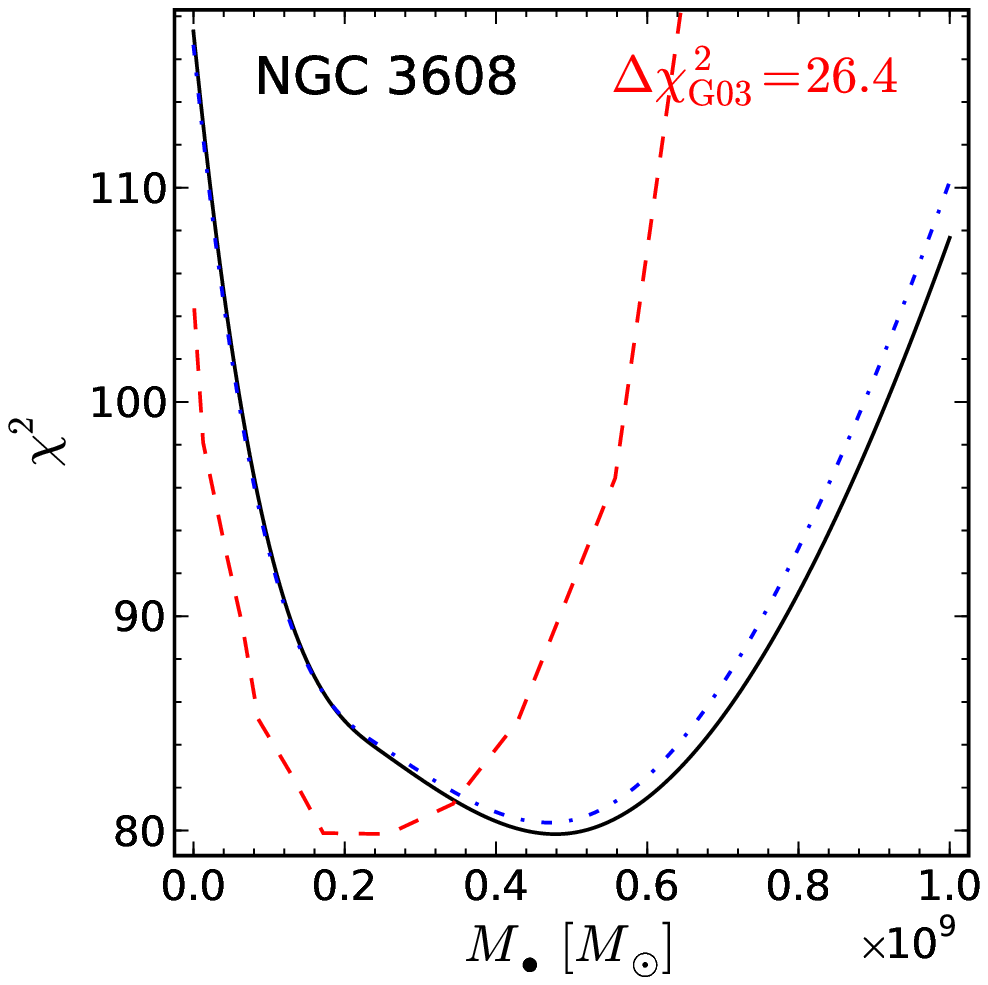}}
\put(104,104){\includegraphics[width=\wi]{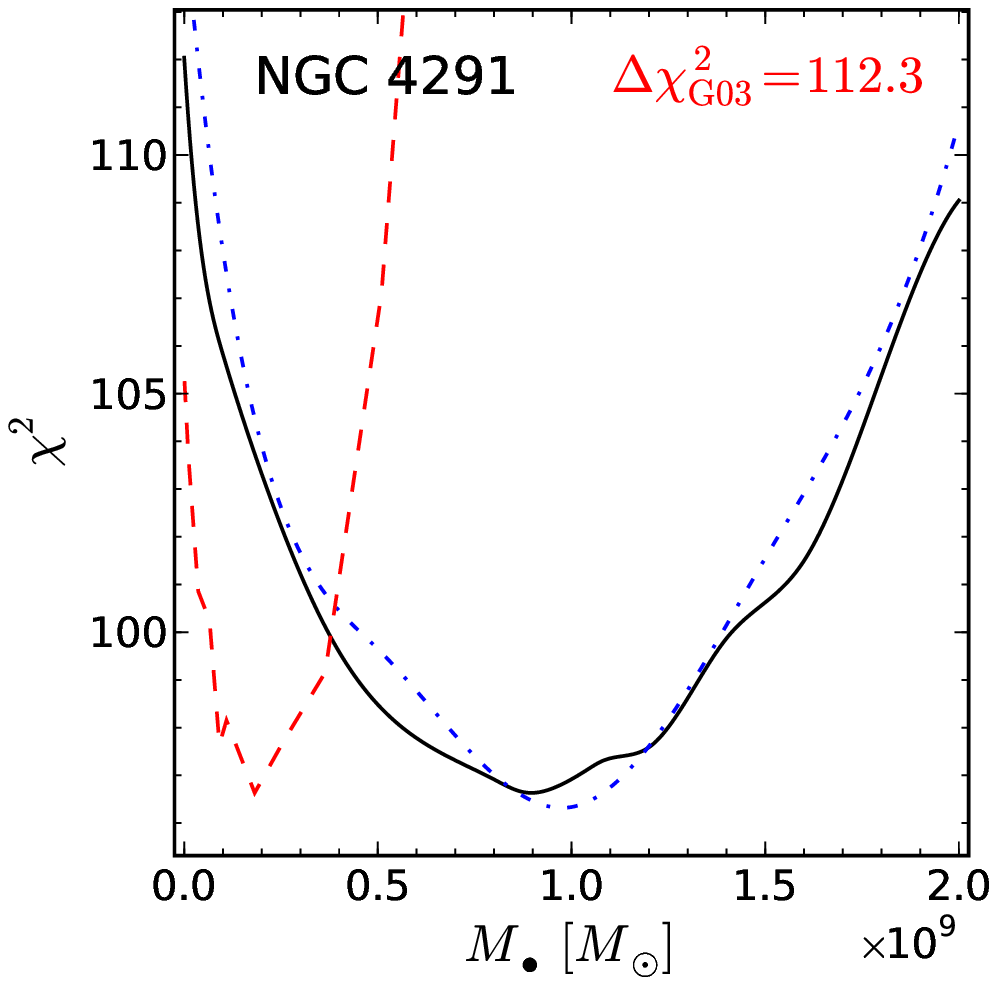}}

\put(0,52){\includegraphics[width=\wi]{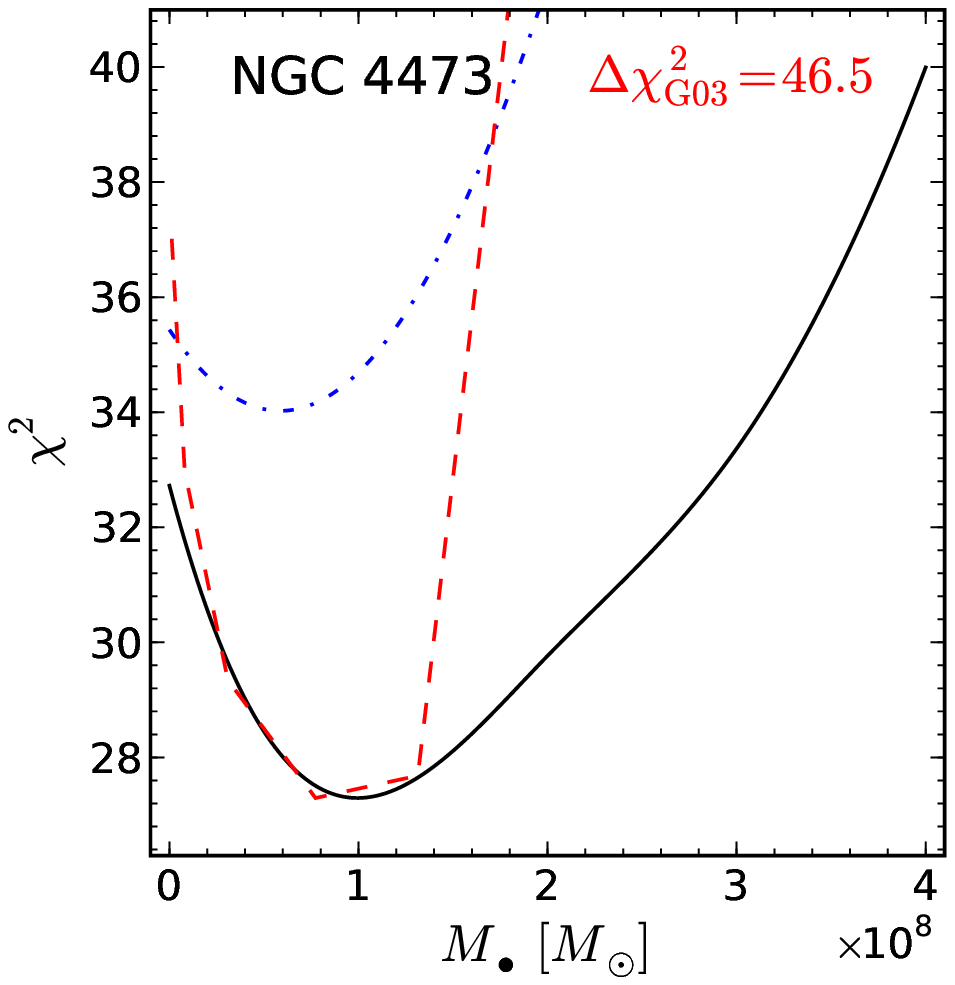}}
\put(52,52){\includegraphics[width=\wi]{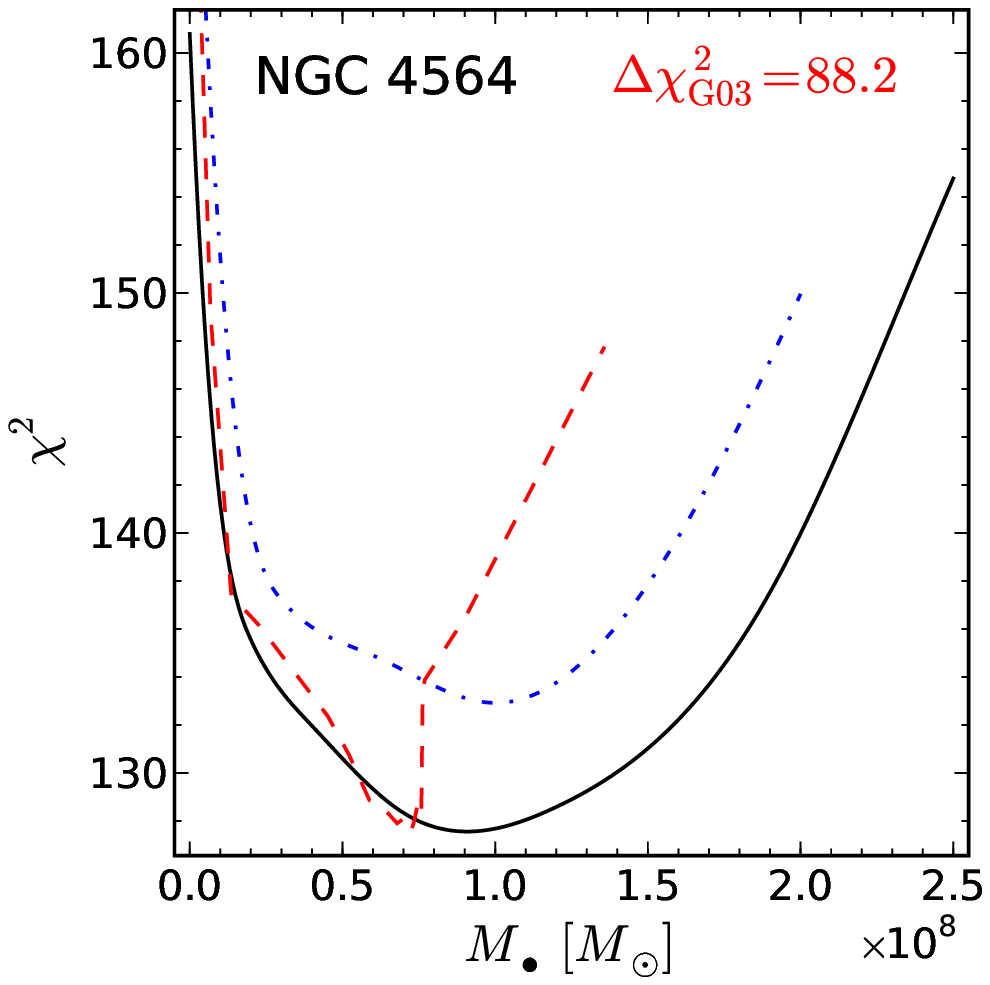}}
\put(104,52){\includegraphics[width=\wi]{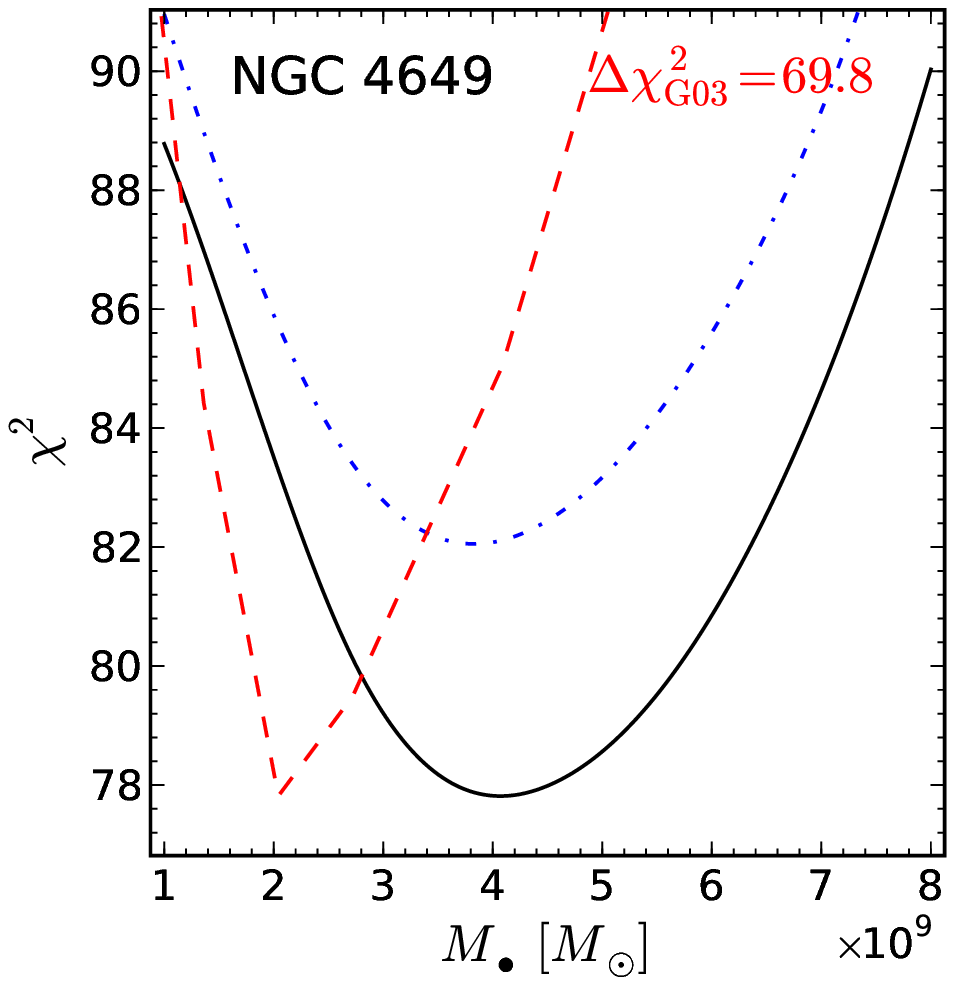}}

\put(0,0){\includegraphics[width=\wi]{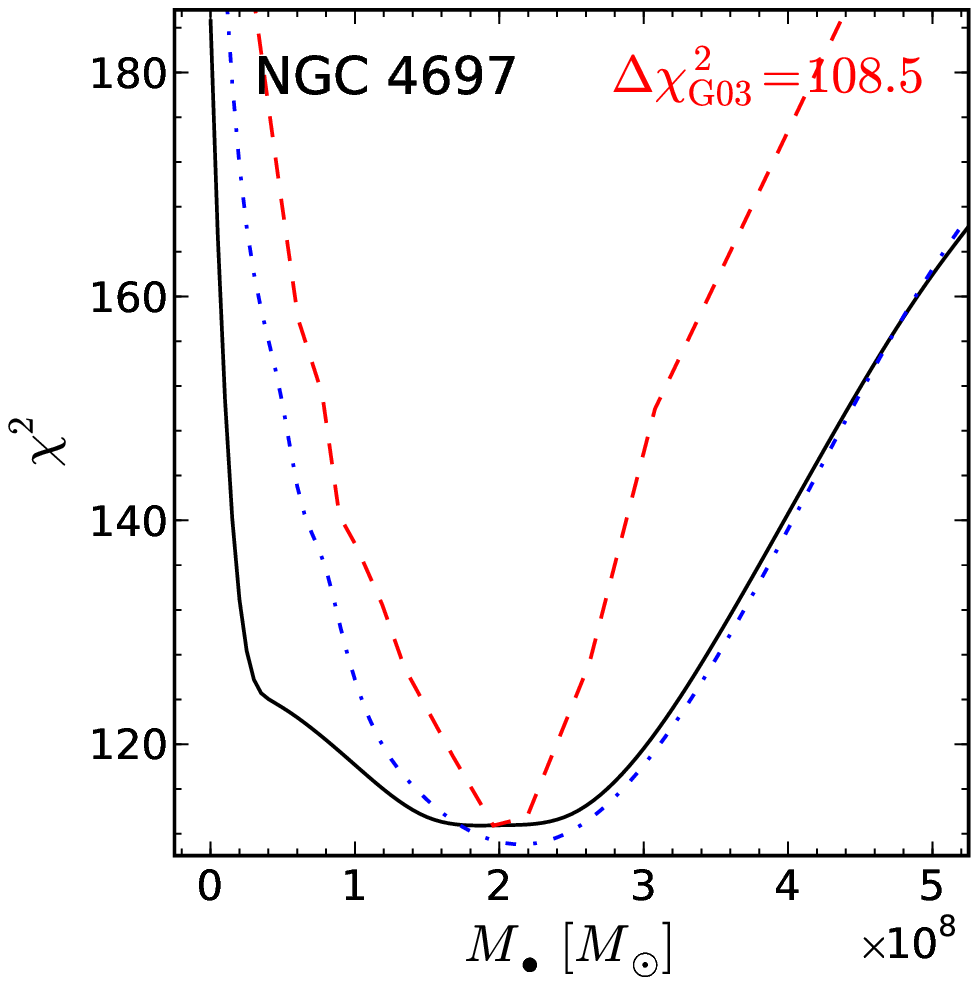}}
\put(52,0){\includegraphics[width=\wi]{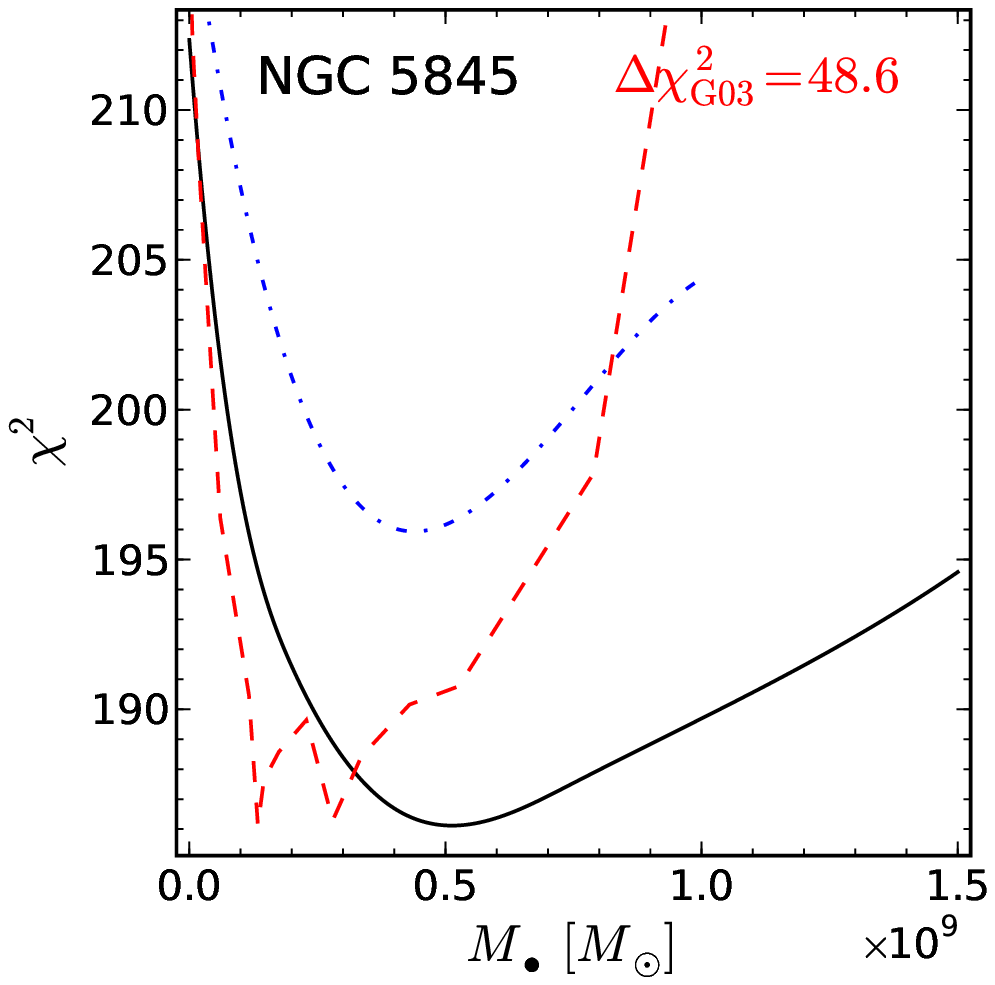}}
\put(104,0){\includegraphics[width=\wi]{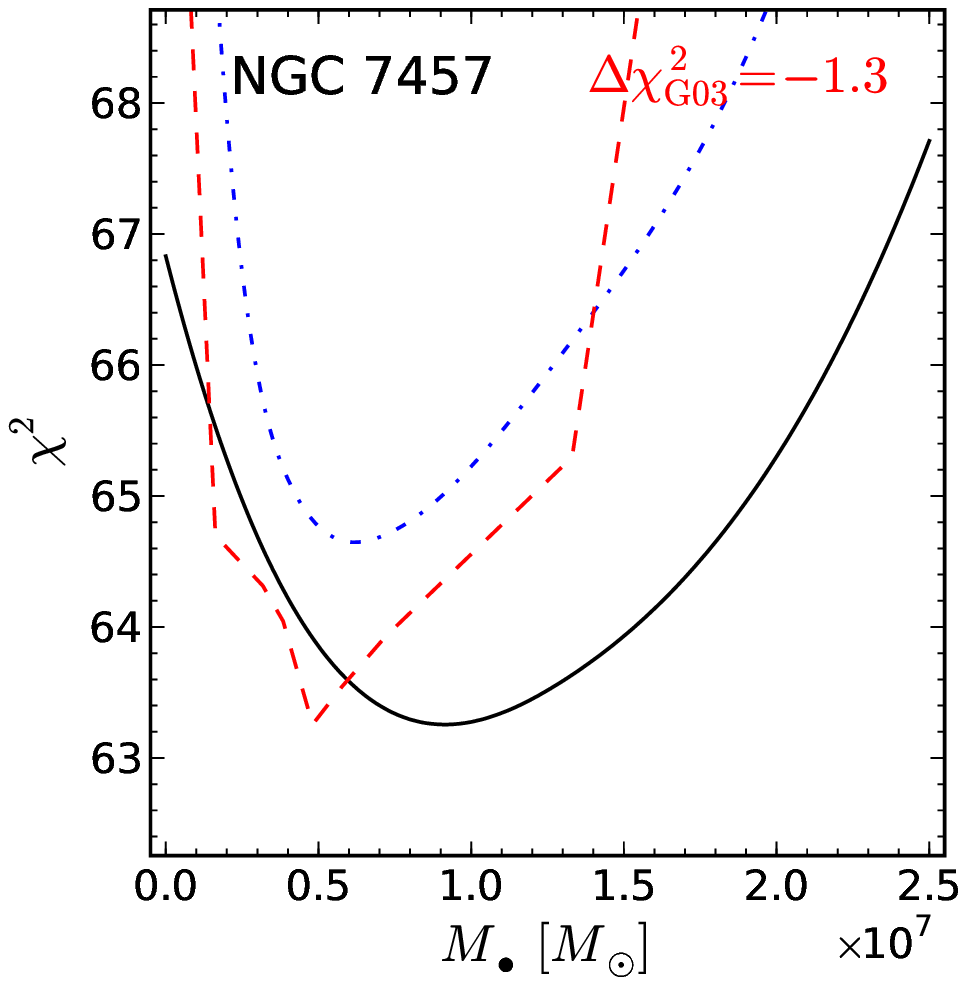}}
\end{picture}
\caption{Comparison of the $\chi^2$ distributions as a function of $M_\bullet$ (marginalized over $M/L$). The black solid line shows the models including a DM halo. Their $\chi^2$ values always show the actual modeling result. The $\chi^2$ distribution for the models without a DM halo is shown as the blue dashed-dotted line. The zero point has been shifted for NGC~2778 and NGC~3384 by an offset given in the figure as $\Delta \chi^2$ (in blue). The $\chi^2$ distribution of G03 is shown as the red dashed line, offset by the value given as $\Delta \chi^2_\mathrm{G03}$ (in red). The distributions have been scaled in $M_\bullet$, to account for the difference in the assumed distance.}
\label{fig:cmpchi}
\end{figure*}

The $\chi^2$ distribution as a function of $M_\bullet$ (marginalized over $M/L$) is shown as the blue line in Figure~\ref{fig:cmpchi}. We determined our stated best fit $M_\bullet$ and $M/L$ from their marginalized $\chi^2$ distributions, using the mid-point of the $\Delta \chi^2=1$ interval, which corresponds to a $1\sigma$ uncertainty. The results are presented in Table~\ref{tab:res}.

\begin{figure*}
\centering
\setlength{\unitlength}{1mm}
\begin{picture}(156,210)
\put(0,156){\includegraphics[width=\wi]{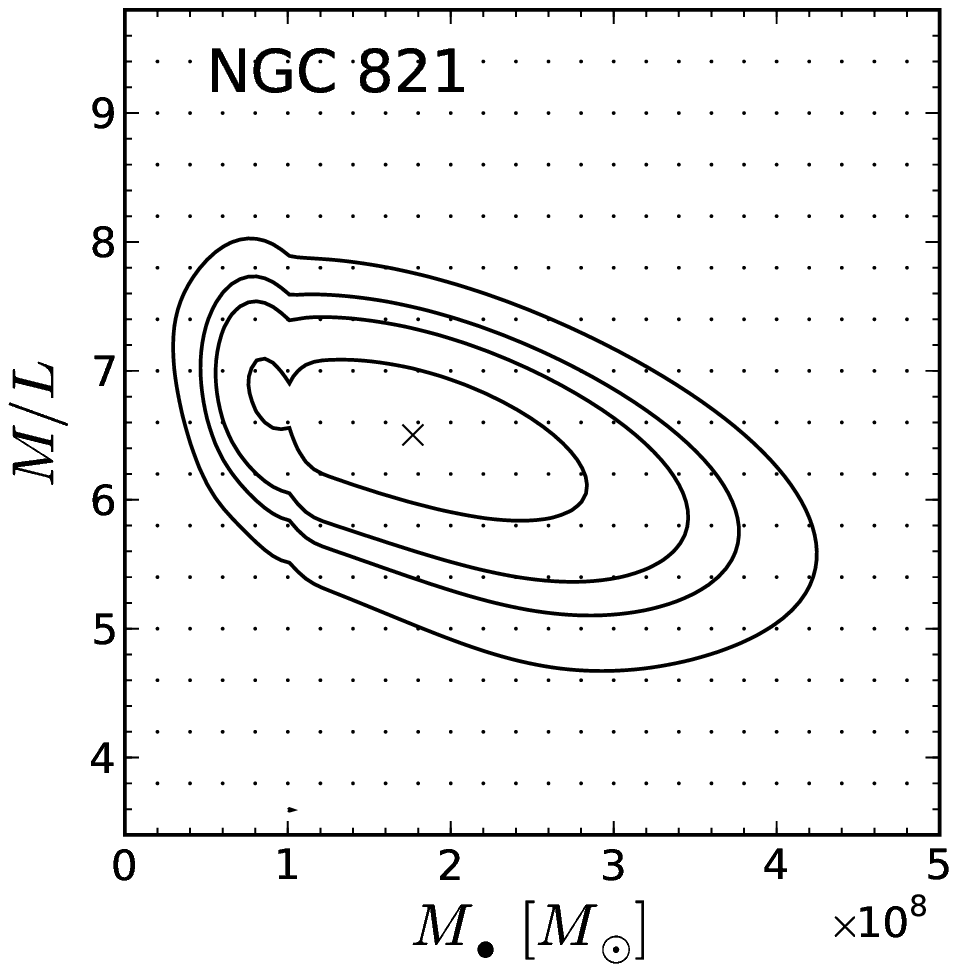}}
\put(52,156){\includegraphics[width=\wi]{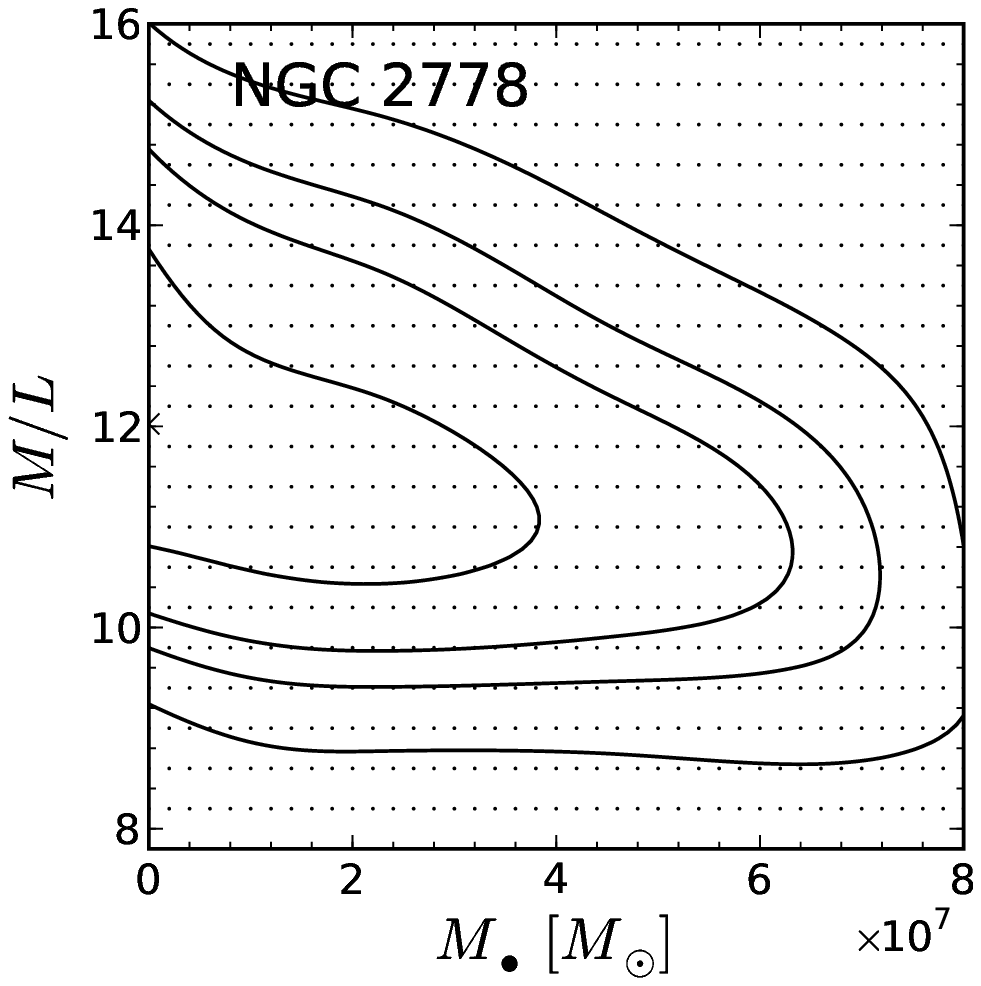}}
\put(104,156){\includegraphics[width=\wi]{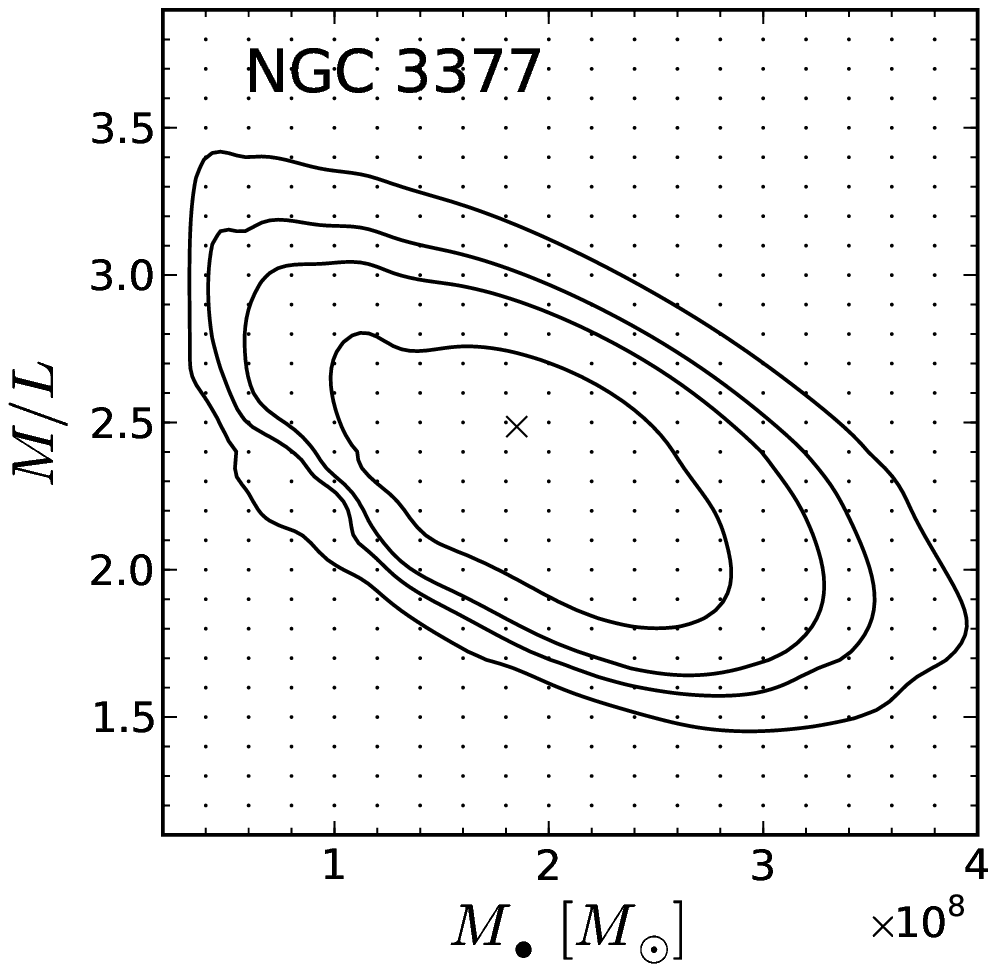}}

\put(0,104){\includegraphics[width=\wi]{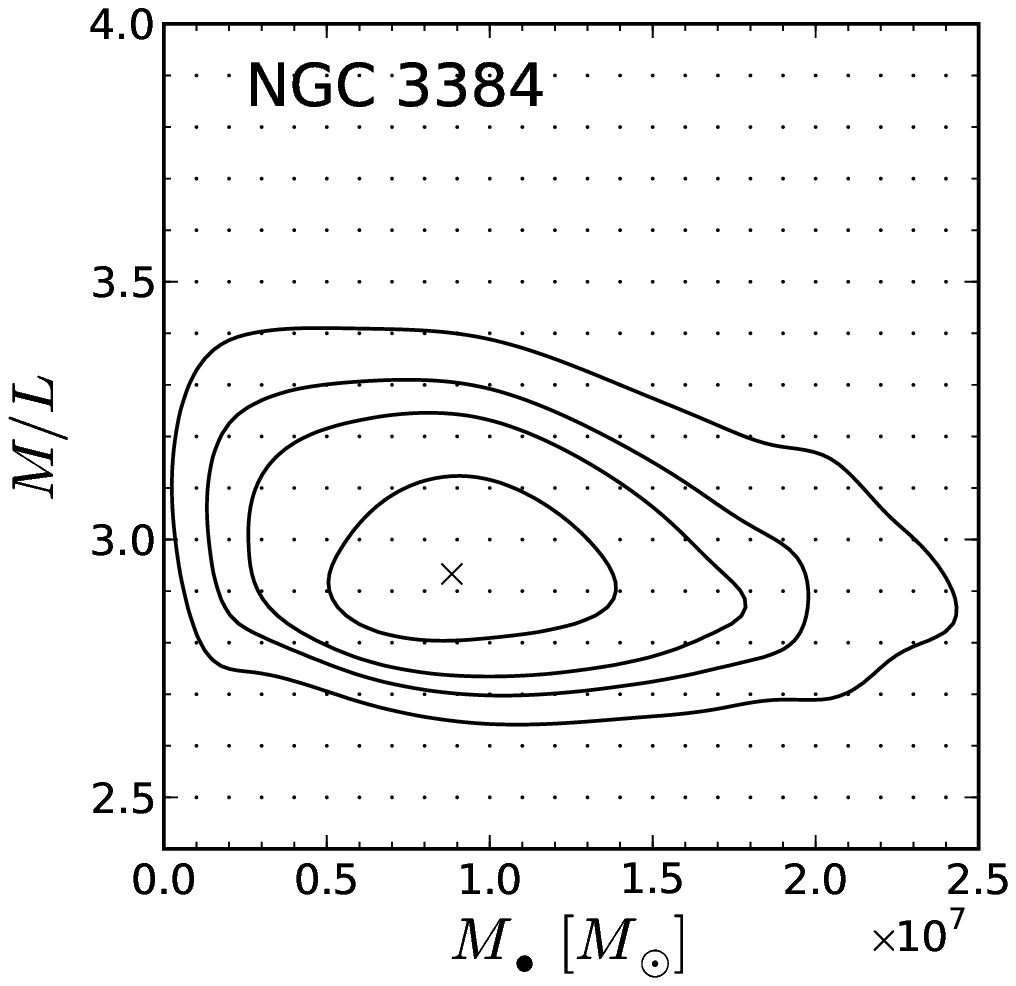}}
\put(52,104){\includegraphics[width=\wi]{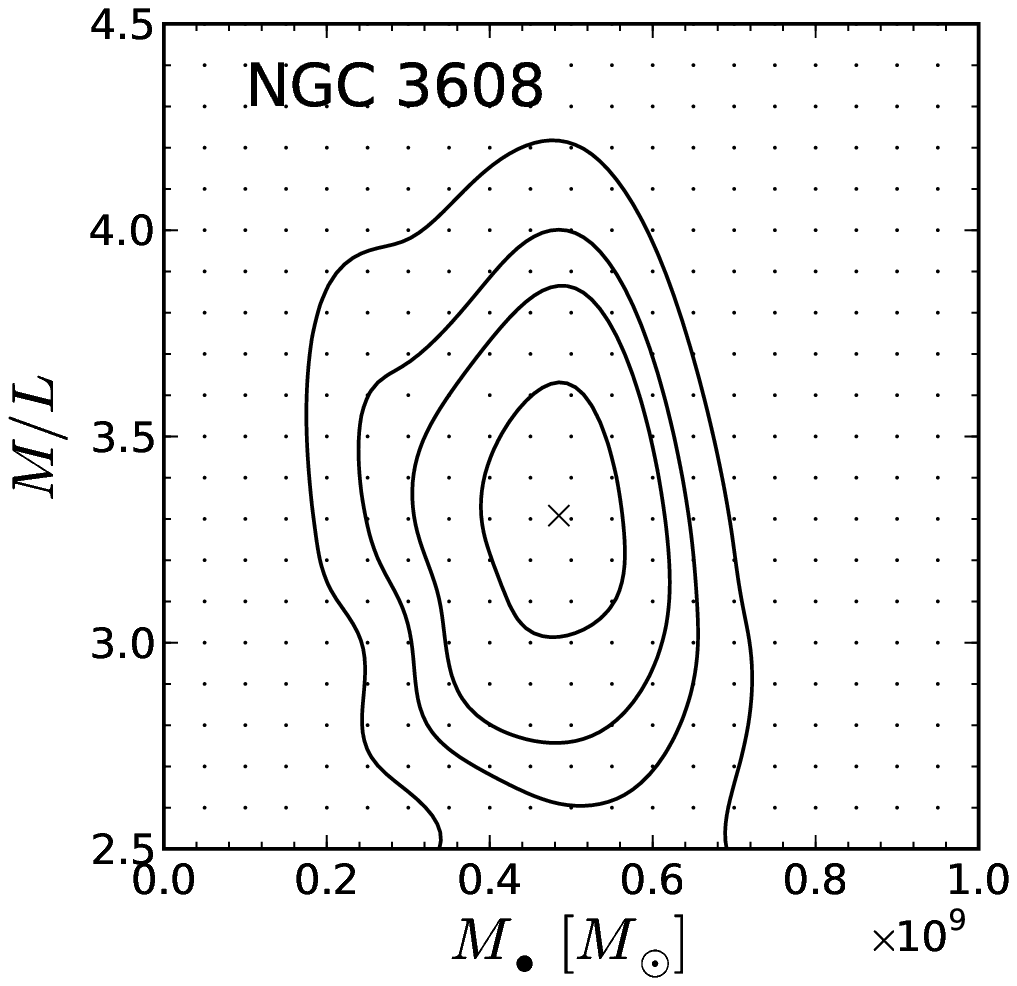}}
\put(104,104){\includegraphics[width=\wi]{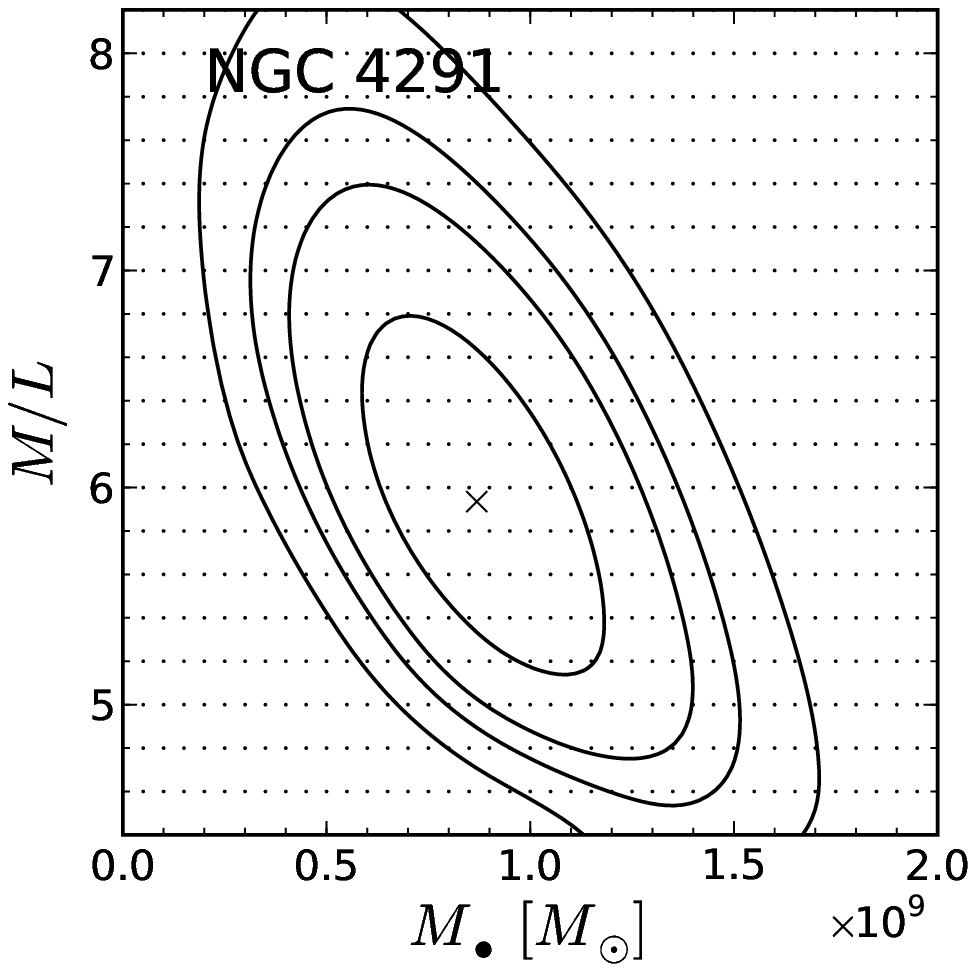}}

\put(0,52){\includegraphics[width=\wi]{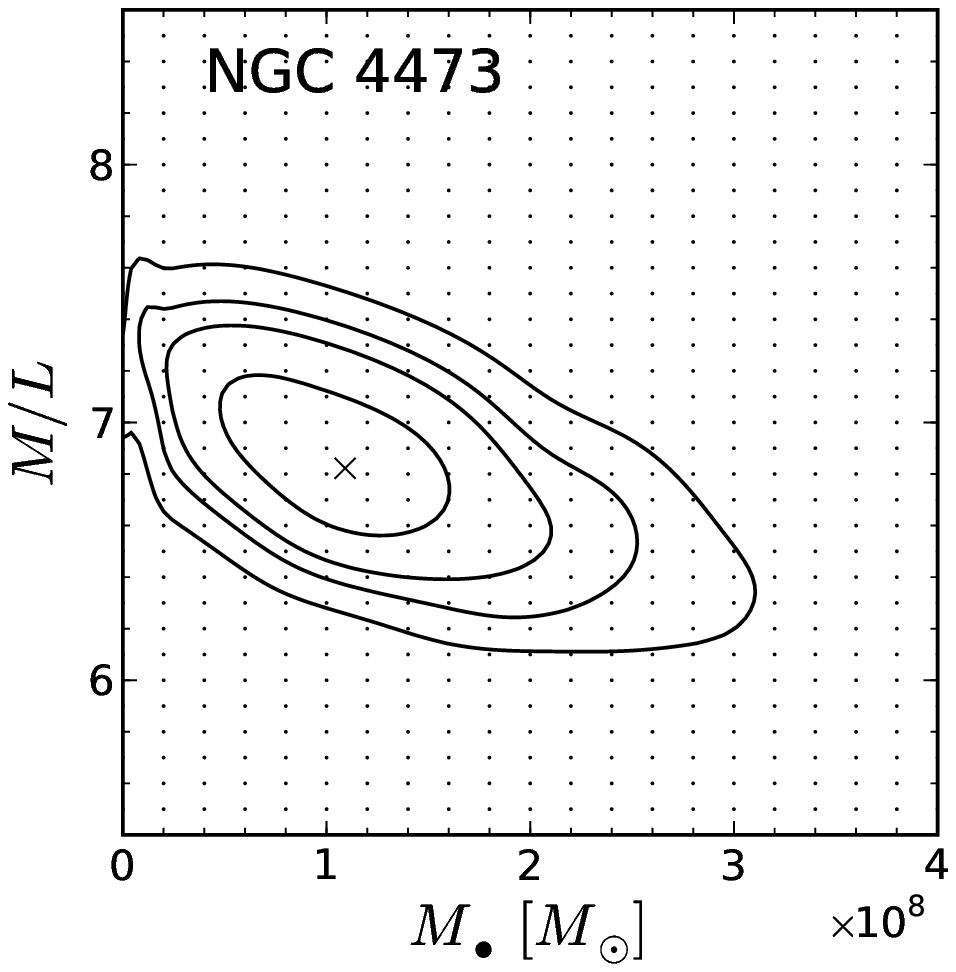}}
\put(52,52){\includegraphics[width=\wi]{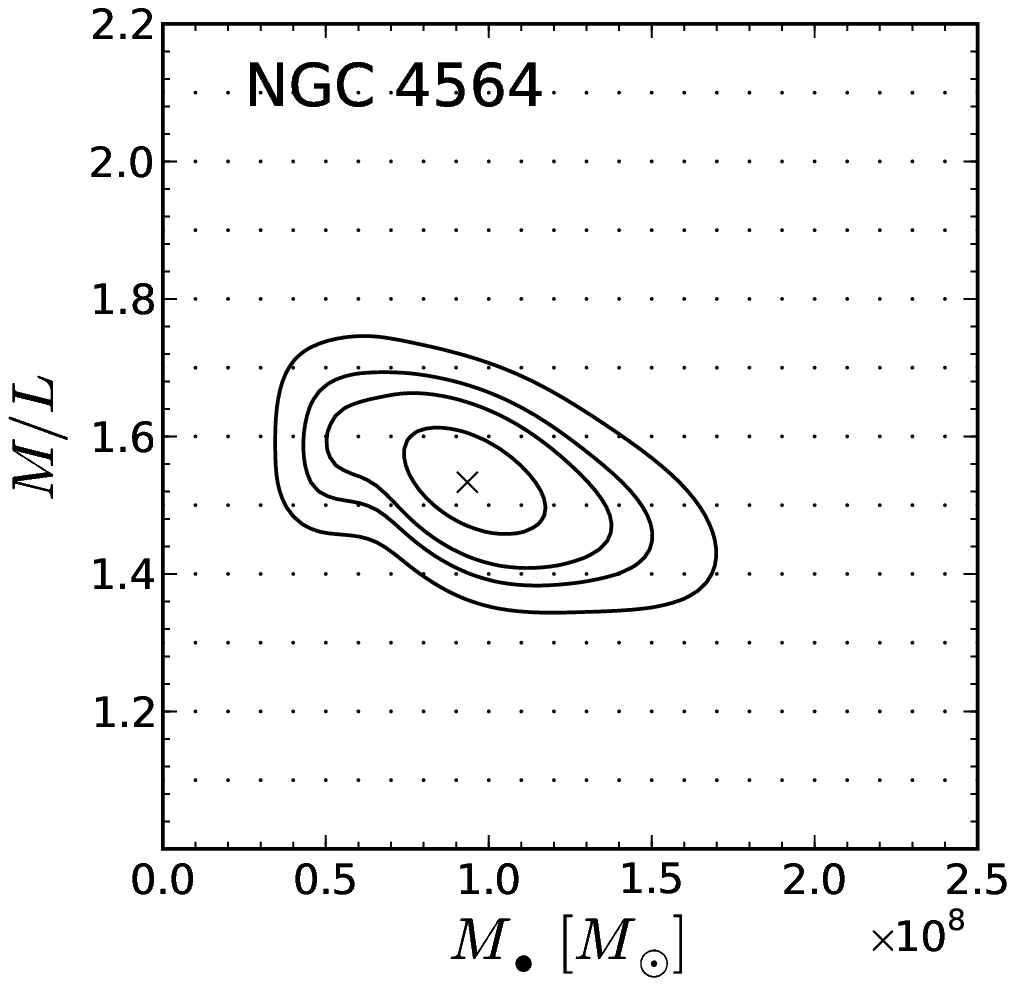}}
\put(104,52){\includegraphics[width=\wi]{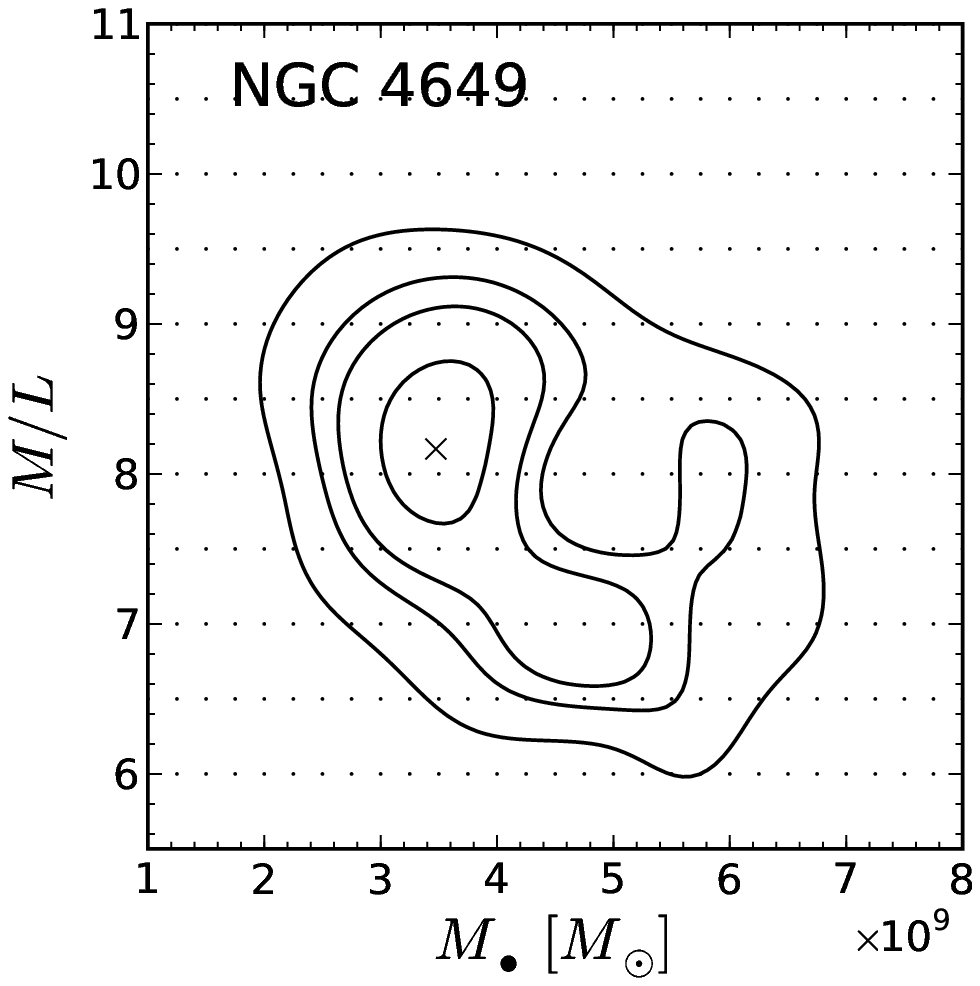}}

\put(0,0){\includegraphics[width=\wi]{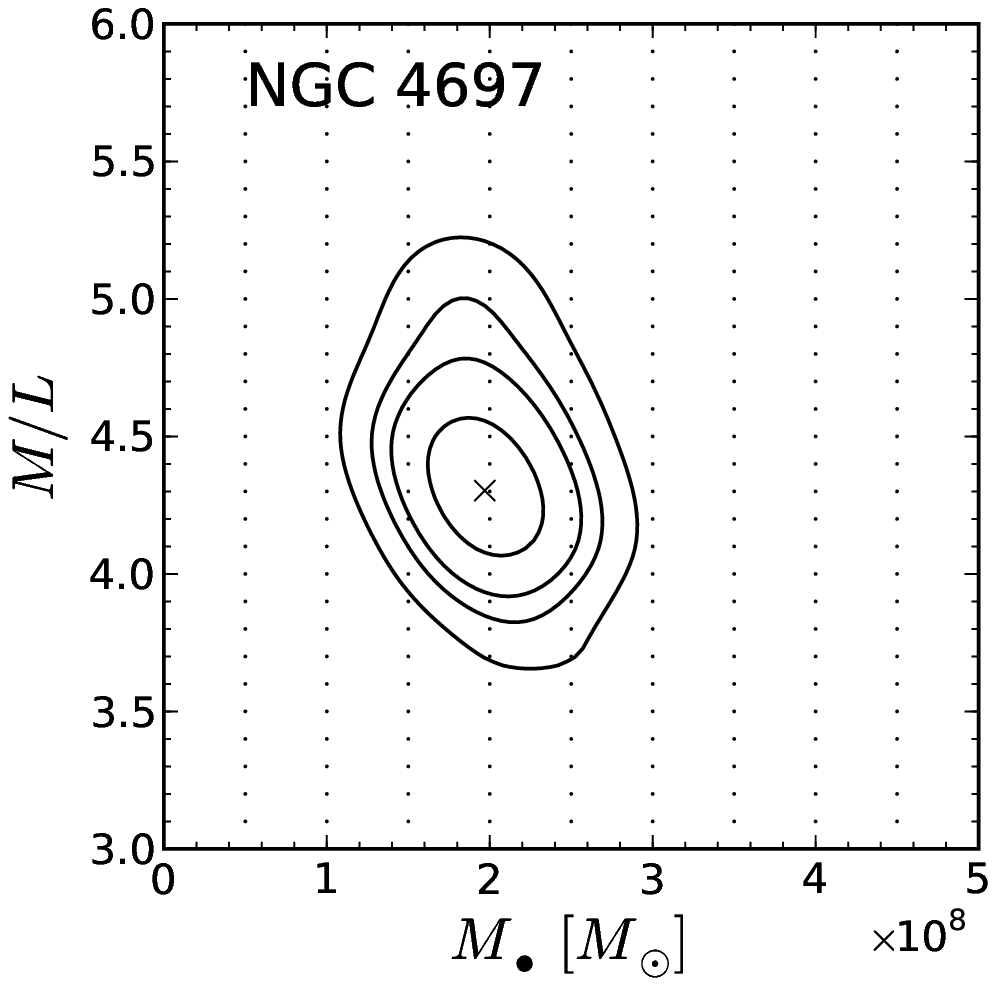}}
\put(52,0){\includegraphics[width=\wi]{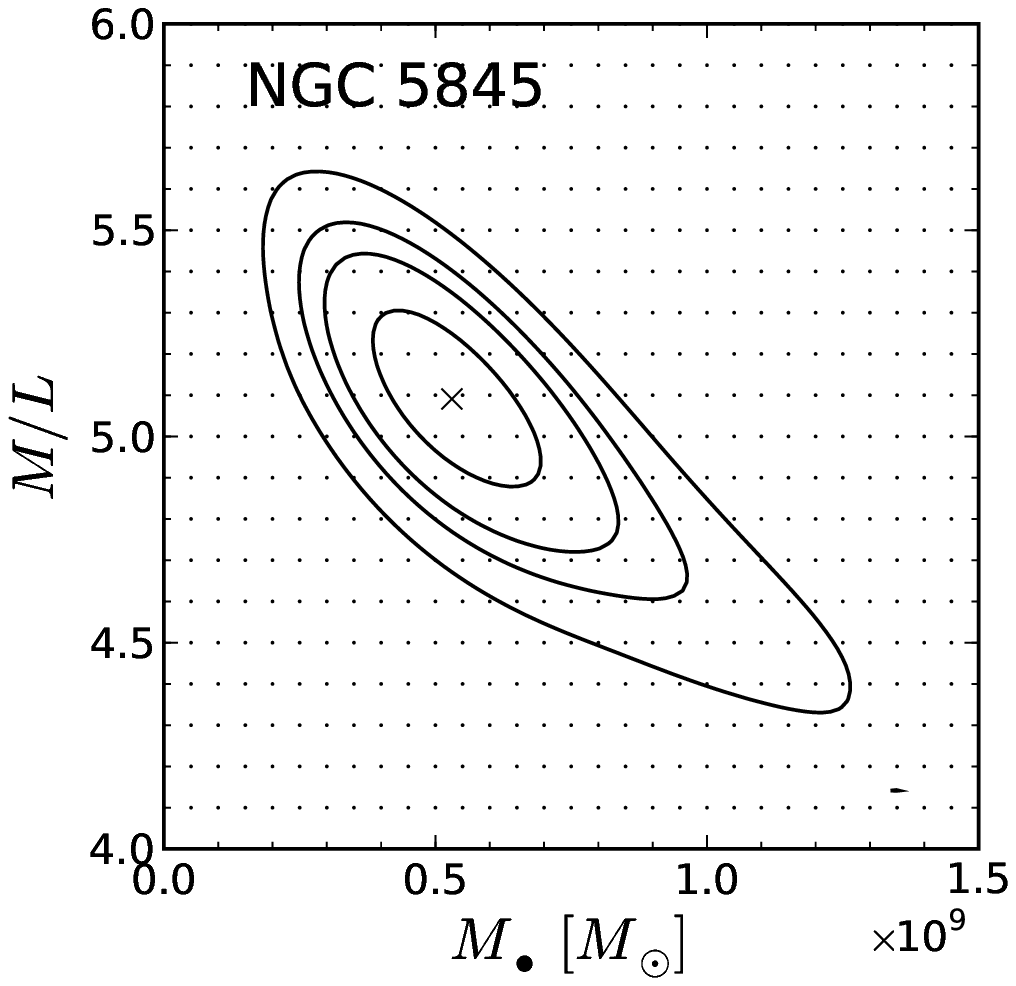}}
\put(104,0){\includegraphics[width=\wi]{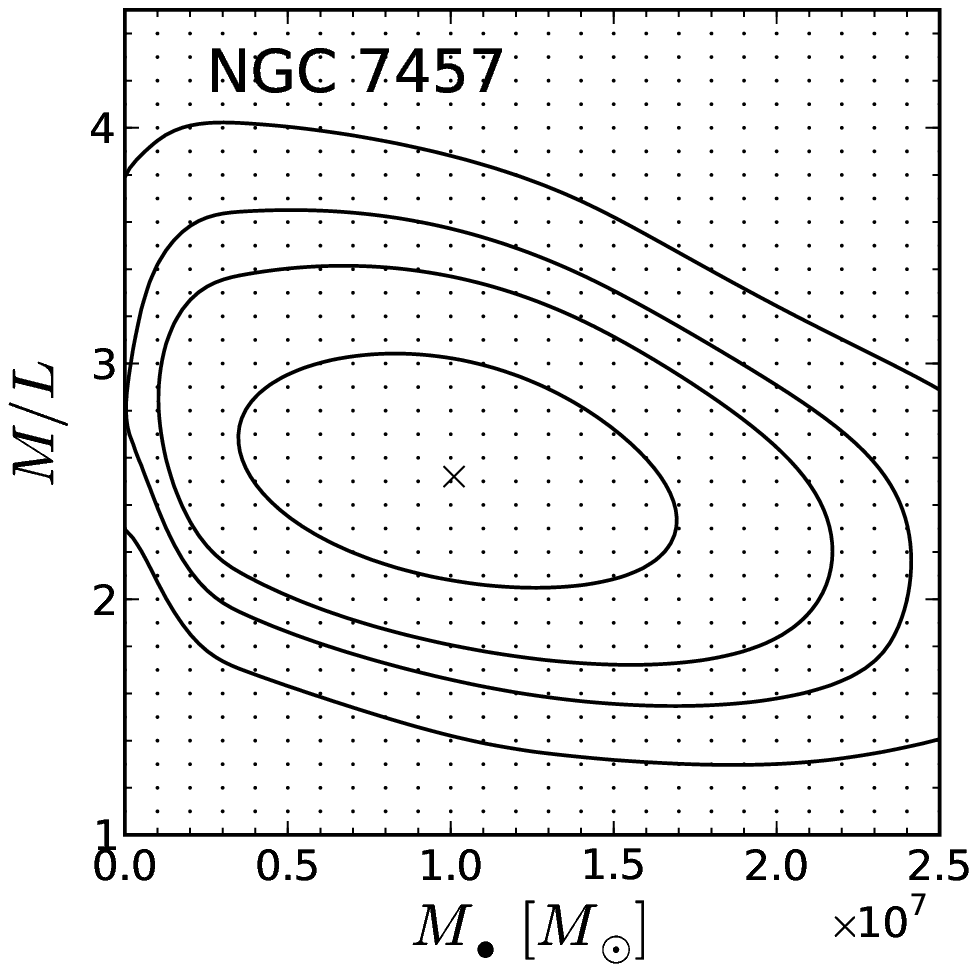}}
\end{picture}
\caption{Two-dimensional contour plot of $\chi^2$ as a function of $M_\bullet$ and $M/L$ for each galaxy. Here the models with a DM halo included are shown. The contour lines show $\Delta \chi^2$ values of $1.0$, $2.71$, $4.0$, and $6.63$ (corresponding to 68\%, 90\%, 95\%, and 99\% confidence for one degree of freedom). The points indicate the individual models we ran. The contours are derived from a smoothing spline to these models. The cross represents the best fit.}
\label{fig:condm}
\end{figure*}

\subsection{Models with a DM halo} \label{sec:dm}
It has been shown that dynamical models are clearly able to detect and constrain the presence of a DM halo, if the data range sufficiently far in radius \citep{Rix:1997,Kronawitter:2000,Thomas:2007,Weijmans:2009,Forestell:2010}. Due to the faintness of the stellar component at large radii, other kinematic tracers such as globular clusters \citep{Zepf:2000,Pierce:2006,Bridges:2006} or planetary nebulae \citep{Mendez:2001,Romanowsky:2003,Coccato:2009} have to be used.
Furthermore, if dynamical coverage of both the central and the outer regions of the galaxy is present, it is possible to constrain $M_\bullet$, $M/L$, and the dark halo parameters by dynamical modeling \citep{Gebhardt:2009,Shen:2010}.

Two common parameterizations for the DM halo are a Navarro-Frenk-White (NFW) profile \citep{Navarro:1996} and a DM distribution based on a cored logarithmic potential \citep{Binney:1987,Thomas:2005}. For a sample of 17 early-type galaxies, \citet{Thomas:2007} found both profiles to give consistent results, with tentative evidence to favor a logarithmic dark halo. \citet{Gebhardt:2009} confirmed this result for M87, and \citet{McConnell:2010} found consistent results for $M_\bullet$ using either a logarithmic dark halo or an NFW profile. In the following, we will use a DM halo with a cored logarithmic potential, whose density profile is given by
\begin{equation}
 \rho_\mathrm{DM}(r) = \frac{V_c^2}{4\pi G}\frac{3r_c^2 +r^2}{(r_c^2+r^2)^2} \ ,
\end{equation} 
where $V_c$ is the asymptotically constant circular velocity and $r_c$ is the core radius, within which the DM density is approximately constant.

Our data in general do not constrain the DM profile, as we are lacking kinematic information at large radii. While for a few galaxies in our sample, large radii kinematic information for the stars, globular clusters or planetary nebulae exist in the literature, we do not include them in this analysis. In this work, we are not aiming at constraining the DM halo itself, but we are mainly interested in the effect of including such a halo for the recovered black hole mass. We leave a more detailed investigation of the combined DM halo and black hole properties for these individual galaxies to future work. This also allows a better direct comparison to the work of G03 and the models without a DM halo, presented in the previous section.

Therefore, we assume a fixed DM halo, with fixed parameters $V_c$ and $r_c$. These are taken from the scaling relations presented by \citet{Thomas:2009}, based on the galaxy luminosity:
\begin{equation}
 \log r_c = 1.54 + 0.63 ( \log (L_B/L_\odot) -11)
\end{equation} 
\begin{equation}
 \log V_c = 2.78 + 0.21 ( \log (L_B/L_\odot) -11) \ ,
\end{equation} 
and given in Table~\ref{tab:res}.
These scaling relations have been established based on a sample of 12 early-type galaxies in the Coma cluster with old stellar populations. While our sample does not have to follow these scaling relationships exactly, they at least provide well motivated parameters for the DM halo. Younger early-type galaxies and disk galaxies have been found to have on average a less massive halo, thus our approach tends to maximize the DM contribution. We investigate the effect of changing the assumed DM halo on the central black hole mass further below. 

Thus, for each galaxy, we ran a grid of models for varying $M_\bullet$ and $M/L$ with fixed DM halo.
We show the two-dimensional distribution of $\chi^2$ as a function of $M_\bullet$ and $M/L$ in Figure~\ref{fig:condm}.  The contours are based on the $\chi^2$ values of the underlying grid points, applying a two-dimensional smoothing spline \citep{Dierckx:1993}. The marginalized $\chi^2$ distribution as a function of $M_\bullet$ is shown as the black line in Figure~\ref{fig:cmpchi}. We again determined the best-fit values for $M_\bullet$ and $M/L$ from the marginalized distribution and have given them in Table~\ref{tab:res}.

\section{Comparison of black hole masses}
\subsection{Comparison with \citet{Gebhardt:2003}}
As we are using the same data as those in G03, the only difference between the work presented in Section~\ref{sec:nodm} and in G03 is the improved modeling code. Thus, we would expect to recover the same black hole masses as in G03.

For a comparison with G03, their masses are first increased by a factor of 1.09, due to a unit conversion error \citep{Siopis:2009}, and then are rescaled, according to the difference in the adopted distance, assuming $M_\bullet \propto d$. These masses are listed in \citet{Gultekin:2009}, apart from NGC~821. This galaxy has an erroneous black hole mass in G03, corrected in \citet{Richstone:2004}. After accounting for the factor of 1.09 and the distance difference, the black hole mass for NGC~821 is $M_\bullet=9.9\times 10^7\, M_\odot$.

\begin{figure}
\centering
\resizebox{\hsize}{!}{\includegraphics[clip]{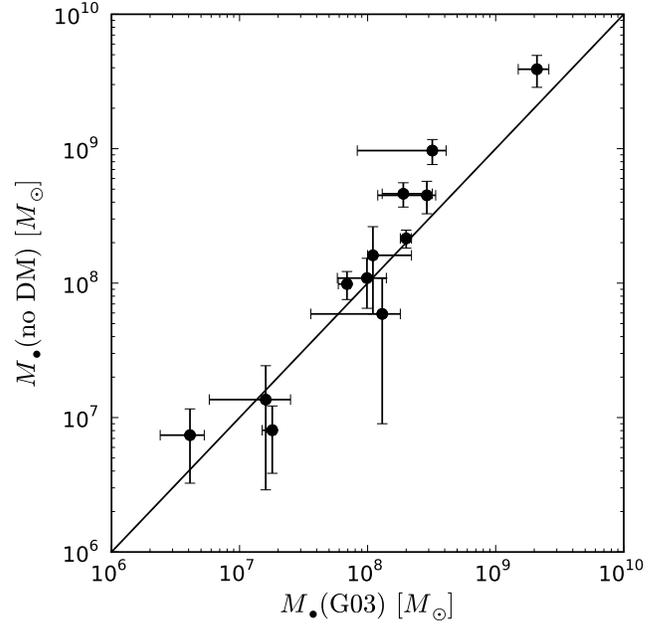}}
\caption{Black hole masses given by \citet{Gebhardt:2003} vs. the black hole masses determined in this work (without a dark halo). The solid line shows a one-to-one correspondence.}
\label{fig:cmpmass}
\end{figure}

\begin{figure}
\centering
\resizebox{\hsize}{!}{\includegraphics[clip]{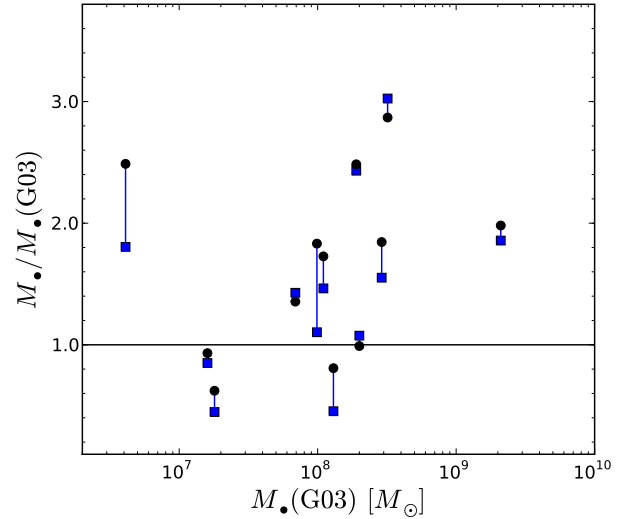}}
\caption{Ratio between the $M_\bullet$ determined in this work and the $M_\bullet$ given in \citet{Gebhardt:2003}, as a function of the G03 $M_\bullet$. The blue squares are for the models without a dark halo, while the black circles correspond to the result, when a dark halo is included in the models. The solid line is a one-to-one correspondence to the G03 values.}
\label{fig:deltamass}
\end{figure}

In Figure~\ref{fig:cmpmass} we compare the black hole masses, determined without including a DM halo, with the black hole masses given in G03. In Figure~\ref{fig:deltamass} we show as blue squares the ratio between both mass determinations as a function of the G03 mass. The marginalized $\chi^2$ distributions for the individual objects are shown in Figure~\ref{fig:cmpchi}, as blue dashed-dotted lines for the current work masses and as red dashed lines for the G03 distributions. Note that the G03 distributions are offset in $\chi^2$, such that the minimum corresponds to the minimum of the  $\chi^2$ distribution including a dark halo, shown in black. The reason for the offset in $\chi^2$ is mainly due to the larger number of orbits used in the current modeling, compared to G03.

For three objects (NGC~821, NGC~2778, and NGC~4697) the difference in $M_\bullet$ is less than 20~\%, thus consistent with our previous work. The internal structures of the dynamical models (as discussed below) are similar for these three galaxies in the old and new models, which explains the reason for lack of change. The small difference is probably due to the presence of numerical noise in the models. This noise is mainly caused by the use of a finite number of individual orbits instead of a smooth orbit distribution function. The comparison of the $\chi^2$ distribution for the three objects shows that they are basically consistent, while the distribution may widen, possibly due to a more complete orbit library. Also, for NGC~3377 and NGC~4564 ,the difference in $M_\bullet$ is within the stated uncertainties.

\begin{figure*}
\centering
\setlength{\unitlength}{1mm}
\begin{picture}(156,210)
\put(0,156){\includegraphics[width=\wi]{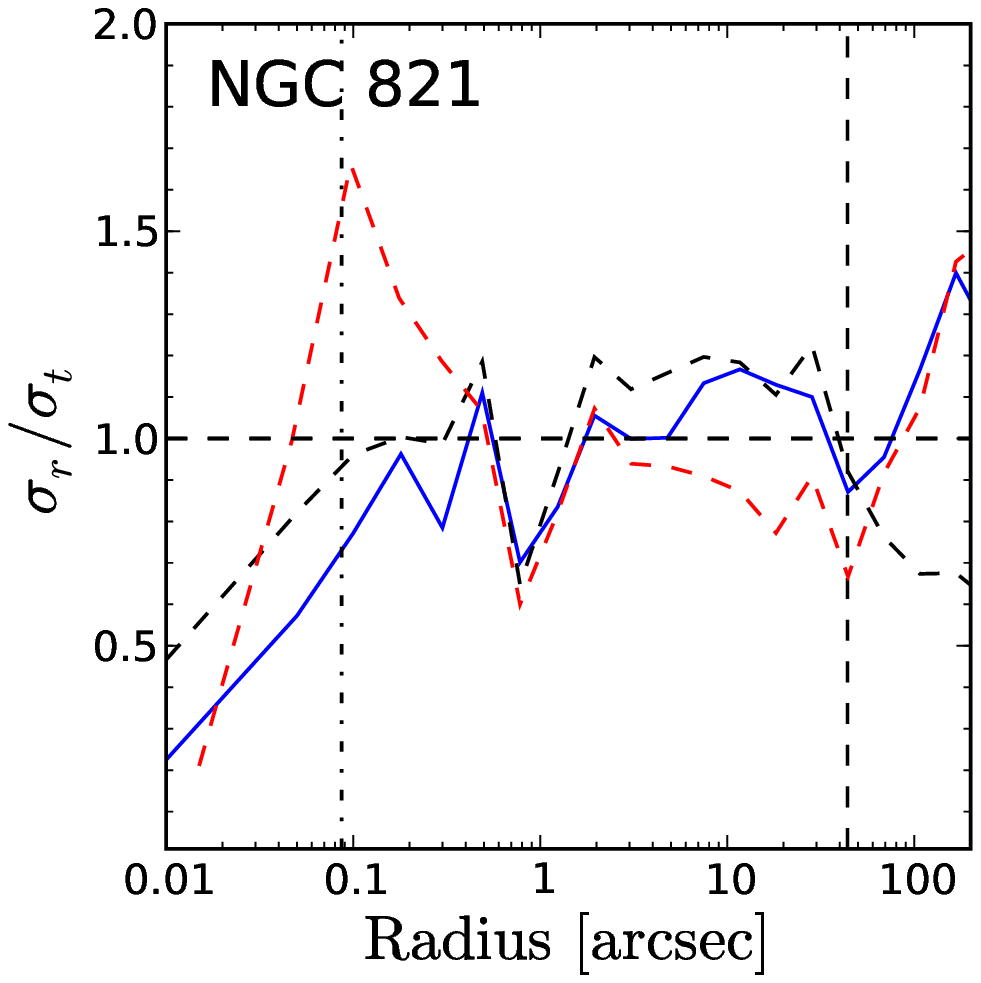}}
\put(52,156){\includegraphics[width=\wi]{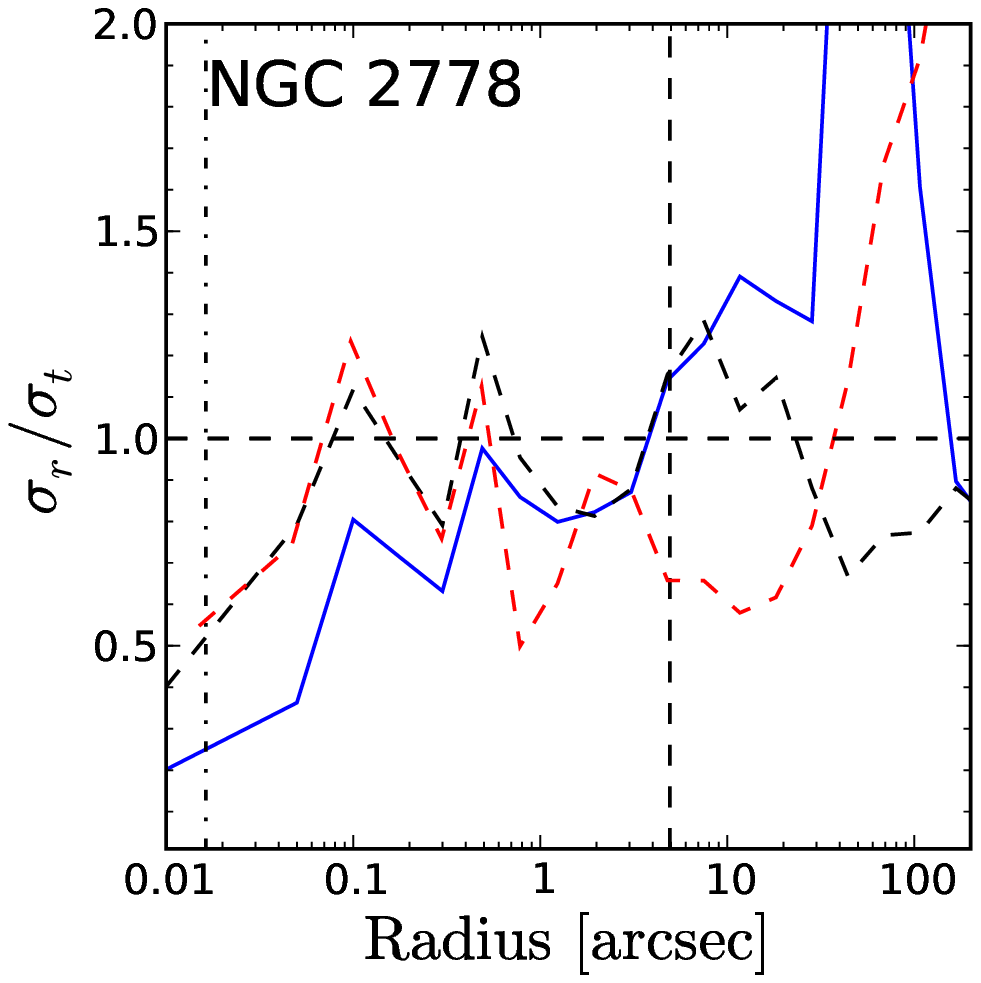}}
\put(104,156){\includegraphics[width=\wi]{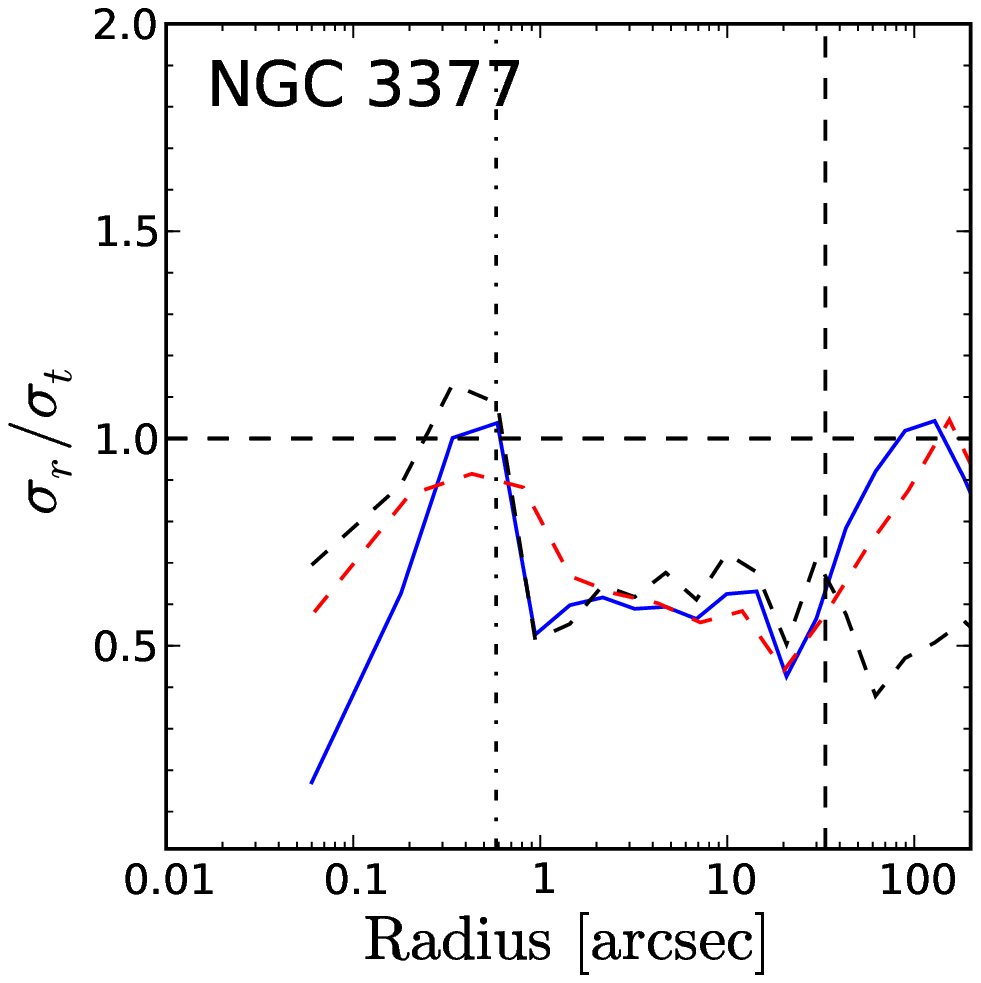}}

\put(0,104){\includegraphics[width=\wi]{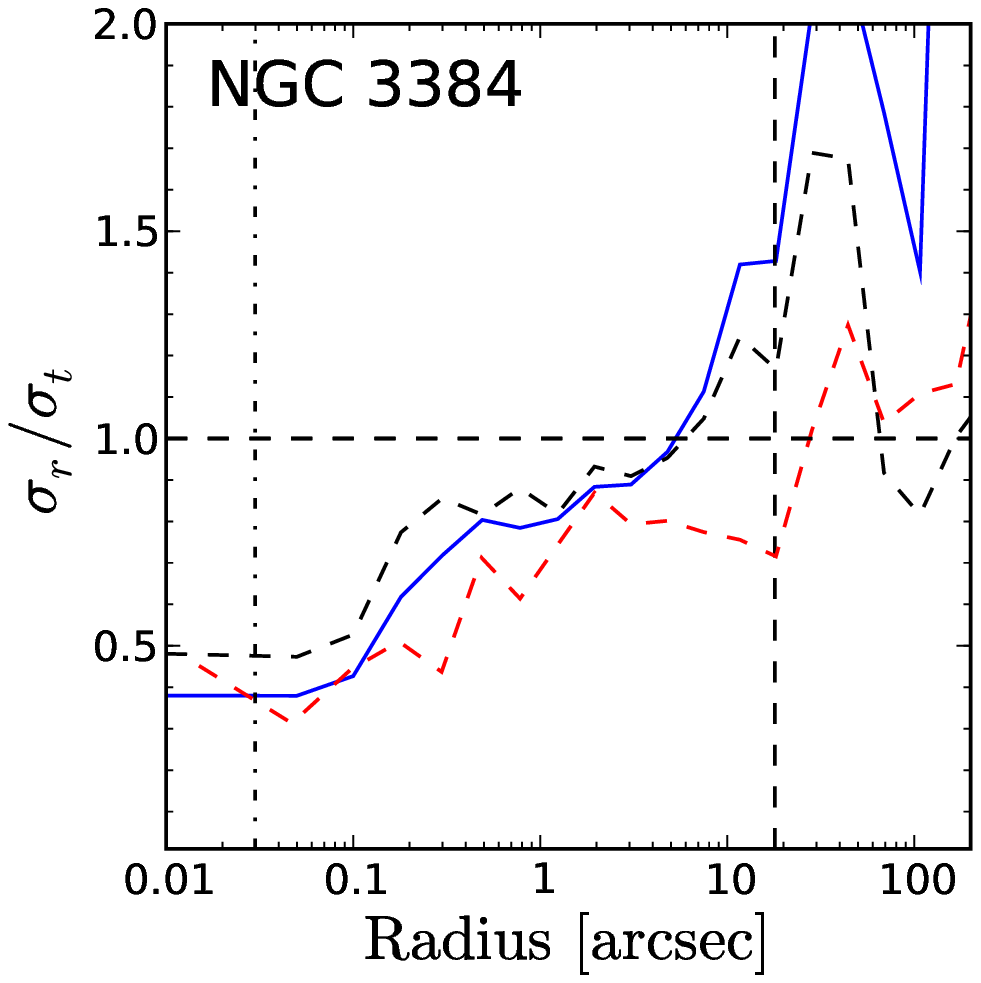}}
\put(52,104){\includegraphics[width=\wi]{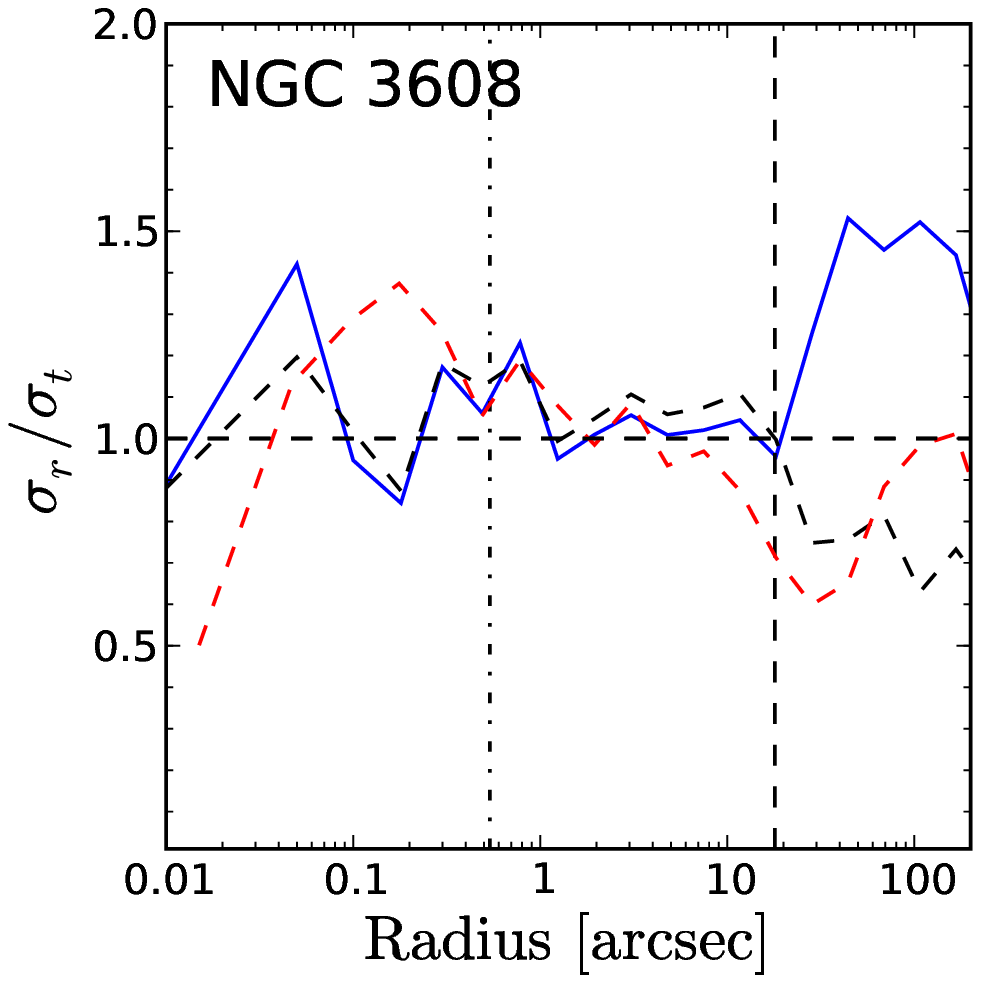}}
\put(104,104){\includegraphics[width=\wi]{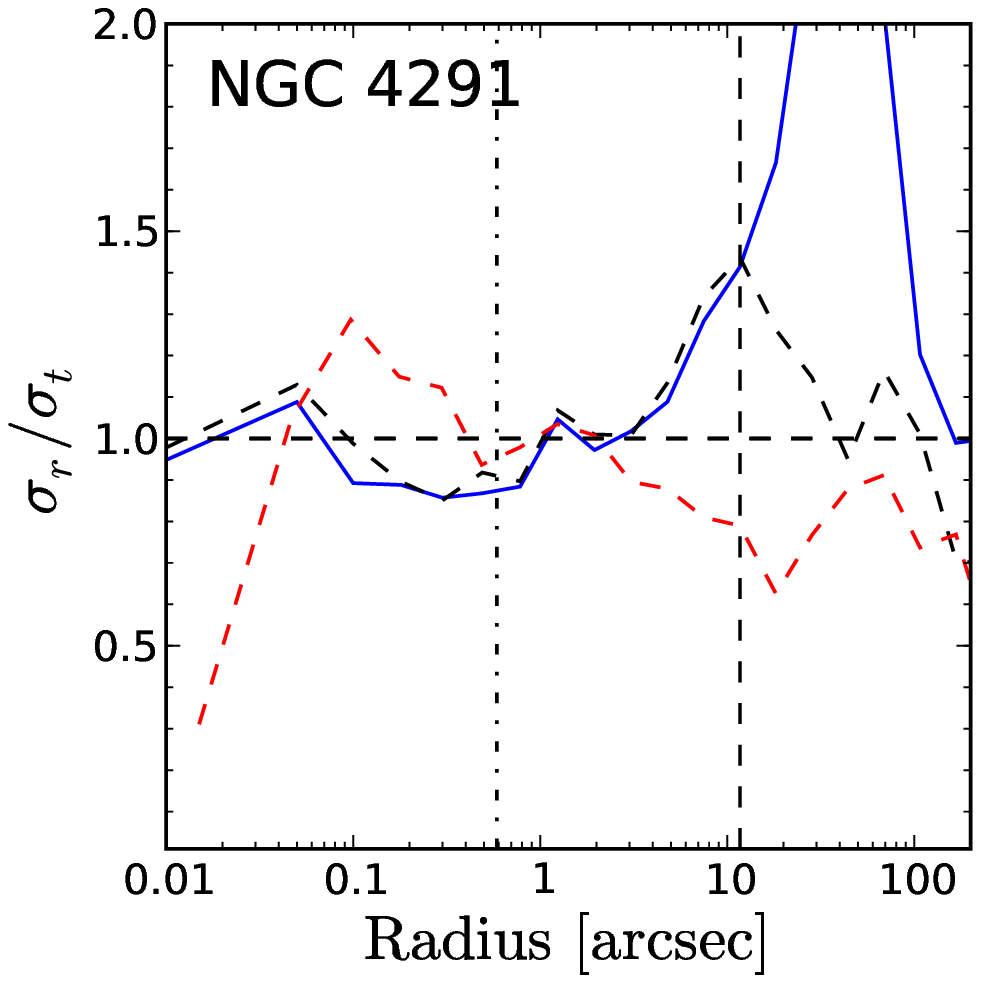}}

\put(0,52){\includegraphics[width=\wi]{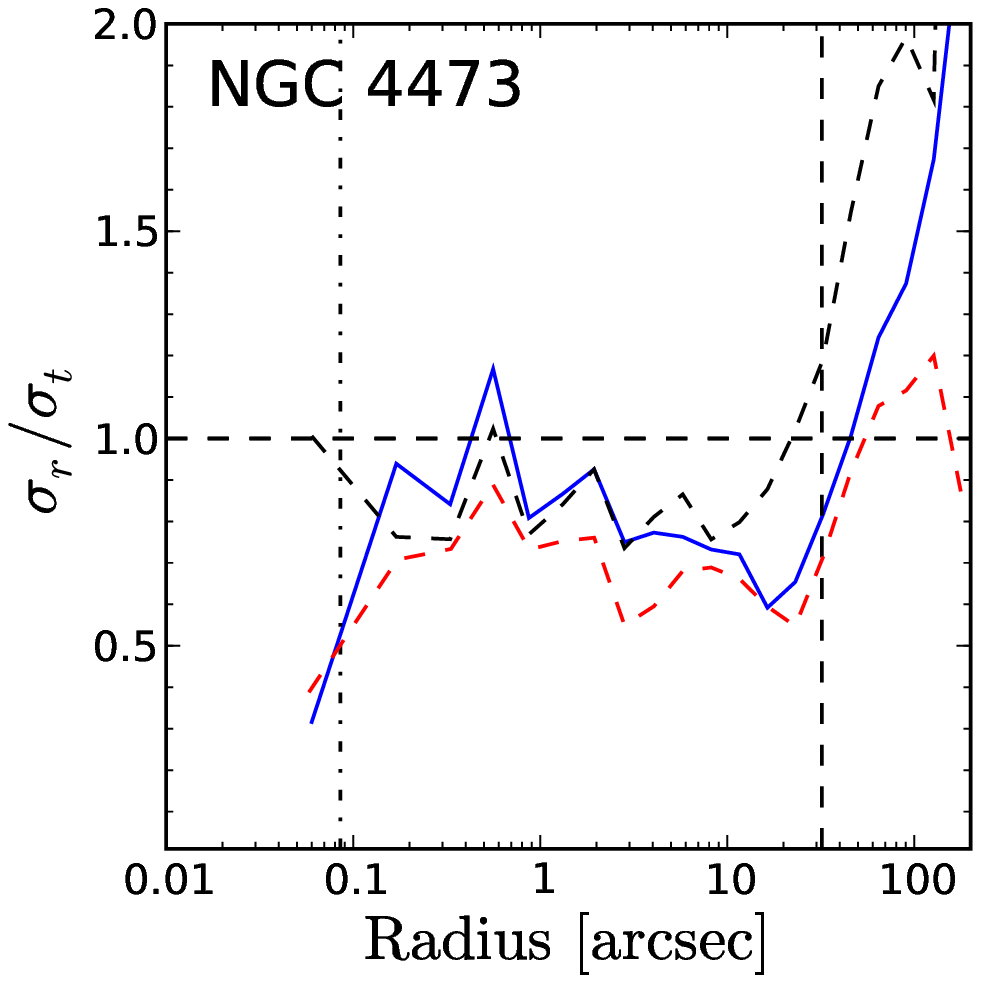}}
\put(52,52){\includegraphics[width=\wi]{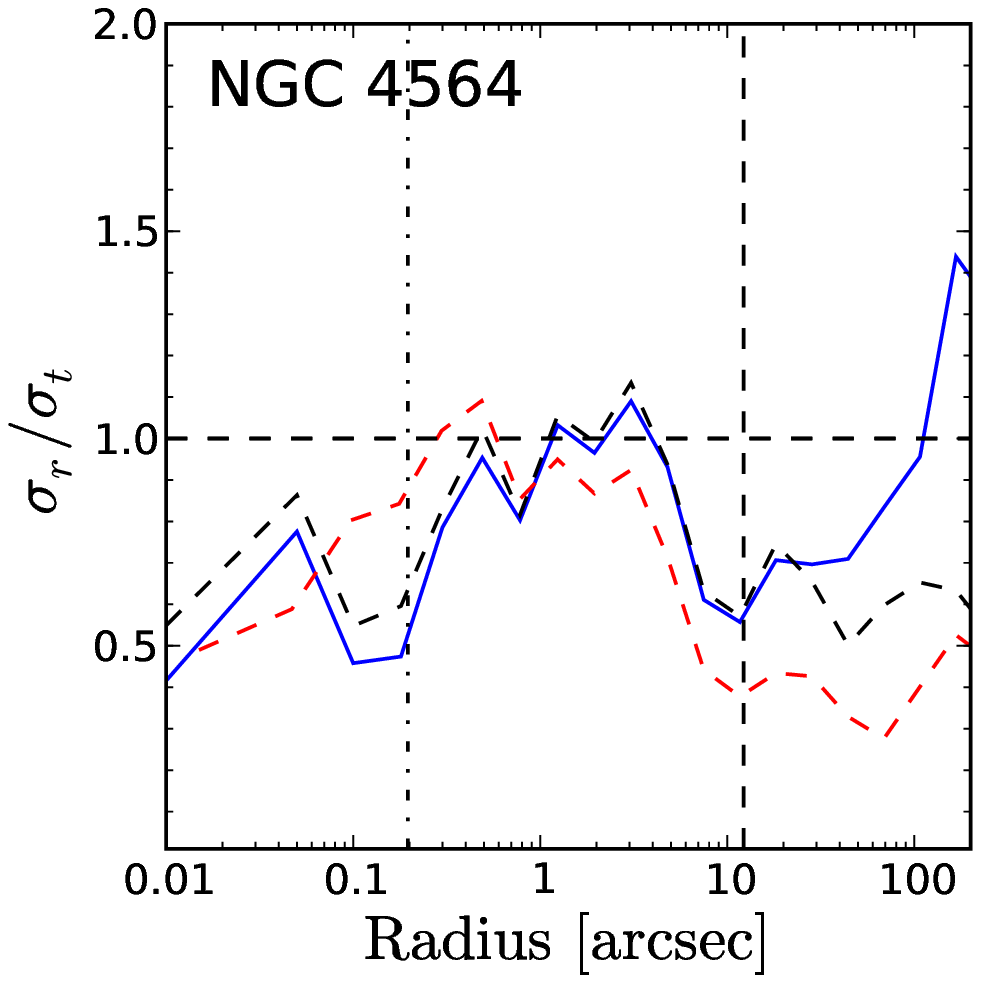}}
\put(104,52){\includegraphics[width=\wi]{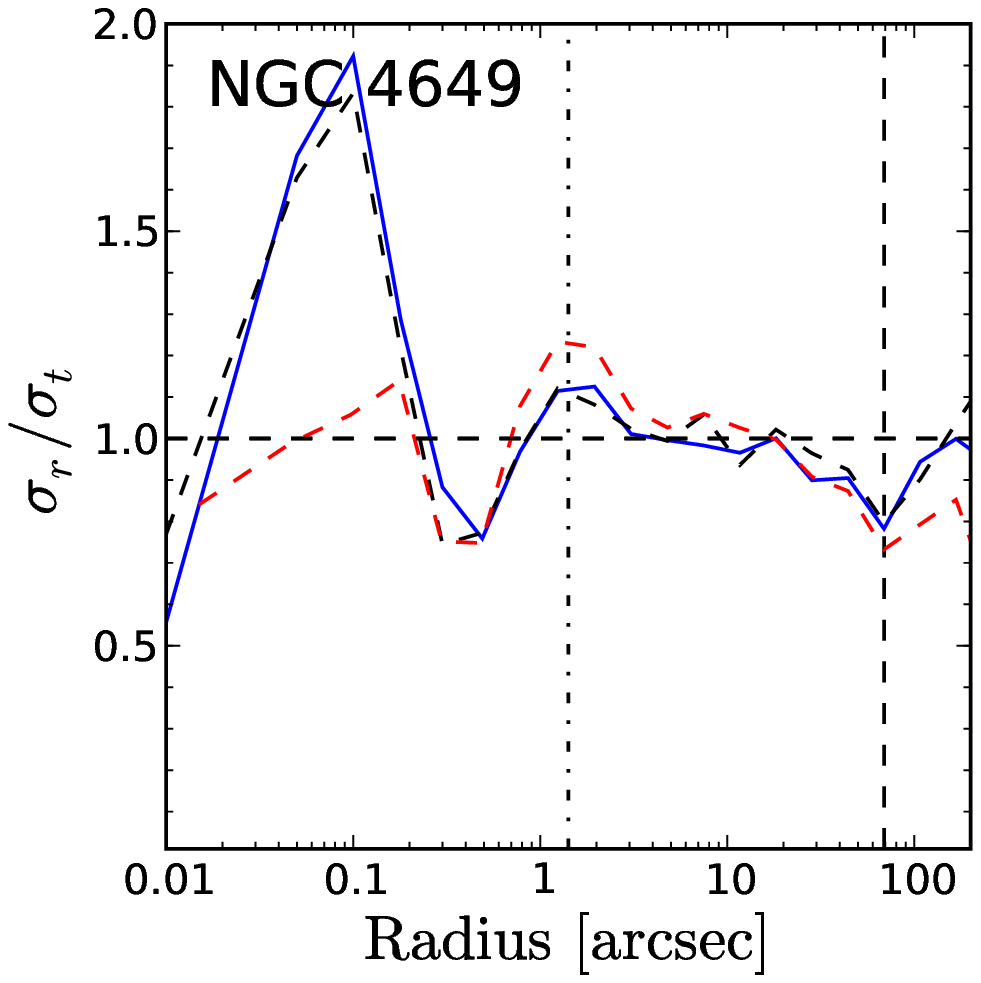}}

\put(0,0){\includegraphics[width=\wi]{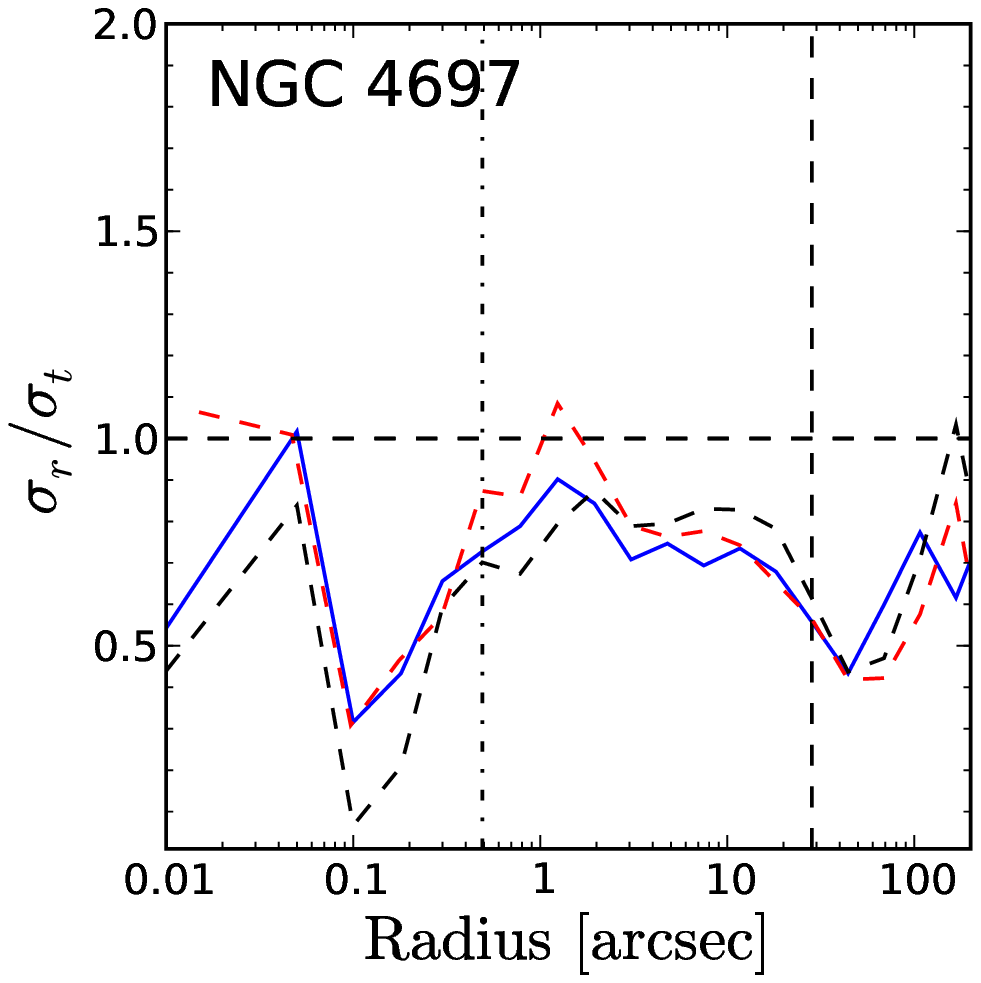}}
\put(52,0){\includegraphics[width=\wi]{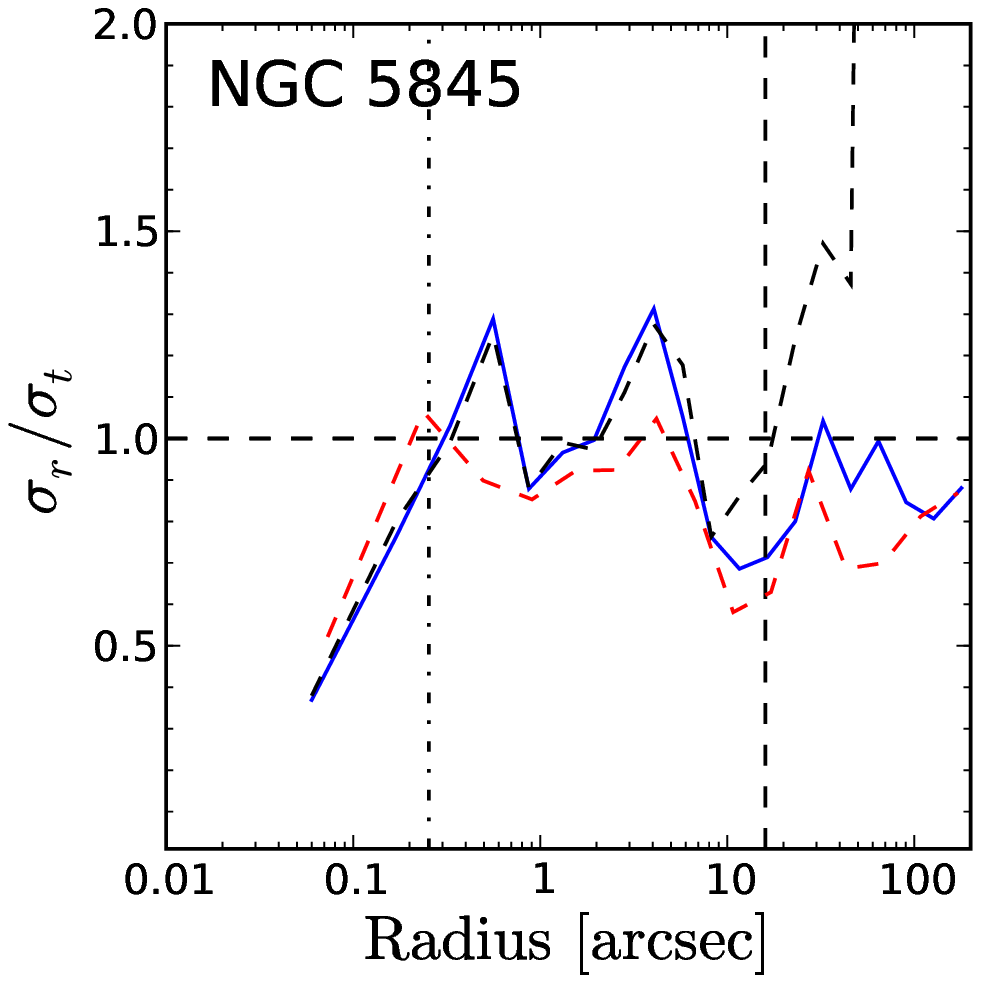}}
\put(104,0){\includegraphics[width=\wi]{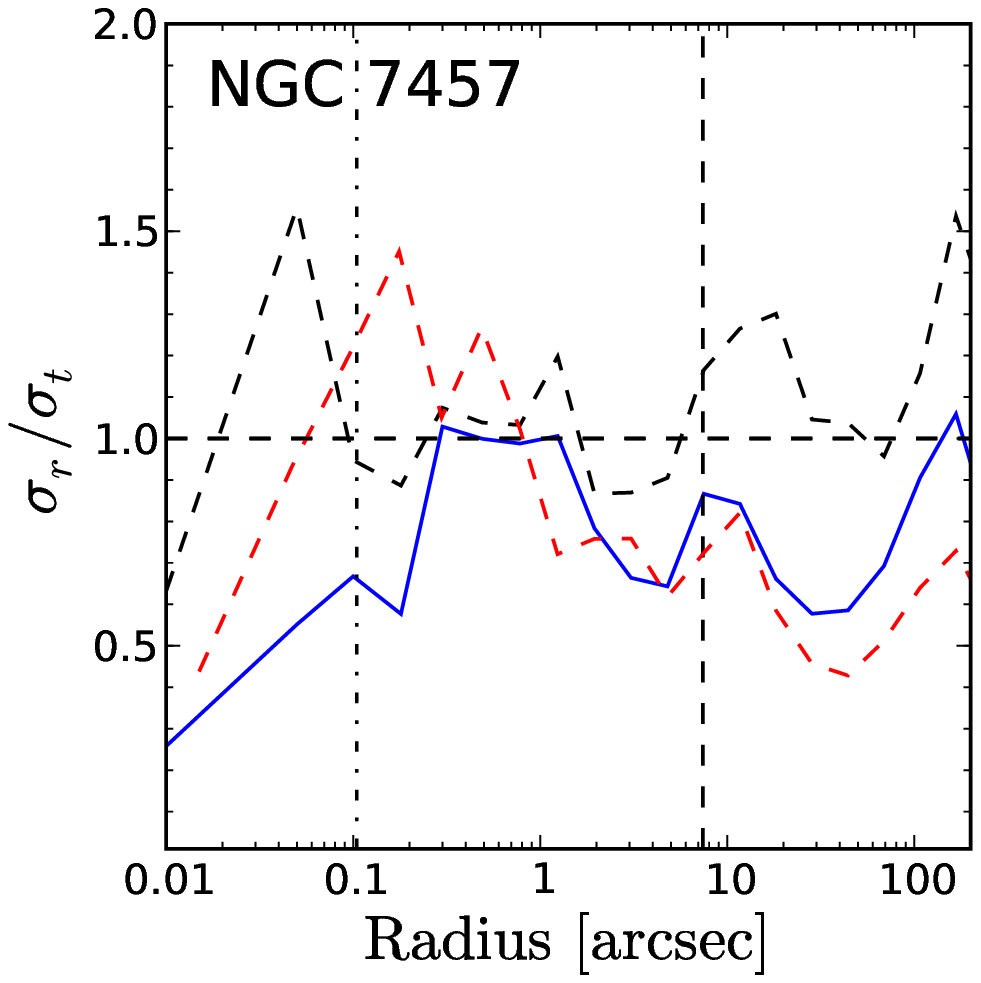}}
\end{picture}
\caption{Ratio of the internal velocity dispersions as a function of radius for the best model of each galaxy. Shown is the ratio $\sigma_r/\sigma_t$ along the major axis. The result with a DM halo, without a DM halo, and the result of \citet{Gebhardt:2003} (without a dark halo) are shown as the black dashed, blue solid and red dashed lines, respectively. The horizontal dashed line corresponds to a non-rotating isotropic model. The vertical dashed-dotted line indicates the black hole's sphere of influence, assuming the black hole mass, determined without including a DM halo. The vertical dashed lines show the radial extent of the ground-based data.}
\label{fig:disp}
\end{figure*}

However, for the rest of the objects the new $M_\bullet$ is significantly offset from the previous estimate, not simply explained by numerical noise. For two objects, NGC~3384 and NGC~4473, $M_\bullet$ decreases; for the rest there is an increase in $M_\bullet$, by up to a factor of three. The mean increase for this sample is a factor of $1.46$ with the standard deviation of $0.73$.
This result is in line with the findings of \citet{Shen:2010}, who found an increase in the mass by a factor of two for NGC~4649. This object is included in Figures~\ref{fig:cmpmass} and~\ref{fig:deltamass} as the highest mass point. \citet{Shen:2010} attribute the mass difference to the better orbit sampling in the new models. In particular, they argue that the old orbit sampling lacks high energy, nearly circular orbits, which lead to an underestimate of $M_\bullet$. NGC~4649 is a core galaxy, and it is important to note that all galaxies with an increase in mass by more than a factor of two are core galaxies as well. This seems to indicate that the previous orbit sampling was not able to properly model core galaxies.

To investigate this issue further for our whole sample, we inspect the internal orbit structure, looking for any clear difference between the models. To do so, we examine the shape of the velocity dispersion tensor, represented by the ratio of radial to tangential dispersion $\sigma_r/\sigma_t$. The tangential dispersion includes contributions from random as well as from ordered motion; thus, it is given by $\sigma_t^2 =  \sigma_\theta^2 + \sigma_\phi^2 + V_\phi^2$. In Figure~\ref{fig:disp}, we compare the internal dispersion ratio $\sigma_r/\sigma_t$ for the best-fit models presented here, with and without a DM halo (as blue solid and black dashed dotted lines, respectively), with the ratio for the models in G03, shown as red dashed lines. We also indicate the black hole sphere of influence $R_\mathrm{inf}= G M_\bullet \sigma^{-2}$, assuming the new $M_\bullet$ (without DM halo). The galaxies with consistent black hole masses, such as NGC~3377 and NGC~4697, also exhibit consistent internal structure. On the other hand, the galaxies with the largest mass increase, especially the core galaxies such as NGC~4291 and NGC~3608 show a clear difference in the internal structure. First, there is a strong radial bias at large radii for these galaxies, especially compared to the previous dispersion ratio. However, this radial bias is mainly outside the range for which kinematic data are available and is therefore driven by the maximization of the entropy. We do not expect these orbits to have an influence on the black hole mass determination. Second, there is a stronger tangential bias inside the black hole sphere of influence. In particular, the previous models exhibit a radial bias within $R_\mathrm{inf}$ for the largest outliers. G03 only sampled the zero-velocity curve, instead of the whole phase space, and due to the coarse sampling of drop points they missed the orbits near the pole that are nearly circular. This sampling then causes a radial orbital bias. 
This radial bias is removed in the models presented here, using a better orbit sampling.
An increase in tangential orbits will reduce the projected line-of-sight velocity dispersion and therefore a more massive black hole is required to match the observed velocity dispersion profile.
On the other hand, NGC~4473 which shows a decrease in the determined black hole mass, exhibits a stronger radial bias in the new modeling compared to G03. This radial bias is probably caused by the presence of a nuclear disk in this galaxy.

Thus, we find that the main reason for the change in black hole mass is the different orbit sampling used, as already found by \citet{Shen:2010} for NGC~4649. We now cover the phase space more completely and therefore also include orbits missed by the previous sampling. This issue is of special importance for core galaxies, as they often show a significant tangential orbital bias in their center, i.e., they usually have the largest $\sigma_t$ (G03). 

\begin{figure}
\centering
\resizebox{\hsize}{!}{\includegraphics[clip]{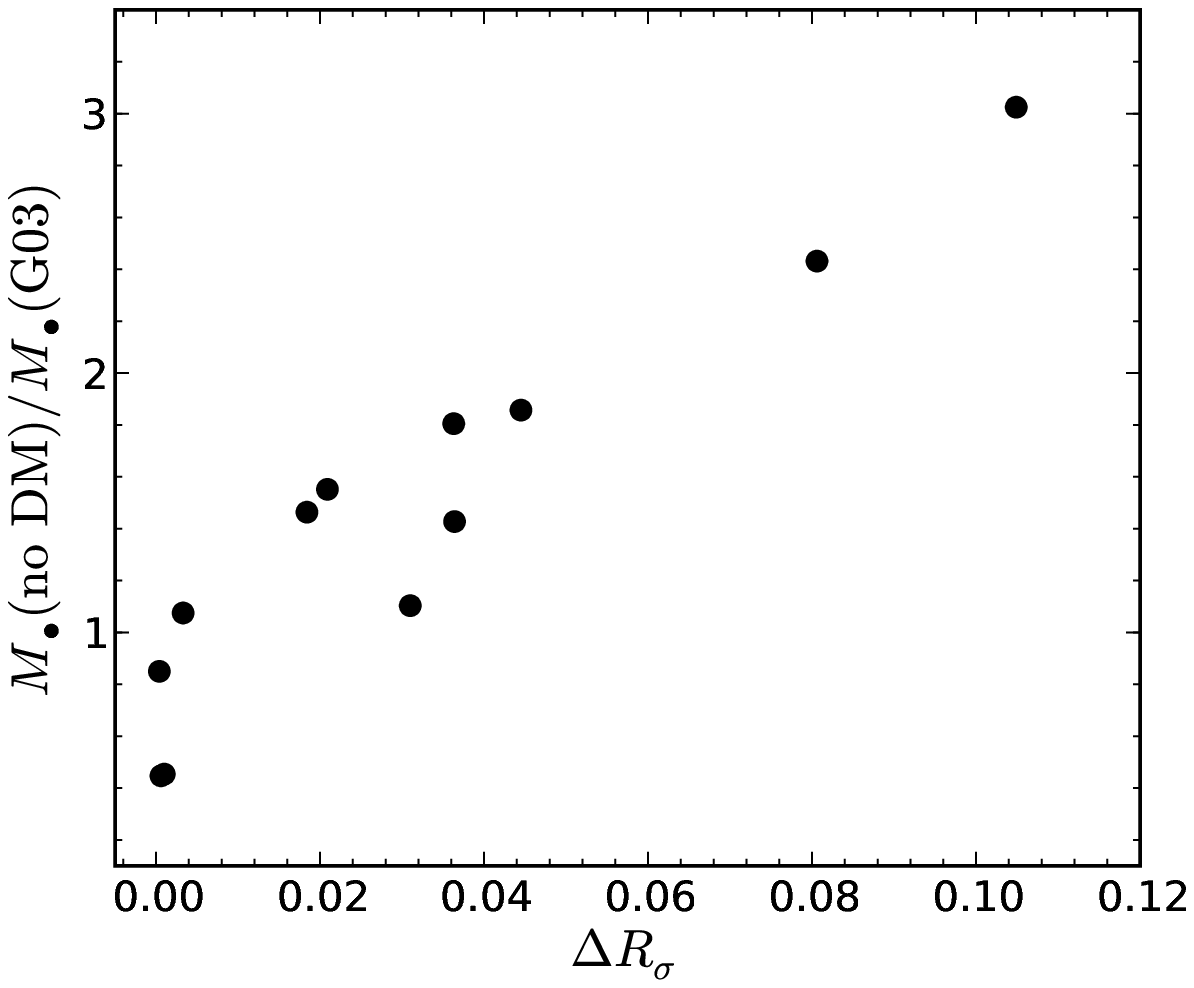}}
\caption{Ratio between the $M_\bullet$ determined in this work and the $M_\bullet$ given in \citet{Gebhardt:2003}, as a function of the excess in radial motion in the internal structure of the models of \citet{Gebhardt:2003}, as defined by Equation~(\ref{eq:momexcess}). A correlation between both quantities is apparent.}
\label{fig:momexcess}
\end{figure}

To illustrate this point, we computed the difference of the dispersion ratio between G03 and this new model in a shell inside the black hole sphere of influence:
\begin{equation}
 \Delta R_\sigma = \int_{r_\mathrm{min}}^{r_\mathrm{max}} \left[  \left( \sigma_r/\sigma_t\right)_\mathrm{G03}  - \left( \sigma_r/\sigma_t\right) \right]  \mathrm{d}r  \ , \label{eq:momexcess}
\end{equation} 
with $r_\mathrm{min}= 0.1 R_\mathrm{inf}$ and $r_\mathrm{max} = R_\mathrm{inf}$. This quantity is just a simple and quick way of quantifying the change in the orbital structure and is just meant to highlight the relation between the change in orbital structure and the change in black hole mass.
In Figure~\ref{fig:momexcess} we plot it against the ratio of the black hole masses $M_\bullet / M_{\bullet,\mathrm{G03}}$. There is a clear correlation between the quantities, confirming our previous argument.
We have also tested the effect of decreasing the number of orbits in the modeling, but saw no clear influence on the best-fit black hole mass. Thus we confirm our previous results in finding that the recovered black hole mass is not affected by the number of orbits \citep{Gebhardt:2004,Richstone:2004,Shen:2010}.

This investigation emphasizes the need for a complete orbital sampling of phase space for dynamical modeling of galaxies, especially of core galaxies.

\subsection{Effect of a dark matter halo on the determined black hole mass}
\begin{figure}
\centering
\resizebox{\hsize}{!}{\includegraphics[clip]{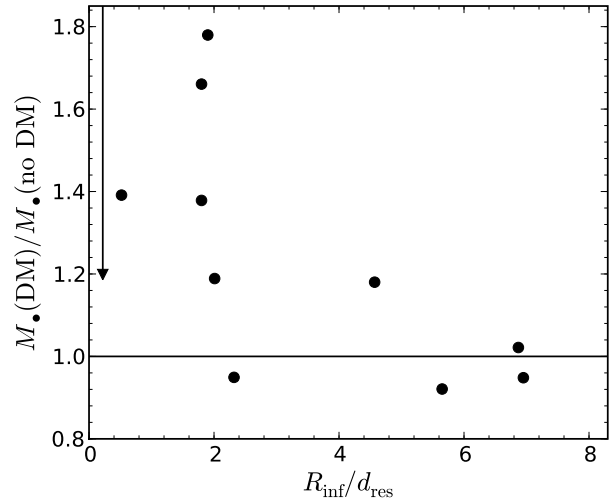}}
\caption{Ratio between the $M_\bullet$ without accounting for the DM halo and the $M_\bullet$ including a DM halo in the model, as a function of the ratio of black hole sphere of influence over the spatial resolution. The arrow indicates the upper limit for the mass of NGC~2778 when DM is included. The solid line shows a one-to-one correspondence between both masses.}
\label{fig:deltamdm}
\end{figure}

The main motivation of this paper is to investigate the effect of the inclusion of a DM halo in the dynamical modeling on the determined black hole mass. In Figure~\ref{fig:deltamdm} we show the difference in black hole mass between the models with and without the inclusion of a DM halo as a function of the resolution of the black hole sphere of influence divided by the spatial resolution of the kinematic observation ($R_\mathrm{inf}/d_\mathrm{res}$). For the computation of $R_\mathrm{inf}$ we used the black hole mass including a DM halo (given in Table~\ref{tab:res}). The spatial resolution is given by the seeing and the aperture of the \textit{HST} kinematic observations; thus, $d_\mathrm{res}=0.08$ for the STIS data and $d_\mathrm{res}=0.15$ for the FOS data.

As expected, there is a general trend  of an increase in $M_\bullet$ when a DM halo is included. For five objects, we find almost no change in $M_\bullet$, while for one object -- NGC~2778 -- the significance of the black hole detection even vanishes, with the minimum $\chi^2$ for no black hole. The other six galaxies show an increase in the measured $M_\bullet$ between 20\% and 80\% when a DM halo is included. The most extreme case is NGC~4473 probably due to the presence of a nuclear disk, with an increase of a factor of 1.8 when a DM halo is included. For the whole sample we find a mean increase of a factor of $1.22$ with standard variation of $0.27$.
This increase is much less than the factor of more than two found for M87. In contrast to M87, our data set contains no stellar kinematic information at large radii but includes \textit{HST} data at small radii. Thus, we are better able to probe the region affected by the presence of the black hole at the center.

In Figure~\ref{fig:deltamdm}, there appears to be a trend of a larger bias for objects where $R_\mathrm{inf}$ is less well resolved, as would be expected. However, due to the black hole mass uncertainties there is no statistically significant relation.
The most massive galaxy in our sample, NGC~4649, is not shown in the figure, as it would appear at $R_\mathrm{inf}/d_\mathrm{res}\approx 20$ with no significant change in black hole mass. M87 would lie at $R_\mathrm{inf}/d_\mathrm{res}\approx 1.5$ and $M_{\bullet, \mathrm{DM}}/M_{\bullet, \mathrm{no\,DM}}\approx2.8$. In contrast to M87, the galaxies in our sample with a less well resolved sphere of influence exhibiting a smaller change in the determined $M_\bullet$ are less massive and probably reside in less massive DM halos. This indicates that especially for massive galaxies properly resolving $R_\mathrm{inf}$ is important to determine $M_\bullet$ under the consideration of DM.

In Figure~\ref{fig:cmpchi}, the marginalized $\chi^2$ distributions for the individual objects with (solid black line) and without (dotted dashed blue line) a DM halo are shown. 
For the five objects with almost no change in $M_\bullet$ (NGC~3377, NGC~3608, NGC~4291, NGC~4649, and NGC~4697), there is also no change in the $\chi^2$ (apart from NGC~4649). For the other galaxies, including a reasonable DM halo improves the fit in terms of $\chi^2$. The most convincing cases are NGC~821 and NGC~3384, where the model without a DM halo is excluded at more than $3\sigma$ significance. Thus, while we are not able to constrain the shape of the DM halo, at least for some galaxies the presence of such a halo is supported. In total, for six galaxies (NGC~821, NGC~2778, NGC~3384, NGC~4473, NGC~4564, and NGC~5845), the model without a DM halo is excluded with at least $2\sigma$ significance. 

The mean increase of the black hole mass goes along with a decrease of the mass-to-light ratio, as expected. This indicates the degeneracy present between the stellar mass-to-light ratio and the DM contribution in dynamical models. For the whole sample we find a decrease in $M/L$ of 6\% with a scatter of 5\%.

\begin{figure*}
\centering
\setlength{\unitlength}{1mm}
\begin{picture}(180,120)
\put(0,80){\includegraphics[width=\wib]{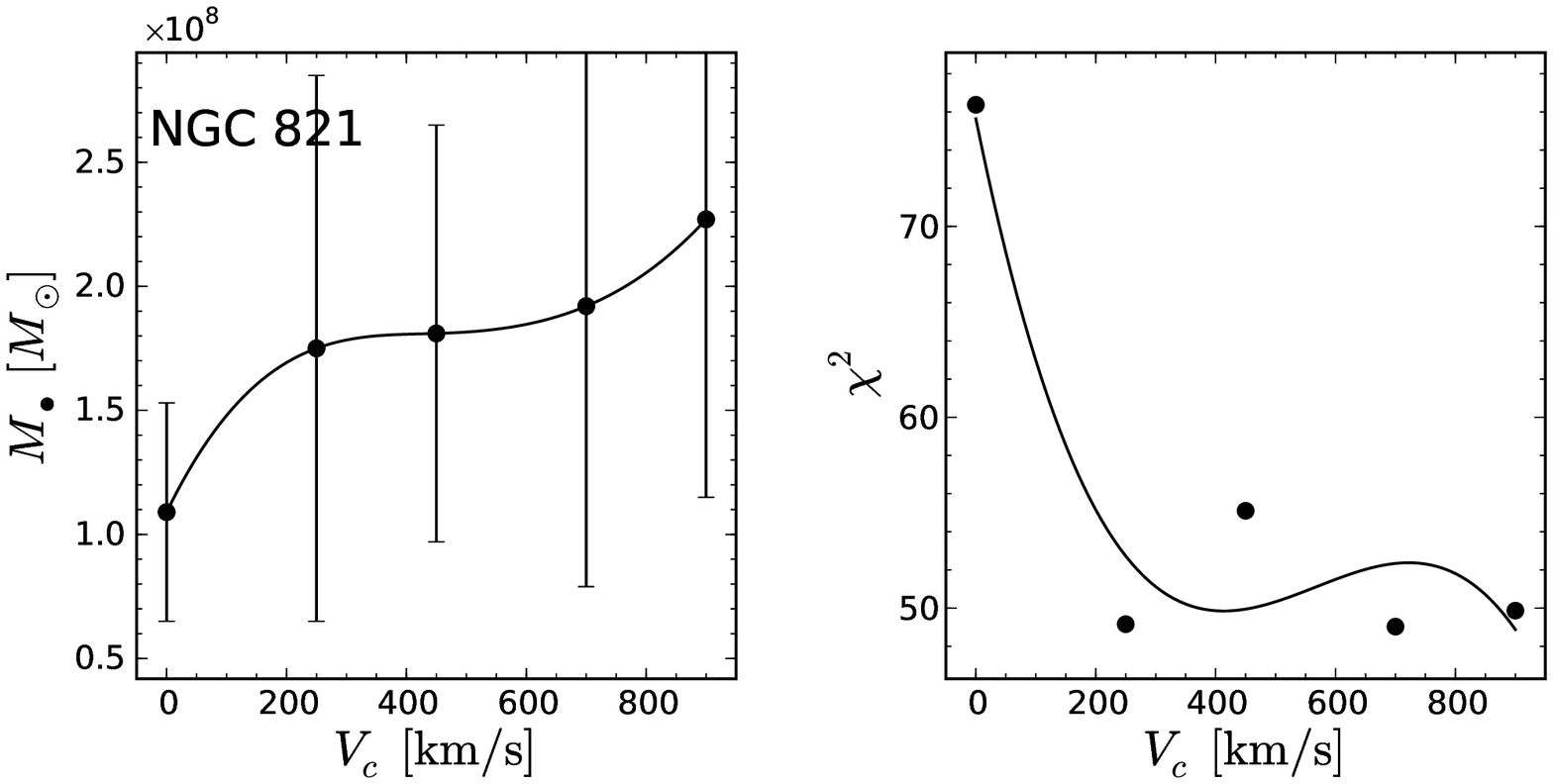}}
\put(90,80){\includegraphics[width=\wib]{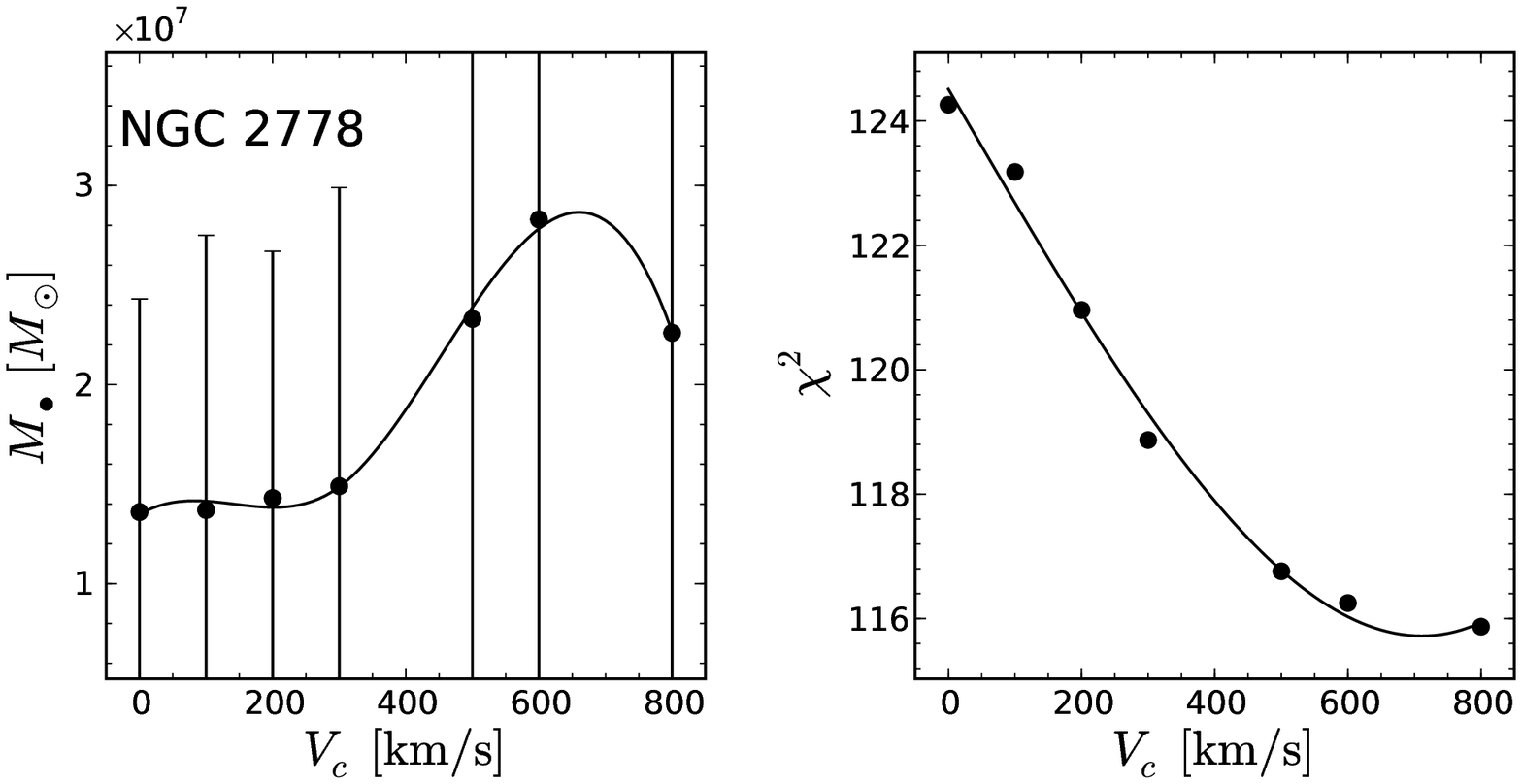}}

\put(0,40){\includegraphics[width=\wib]{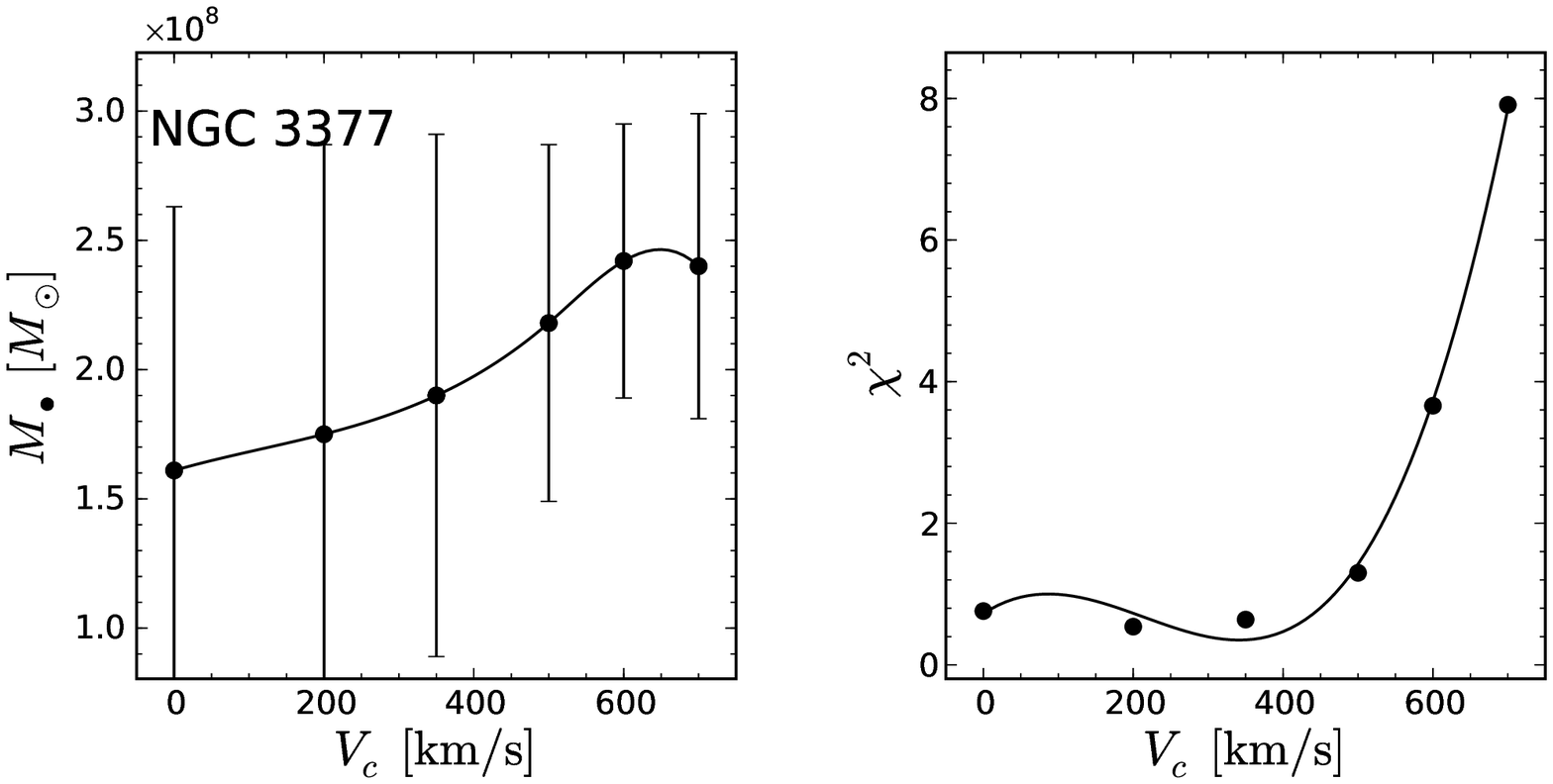}}
\put(90,40){\includegraphics[width=\wib]{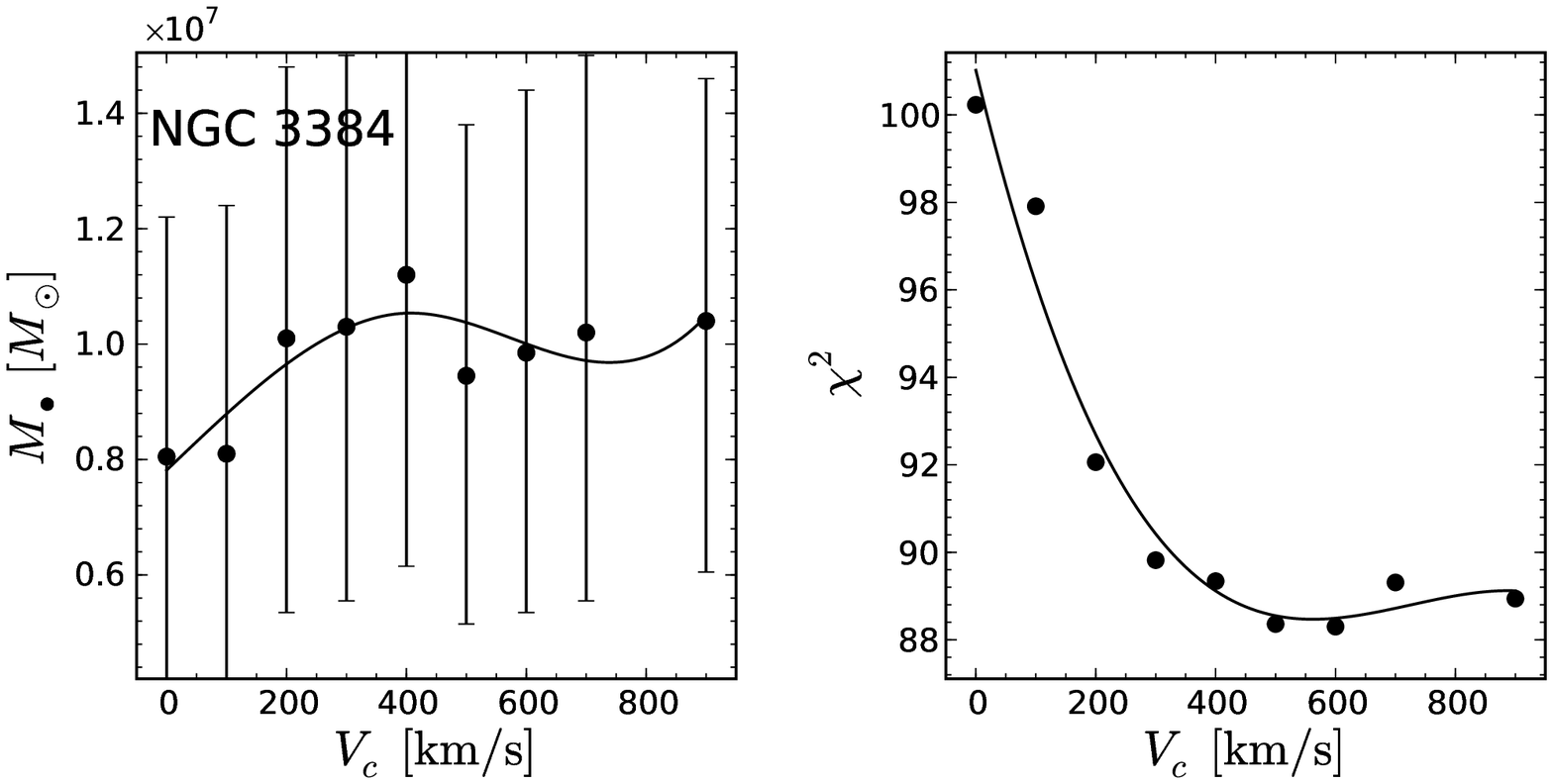}}

\put(0,0){\includegraphics[width=\wib]{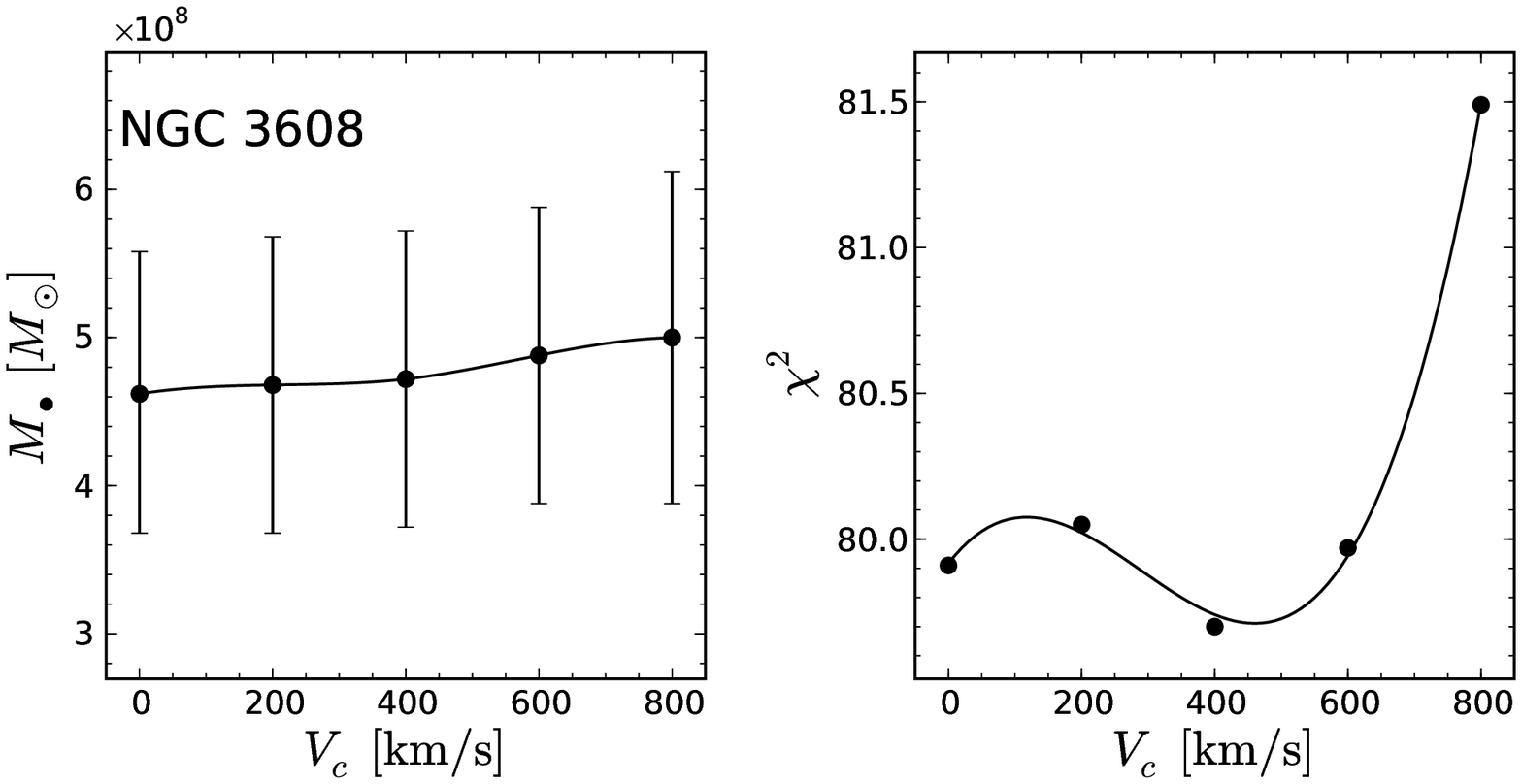}}
\put(90,0){\includegraphics[width=\wib]{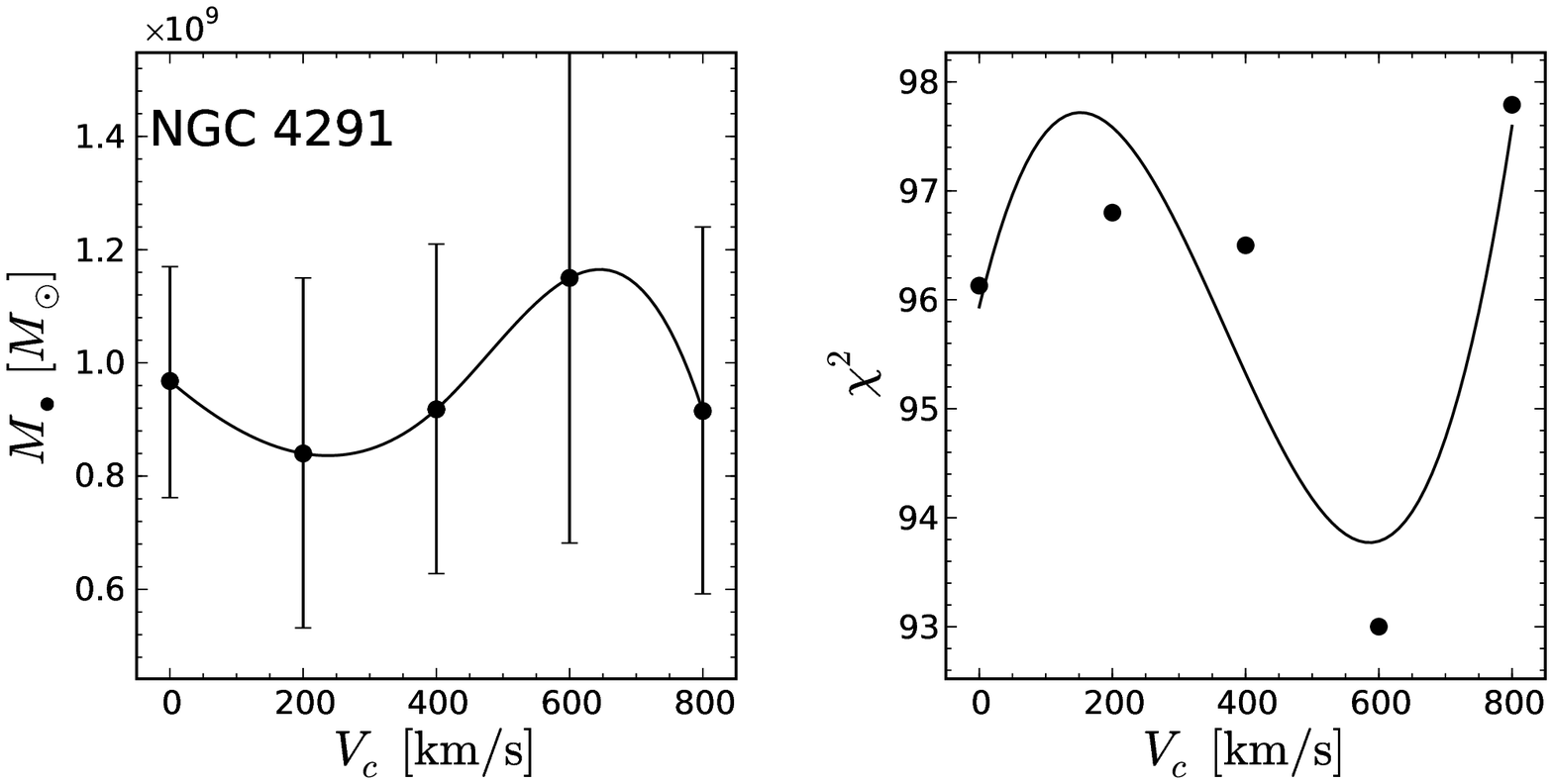}}
\end{picture}
\caption{Variation of the black hole mass and the minimal $\chi^2$ as a function of the assumed DM halo for each galaxy. In the left columns of the subpanels the change in $M_\bullet$ for various values of $V_\mathrm{DM}$ is shown. The error bars correspond to $1\sigma$ (i.e., $\Delta \chi^2=1$). In the right columns the corresponding value of $\chi^2$ is shown. A spline interpolation is shown as the solid line.}
\label{fig:dmtest}
\end{figure*}

\begin{figure*}
\centering
\setlength{\unitlength}{1mm}
\begin{picture}(180,120)
\put(0,80){\includegraphics[width=\wib]{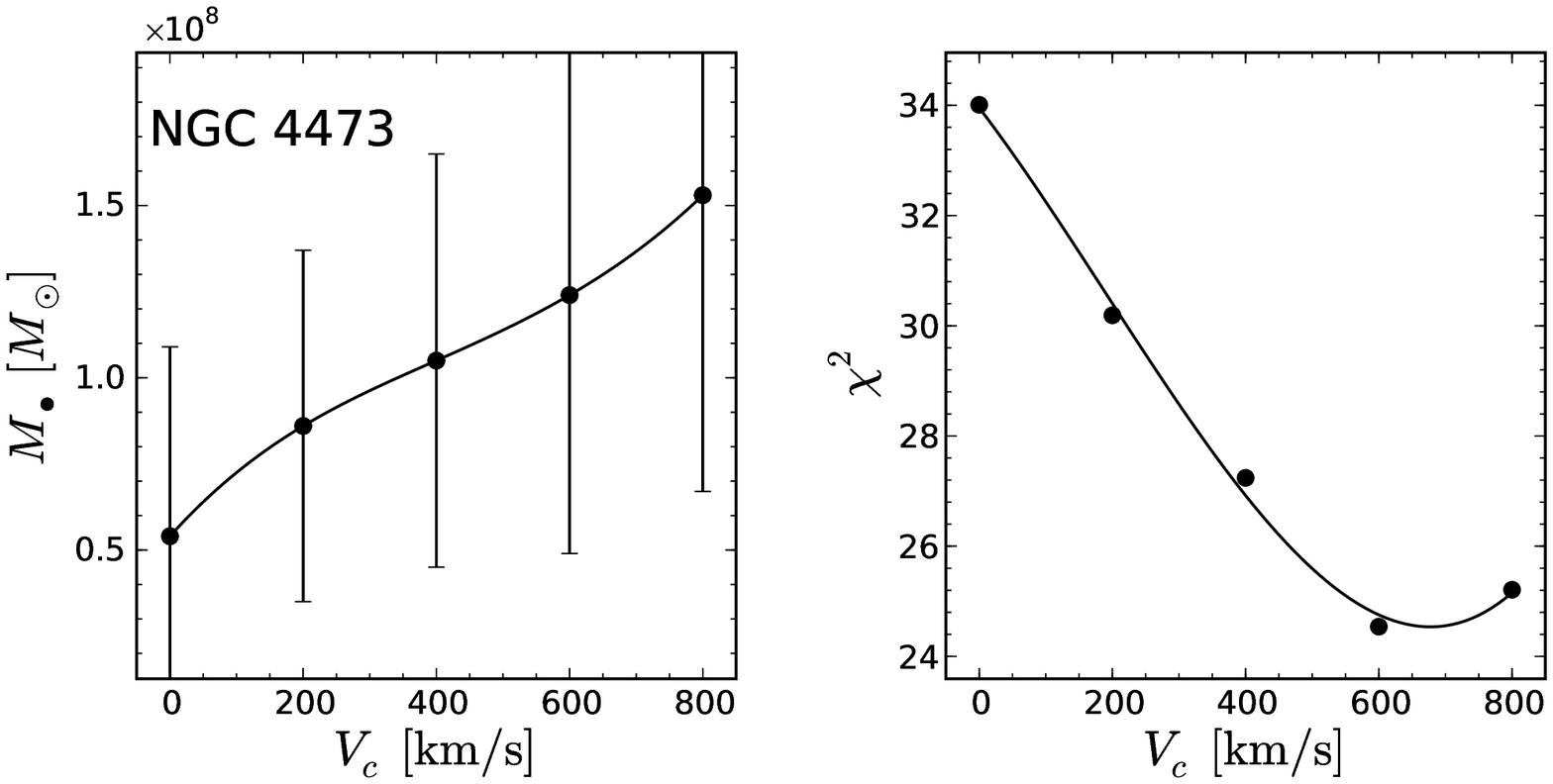}}
\put(90,80){\includegraphics[width=\wib]{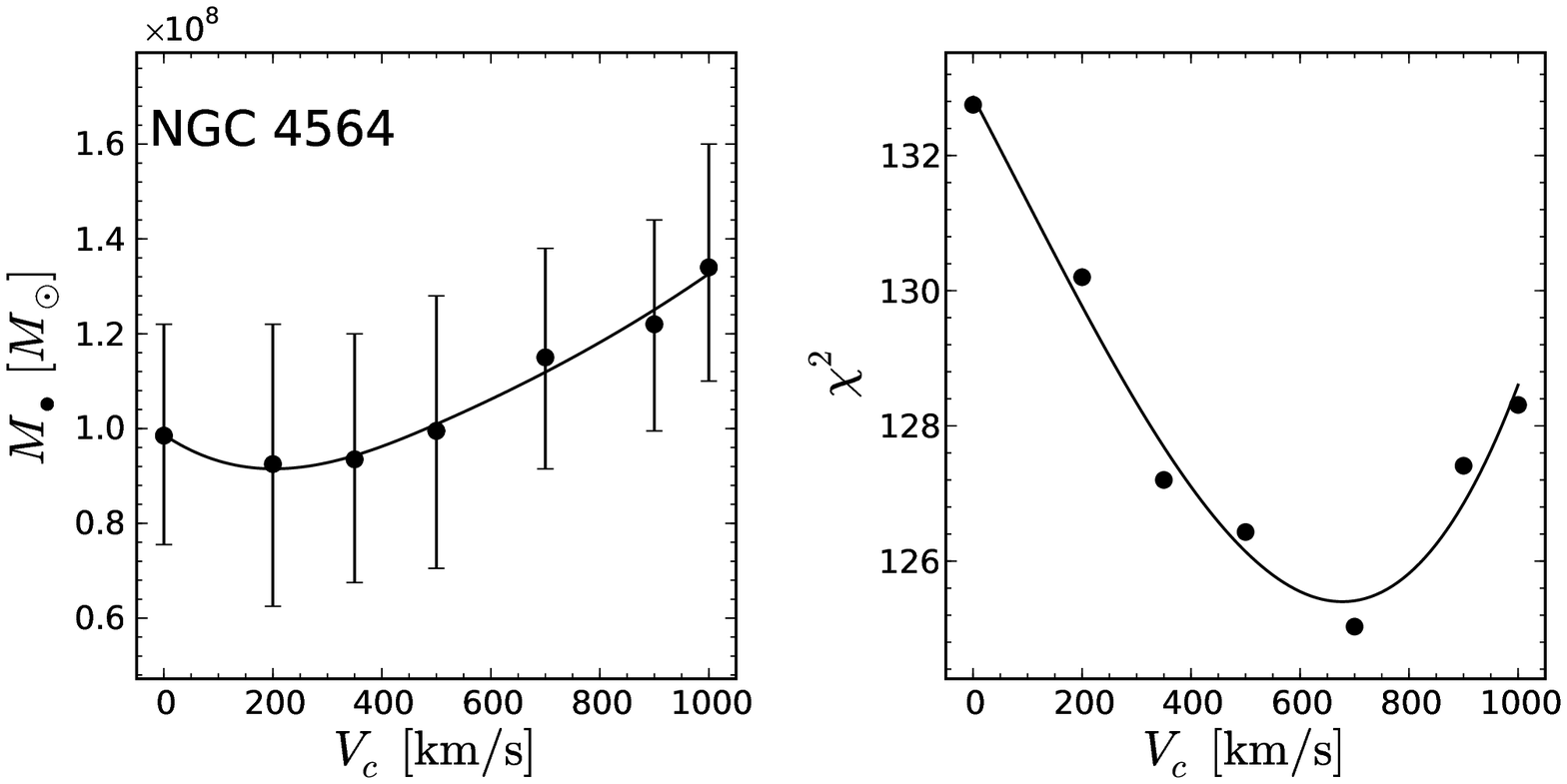}}

\put(0,40){\includegraphics[width=\wib]{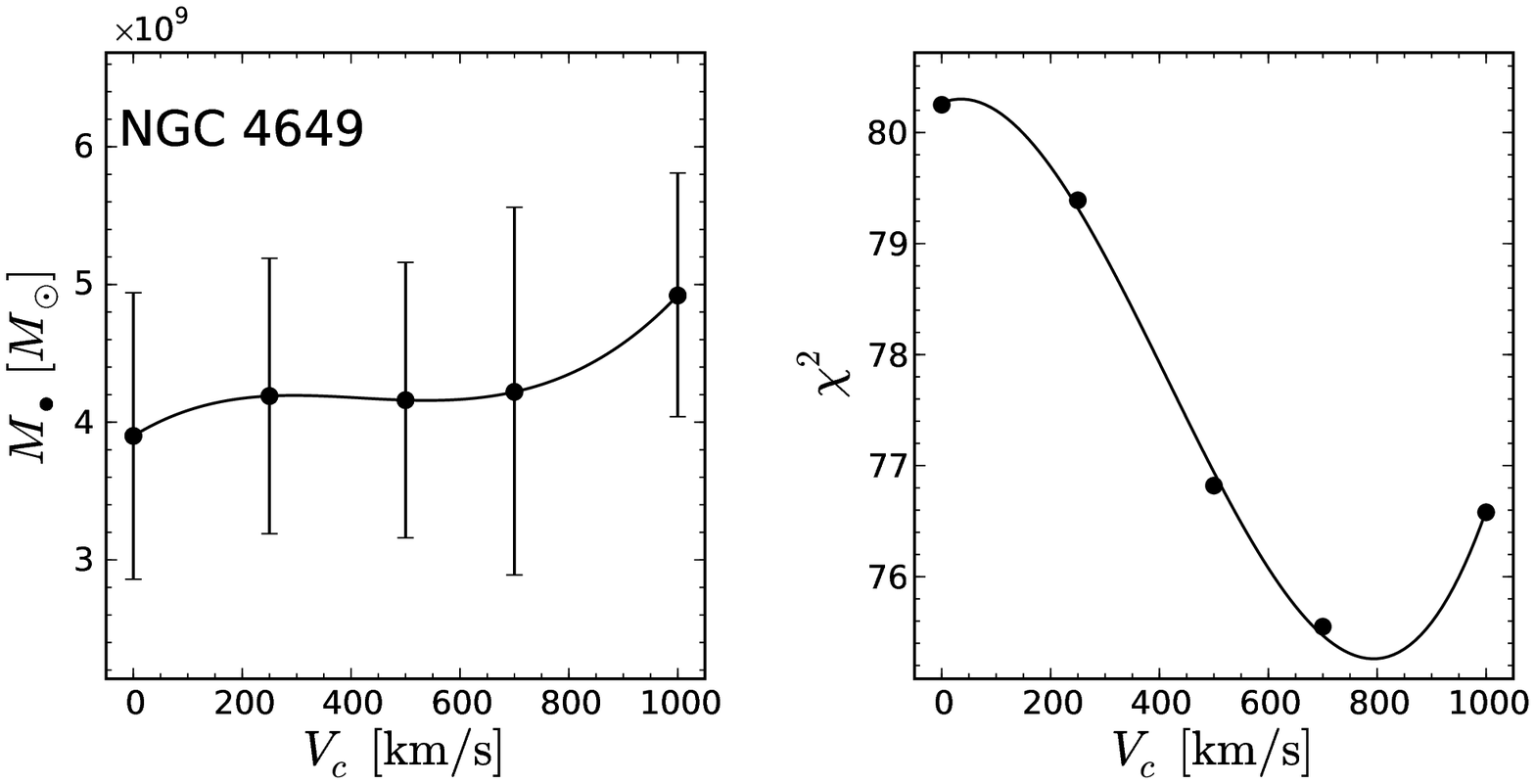}}
\put(90,40){\includegraphics[width=\wib]{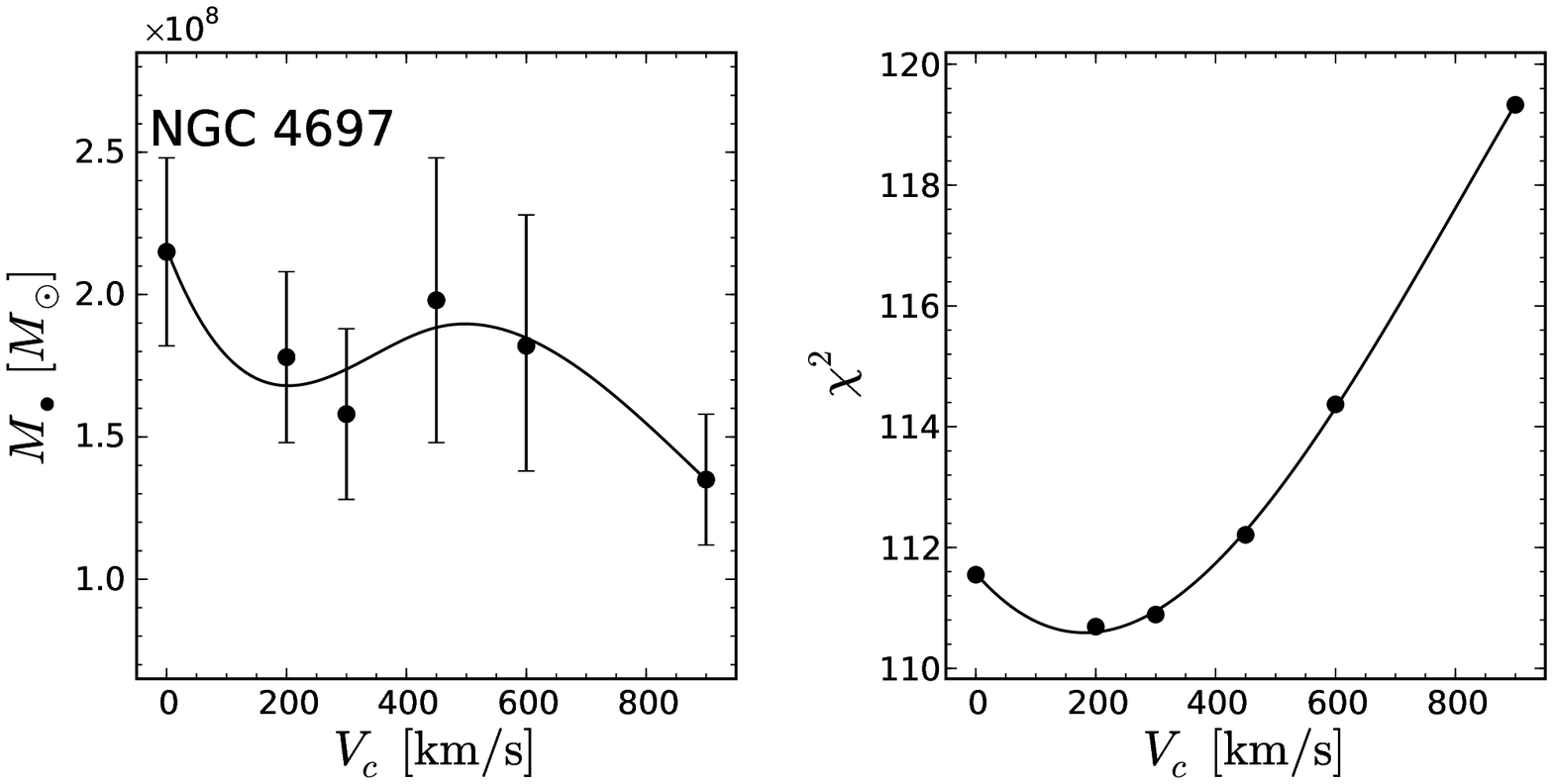}}

\put(0,0){\includegraphics[width=\wib]{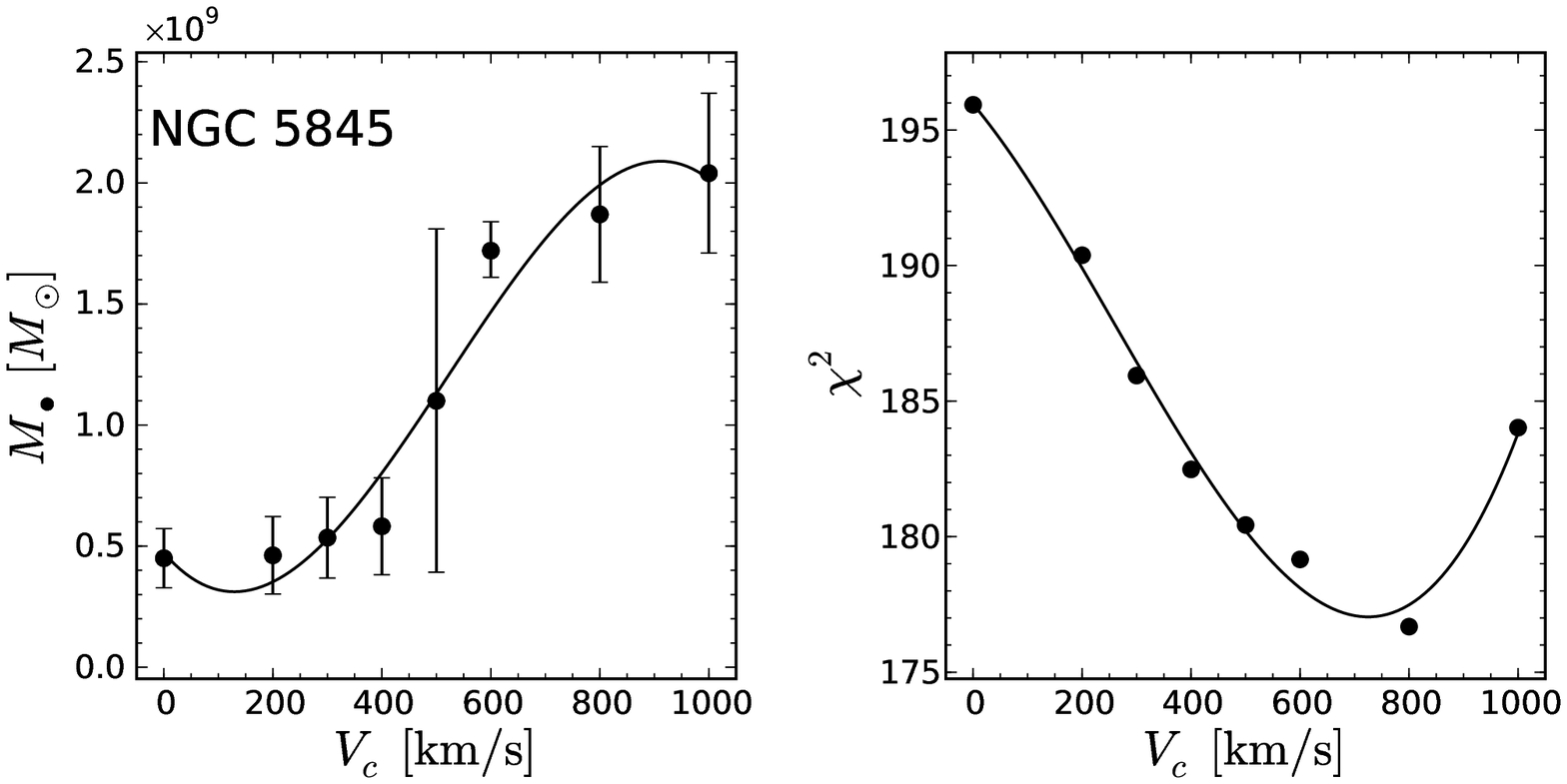}}
\put(90,0){\includegraphics[width=\wib]{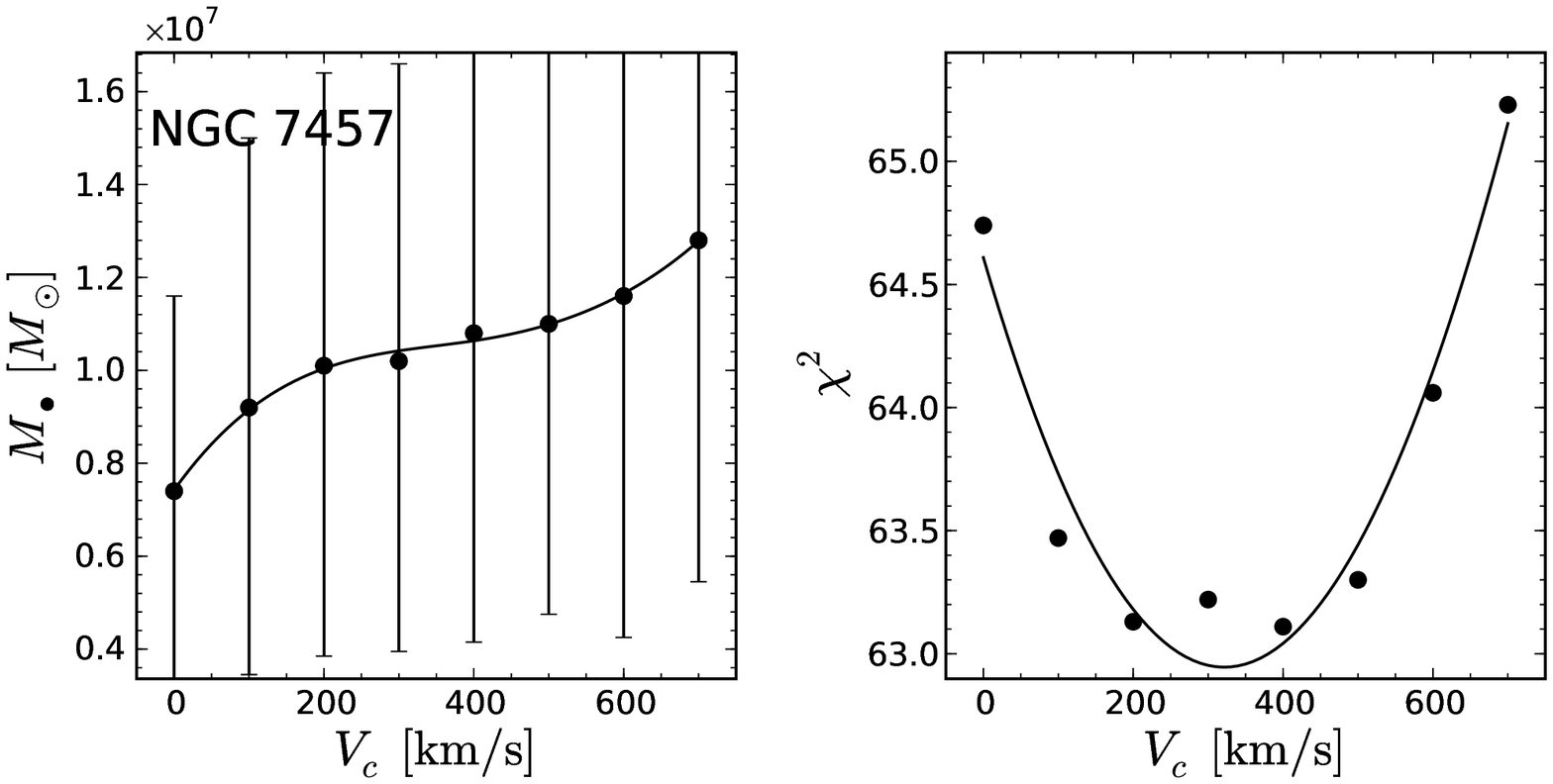}}
\end{picture}
\caption{Same as Figure~\ref{fig:dmtest} for the second half of the galaxy sample.}
\label{fig:dmtest2}
\end{figure*}

Even if our choice of DM halo is well motivated by the scaling relations of \citet{Thomas:2009}, it is basically an ad hoc assumption we had to make as we do not have the data to robustly constrain the DM halo profile. To at least test the effect of changing the assumed DM halo on the black hole mass, we ran a set of models, changing $V_c$ in the logarithmic DM potential. We restrict ourselves to changing only this one parameter, as we want to avoid sampling the whole four-dimensional parameter space. It has also been found that $V_c$ and $r_c$ are degenerate, especially if the large radii coverage is poor \citep{Shen:2010,Forestell:2009}. For each galaxy, we assume a twice as massive DM halo and a DM halo about half as massive, as well as some additional values. The results are shown in Figures~\ref{fig:dmtest} and~\ref{fig:dmtest2}. We confirm the basic trends of an improved $\chi^2$ for a reasonable massive halo and an increase in $M_\bullet$ for a more massive halo. The range of given $M_\bullet$ approximately covers the range consistent with the current data, as long as the DM halo is not constrained for these galaxies.


\subsection{Notes on individual galaxies}
In the following, we provide more detailed information on the black holes for some individual galaxies.

\textit{NGC~821.} There are two recent studies on the DM halo of this galaxy, providing large radii data. \citet{Weijmans:2009} used SAURON data to measure LOSVDs out to $\sim4$ effective radii. Their assumed DM halo gives $M_\mathrm{DM}=9\times10^9 \, M_\odot$ within the effective radius (assuming $R_e=5.1$~kpc), using an NFW profile. \citet{Forestell:2010} used long-slit data from the Hobby-Eberly Telescope to measure the LOSVD out to $\sim 2 R_e$. Assuming their power law fit to the DM halo, we find $M_\mathrm{DM}=8\times10^9 \, M_\times$ within $R_e$. Our assumed DM halo is more than twice as massive. Thus, $M_\bullet$ for the true DM halo should be contained within the range spanned by our no DM and DM solution. Including these large radii data into the dynamical models is beyond the scope of this paper.
The nuclear supermassive black hole in NGC~821 has been detected as a weak X-ray source, implying a very weak level of activity \citep[$L_\mathrm{X}/L_\mathrm{Edd}\sim 10^{-8}$;][]{Pellegrini:2007}. There is also evidence for the presence of a jet \citep{Fabbiano:2004,Pellegrini:2007}.

\textit{NGC~2778.} This galaxy already had the least confident black hole detection in G03. Assuming the value for $M_\bullet$ of our no DM halo model, we do not resolve the black hole's sphere of influence ($R_\mathrm{inf}/d_\mathrm{res}=0.2$). Including a DM halo in the model improves the fit significantly, but the significance of the black hole detection disappears. However, the previous $M_\bullet$ estimate is still fully consistent with the $1\sigma$ upper limit of $M_{\bullet,\mathrm{up}}=2.99\times 10^7 \, M_\odot$ that we derive for NGC~2778 under the presence of a DM halo. This behavior might indicate the need to properly resolve $R_\mathrm{inf}$ when a DM halo is included to properly determine $M_\bullet$.

\textit{NGC~3377.} The black hole mass for this galaxy increased by $\sim70$\%, compared to G03, mainly caused by the stronger widening of the confidence contours at the high-mass end than at the low-mass end. The previous value is still fully consistent within $1\sigma$. \citet{Copin:2004} reported a black hole mass of $M_\bullet=8.3\times 10^7 \, M_\odot$ (for our assumed distance) based on Integral Field Unit (IFU) observations with SAURON and OASIS, also still consistent with our results within $1\sigma$. 
The first detection of a black hole in NGC~3377 has been reported by \citet{Kormendy:1998}, based on ground-based observations. Using an isotropic model, they found $M_\bullet=2.1\times 10^8 \, M_\odot$ and $M/L_V =2.0$ (for our assumed distance), in good agreement with our results. NGC~3377 is a rapid rotator and close to isotropy, justifying the isotropic assumption for this galaxy.
NGC~3377 exhibits a nuclear X-ray source, showing a jet like feature \citep{Soria:2006}. 

\textit{NGC~3384.} Besides NGC~2778, this is the only other galaxy for which the sphere of influence is not resolved. While for NGC~2778 the new code does not lead to a change of the $\chi^2$ distribution, for NGC~3384 $M_\bullet$ decreases. This might indicate a larger uncertainty in the determination of $M_\bullet$ using different modeling codes when $R_\mathrm{inf}$ is not resolved.
NGC~3384 is the galaxy with the strongest constraints on the presence of a DM halo. For this galaxy, we ran a grid of models changing $r_c$ as well as $V_c$, but we found no change in $\chi^2$ for different values of $r_c$. However, we are able to set a lower limit on $V_c$ with $V_c > \sim 350$~km s$^{-1}$ at $1\sigma$ confidence. The no DM halo model is excluded at more than $3\sigma$ confidence (see Figure~\ref{fig:dmtest}).

\textit{NGC~4473.} NGC~4473 shows evidence for a central stellar disk both in the imaging and the kinematics, as discussed by G03. We followed G03 and include a central exponential disk and also assumed a galaxy inclination of $71^\circ$, as found for the disk component. Thus this galaxy is the only case in our sample not modeled with an edge-on inclination. The presence of the disk has a distinct influence on the measured black hole mass causing a relatively large difference between the models with and without a DM halo.

\textit{NGC~4564.} This galaxy is known to have a nuclear X-ray source \citep{Soria:2006}, indicating the presence of an extremely sub-Eddington accreting AGN.

\textit{NGC~4649.} This object has recently been studied by \citet{Shen:2010} including a DM halo in the models. In addition to the stellar kinematics used in this work, they included globular cluster velocities from \citet{Hwang:2008}. Thus, our results are not directly comparable. They report values of $M_\bullet=(4.5\pm1.0)\times 10^9$ when including a DM halo in the models and $M_\bullet=(4.3\pm0.7)\times 10^9$ without a DM halo. Our results for $M_\bullet$ are consistent with their work.

\textit{NGC~4697.} The black hole mass for this galaxy is basically unchanged using the modified code and including a DM halo. This result is consistent with \citet{Forestell:2009}. She used the same data and model code as we did, but augmented by kinematics of planetary nebulae at large radii \citep{Mendez:2001,Mendez:2008,Mendez:2009}, constraining $M_\bullet$ and the DM halo at the same time. Her best-fit model has $M_\bullet=2.1\times 10^8 \, M_\odot$, $M/L = 4.35$, $V_c = 388$~km s$^{-1}$, and $r_c=9$~kpc, assuming a logarithmic halo. We find identical values for $M_\bullet$ and $M/L$, using slightly different DM halo parameters.
NGC~4697 has a nuclear point source detected in X-rays \citep{Soria:2006}, showing that its black hole is active at a low rate.

\textit{NGC~5845.} While there is a moderate increase in $M_\bullet$ when a DM halo is included in the model, the $\chi^2$ distribution flattens at the high-mass end, due to an increased degeneracy between $M_\bullet$ and $M/L$. Increasing the mass of the DM halo strongly enhances this degeneracy, leading to an almost unconstrained $M_\bullet$ over a wide mass range, until for $V_c \approx 600$~km s$^{-1}$ the minimum switches to  $M_\bullet \approx 1.7\times 10^9 \, M_\odot$, still with strong degeneracy between $M_\bullet$ and $M/L$. Kinematic data at large radii, to better constrain the DM halo and $M/L$ would be desirable for this galaxy. The model without a DM halo is excluded with more than $3\sigma$ significance for NGC~5845.
There is a nuclear X-ray source here as well \citep{Soria:2006}. There is evidence for obscuration of the black hole by a dusty disk, with the X-ray emission originating from scattering of the AGN  continuum emission on the surrounding plasma. 

\begin{figure*}
\centering
\includegraphics[width=8cm,clip]{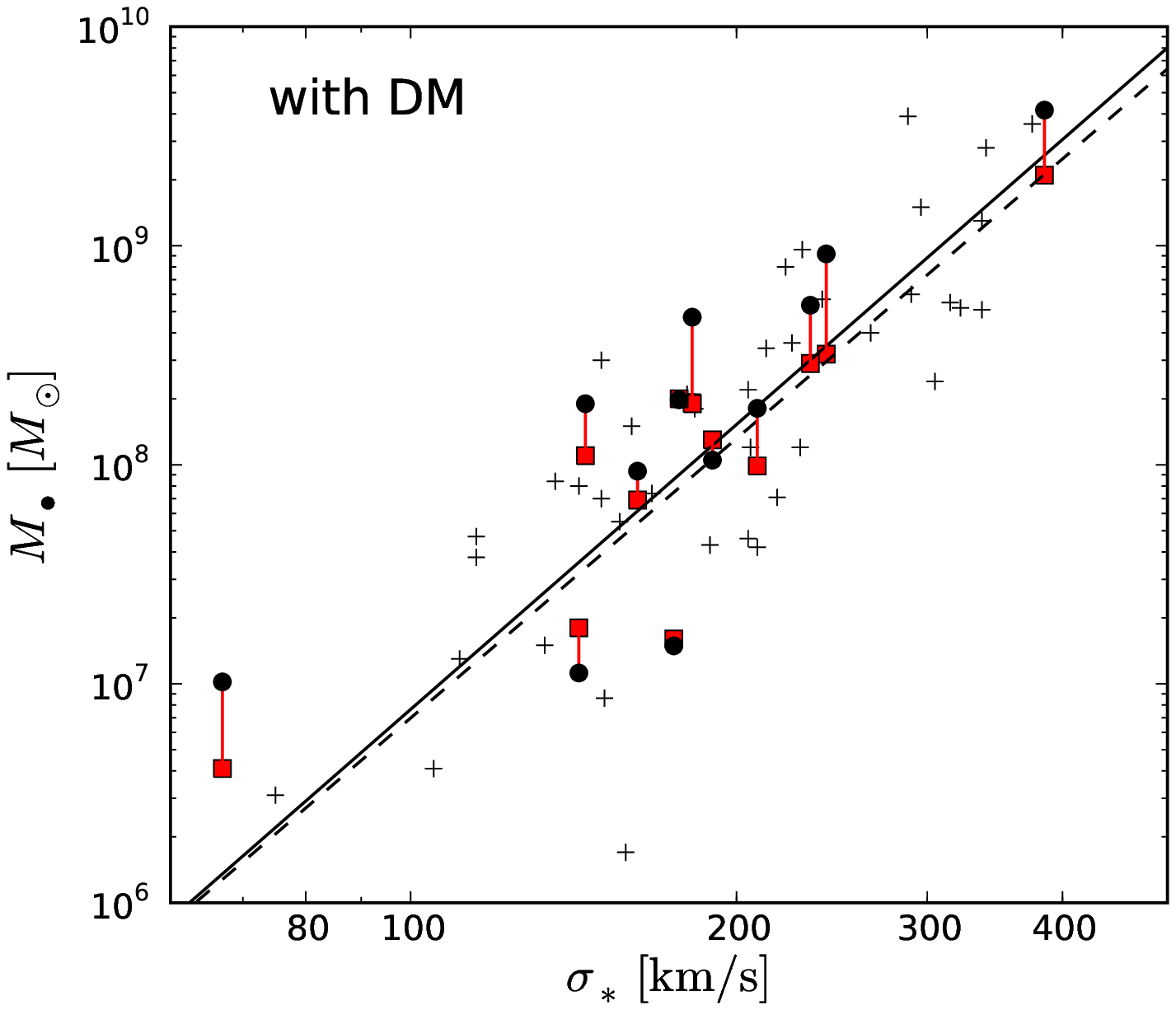}
\includegraphics[width=8cm,clip]{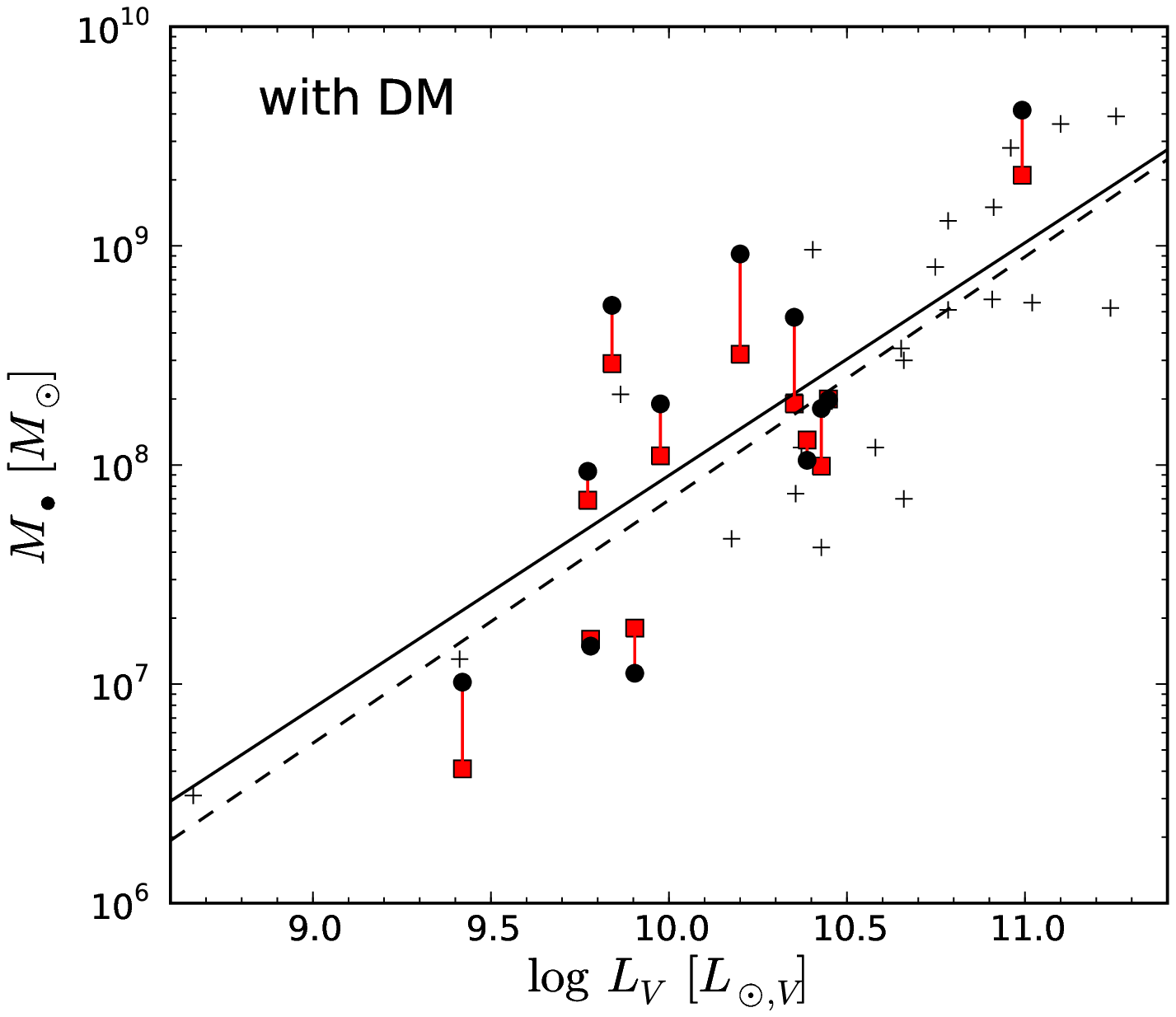}
\caption{Left panel: $M_\bullet-\sigma_\ast$ relationship. The red squares show our sample with the $M_\bullet$ values given by \citet{Gebhardt:2003}, the black circles give the $M_\bullet$ determined in this work with the inclusion of a DM halo. The black solid line shows our updated result for the $M_\bullet-\sigma_\ast$ relation, while the dashed line shows the relation by \citet{Gultekin:2009}. The crosses show the rest of their sample.  Right panel: $M_\bullet-L_V$ relationship. The symbols are the same as in the left panel. The solid line is our best-fit when including a DM halo for our 12 galaxies. The dashed line and crosses are again from \citet{Gultekin:2009}. }
\label{fig:bhbu}
\end{figure*}

\section{The black hole-bulge relations}
As our sample constitutes a significant fraction of the galaxy sample for which dynamical black hole masses are available, it is worth looking at the effect of these new black hole mass measurements on the black hole-spheroid relations, namely the $M_\bullet-\sigma_\ast$ and $M_\bullet-L_V$ relationships. We used the sample of \citet{Gultekin:2009} as the reference sample, containing 49 $M_\bullet$ measurements and 18 upper limits, including our 12 objects.

For the fitting, we used a generalized maximum likelihood method as described by \citet[][; see also \citet{Woo:2010}]{Gultekin:2009}. We minimize the likelihood function $S=-2\ln \mathcal{L}$, with $\mathcal{L}= \prod_i l_i(\mu_i, s_i)$ being the product of the likelihoods for the individual measurements of black hole mass $\mu=\log M_\bullet$ and bulge property $s=\log \sigma_\ast$ or $s=\log L_V$. The likelihood for measuring the mass $\mu_i$ and bulge property $s_i$ for given true mass $\mu$ and true bulge property $s$ is:
\begin{equation}
 l_i(\mu_i, s_i) = \int Q_\mu(\mu_i \mid \mu) Q_s(s_i \mid s) P( \mu \mid s)\, d\mu\,ds \ .
\end{equation}  
We assume $Q_\mu$, $Q_s$ and $P$ to have a log-normal form, with $\sigma_{Q_\mu}$ and $\sigma_{Q_s}$ corresponding to the measurement uncertainty in the black hole mass and bulge property, and $\sigma_{P}= \epsilon_0$ is the intrinsic scatter in the black hole mass-bulge property relation. Upper limits are incorporated in the fit, following \citet{Gultekin:2009}.
Thus, we minimize
\begin{equation}
 S = \sum_{i=1}^N \left[ \frac{ \left( \mu_i -\alpha -\beta s_i \right)^2}{\epsilon_{\mathrm{tot},i}^2} + 2 \ln \epsilon_{\mathrm{tot},i} \right] + 2 \sum_{j=1}^M \ln l_{\mathrm{ul},j} \ ,
\end{equation}
with $\alpha$ and $\beta$ being the normalization and the slope of the black hole-bulge relations, $\epsilon_{\mathrm{tot},i}^2= \sigma_{Q_\mu,i}^2 + \sigma_{Q_s,i}^2 + \epsilon_0^2$, $N$ is the number of black hole measurements, $M$ is the number of upper limits, and $l_{\mathrm{ul},j}$ is the likelihood of the upper limit as in \citet{Gultekin:2009}. 

We first fit the $M_\bullet-\sigma_\ast$ and $M_\bullet-L_V$ relationships using the sample of \citet{Gultekin:2009}, finding identical results. 
We then updated their black hole masses with our new values for the 12 objects in our sample. We find
\begin{equation}
 \log (M_\bullet / M_\odot ) = (8.18\pm0.06)\, + \,(4.32\pm0.31) \log (\sigma_\ast / 200\,\mathrm{km\,s}^{-1}) 
\end{equation} 
with intrinsic scatter $\epsilon_0=0.44\pm0.06$ and
\begin{equation}
 \log (M_\bullet / M_\odot ) = (9.01\pm0.10)\, + \,(1.06\pm0.15) \log (L_V / 10^{11}\,L_{\odot,V}) 
\end{equation} 
with intrinsic scatter $\epsilon_0=0.41\pm0.04$. They are shown in Figure~\ref{fig:bhbu}.

Note that both relations are not based on exactly the same samples. As in \citet{Gultekin:2009}, for the determination of the $M_\bullet-L_V$ relationship we restricted the sample to elliptical and S0 galaxies with reliable bulge$-$disk decomposition. When using the same restricted subsample for the $M_\bullet-\sigma_\ast$ relationship, we find a shallower slope ($\beta=3.80\pm0.33$) and a reduced intrinsic scatter ($\epsilon_0=0.34\pm0.05$), lower than for the $M_\bullet-L_V$ relationship for the same sample. Restricting the sample in this manner is supported by observations that suggest that spiral galaxies do not follow the $M_\bullet-\sigma_\ast$ relation of ellipticals \citep{Greene:2010}.

We also used a generalized least squares method to incorporate measurement uncertainties in both variables and intrinsic scatter as described in \citet{Tremaine:2002}, omitting the upper limits, which yields consistent results.
Compared to \citet{Gultekin:2009}, we find only a slight change for the best-fit. While the slope of the relation is consistent, the normalization increased slightly as well as the intrinsic scatter in both relations.
We also fitted the sample of  \citet{Gultekin:2009} with our updated black hole masses, without accounting for a DM halo. Most of the change in the $M_\bullet-\sigma_\ast$ and $M_\bullet-L_V$ relationships is caused by the improved masses. The effect of the inclusion of a DM halo on these relationships is marginal.

However, this is not a full correction of the black hole-bulge relationships for the effect of a DM halo on the black hole masses, as it is restricted to our sample of 12 galaxies. The rest of the galaxies with stellar dynamical black hole mass measurements potentially suffer from the same systematic bias. 
Ideally, a correction would consist of a re-modeling of these galaxies including a DM halo, as performed in this work for the sample of G03. However, we can use Figure~\ref{fig:deltamdm} as a guideline for an average correction. Figure~\ref{fig:deltamdm} indicates that the correction factor depends on the resolution of the sphere of influence. For $R_\mathrm{inf}/d_\mathrm{res}\gtrsim 3$ including or ignoring a DM halo in the modeling gives consistent results, while for lower values there is on average a systematic bias with a mean $\left\langle M_{\bullet, \mathrm{DM}}/M_{\bullet, \mathrm{no\,DM}}\right\rangle =1.5$ for our sample.

To estimate the effect on the black hole-bulge relations, we increased all stellar dynamical black hole mass measurements in the sample of \citet{Gultekin:2009} with $R_\mathrm{inf}/d_\mathrm{res}< 3$ by this average factor and re-fitted the relations. To investigate the pure change due to the DM halo, we also fitted the black hole-bulge relations to the sample of \citet{Gultekin:2009}, but with $M_\bullet$ of the 12 galaxies of our work replaced by our results without a DM halo. Compared to the best-fit to this sample, we found a slightly increased slope, a consistent intrinsic scatter, and an increase in normalization by $0.04$~dex. We found a normalization, slope, and intrinsic scatter of (8.21,4.38,0.42) for the $M_\bullet-\sigma_\ast$ and (9.05,1.07,0.41) for the $M_\bullet-L_V$ relationship.

Additionally, we fitted only our sample with the values for $M_\bullet$ with and without including a DM halo in the models. We recovered an increase in the normalization of $\sim 0.07$~dex, corresponding to the mean increase in $M_\bullet$ in the sample, while the slope is consistent and the intrinsic scatter decreases.

\newpage

\section{Conclusions}
We investigate the influence of accounting for the presence of a DM halo in the stellar dynamical modeling of galaxies on the measured black hole masses. We use a sample of 12 galaxies, already analyzed by \citet{Gebhardt:2003}, which have ground based as well as high-resolution \textit{HST} observations of the stellar kinematics to address this issue. 

We model these galaxies without and with the presence of a DM halo. In the first case we found a significant difference of the measured black hole masses between our previous results for a large fraction of the sample. For most of the objects the mass increased compared to the values given by \citet{Gebhardt:2003}. We ascribe this difference to the improved code, exhibiting a better coverage of the phase space for the generated orbit library. This shows the importance of a dense coverage of phase space in the dynamical models.

Second, we include a reasonable DM halo into the models, using a scaling relationship based on the galaxy luminosity \citep{Thomas:2009}. We find an increase of the measured black hole mass, but much less than what has been found for M87 and NGC~6086. For these two galaxies, kinematic information is available only at large radii, whereas for our sample we have high-resolution data covering the central parts of the galaxies. Thus, the black hole mass is better constrained by central kinematic observations and less affected by the presence of a DM halo in the models.

Using different massive DM halos for the same galaxy, we confirm the trend of an increase of the recovered black hole mass for a more massive halo as well as a decrease of the mass-to-light ratio. Based on a $\chi^2$ analysis, the presence of a DM halo is implied for five of the 12 galaxies with at least $2\sigma$ significance, although we are not able to constrain the shape of the DM halo.

We study the consequence of our new black hole mass measurements on the $M_\bullet-\sigma_\ast$ and $M_\bullet-L_V$ relationships, updating the sample of \citet{Gultekin:2009} with our results. We found only a mild change in the best-fit values, still consistent with the previous estimate, with a slight increase in the normalization and the intrinsic scatter. We estimated the total effect of a black hole mass increase for galaxies studied by stellar dynamics by accounting for a DM halo which will lead to an increase in the normalization by $\sim0.04-0.07$~dex.

Even if our sample shows only a mild influence of the DM halo on the black hole mass, a DM halo is clearly present. Thus it is necessary to take it into account in the modeling of the galaxy to avoid a systematic bias.

\acknowledgements
{We thank Remco van den Bosch for helpful discussions. A.S. thanks the University of Texas at Austin for their hospitality. A.S. acknowledges support by the DAAD, as well as by the Deutsche Forschungsgemeinschaft under its priority programme SPP1177, grant Wi~1369/23-2. K.G. acknowledges NSF grant 0908639. 
We also acknowledge the use of the computational resources at the Texas Advanced Computing Center at The University of Texas at Austin.}


\begin{thebibliography}{73}
\expandafter\ifx\csname natexlab\endcsname\relax\def\natexlab#1{#1}\fi

\bibitem[{{Binney} {et~al.}(1985){Binney}, {Gerhard}, \& {Hut}}]{Binney:1985}
{Binney}, J., {Gerhard}, O.~E., \& {Hut}, P. 1985, \mnras, 215, 59

\bibitem[{{Binney} \& {Tremaine}(1987)}]{Binney:1987}
{Binney}, J., \& {Tremaine}, S. 1987, {Galactic dynamics}, ed. {Binney, J.~\&
  Tremaine, S.}

\bibitem[{{Bridges} {et~al.}(2006){Bridges}, {Gebhardt}, {Sharples}, {Faifer},
  {Forte}, {Beasley}, {Zepf}, {Forbes}, {Hanes}, \& {Pierce}}]{Bridges:2006}
{Bridges}, T., {et~al.} 2006, \mnras, 373, 157

\bibitem[{{Ciotti} \& {Ostriker}(2007)}]{Ciotti:2007}
{Ciotti}, L., \& {Ostriker}, J.~P. 2007, \apj, 665, 1038

\bibitem[{{Coccato} {et~al.}(2009){Coccato}, {Gerhard}, {Arnaboldi}, {Das},
  {Douglas}, {Kuijken}, {Merrifield}, {Napolitano}, {Noordermeer},
  {Romanowsky}, {Capaccioli}, {Cortesi}, {de Lorenzi}, \&
  {Freeman}}]{Coccato:2009}
{Coccato}, L., {et~al.} 2009, \mnras, 394, 1249

\bibitem[{{Copin} {et~al.}(2004){Copin}, {Cretton}, \& {Emsellem}}]{Copin:2004}
{Copin}, Y., {Cretton}, N., \& {Emsellem}, E. 2004, \aap, 415, 889

\bibitem[{{Cretton} {et~al.}(1999){Cretton}, {de Zeeuw}, {van der Marel}, \&
  {Rix}}]{Cretton:1999}
{Cretton}, N., {de Zeeuw}, P.~T., {van der Marel}, R.~P., \& {Rix}, H. 1999,
  \apjs, 124, 383

\bibitem[{{Dalla Bont{\`a}} {et~al.}(2009){Dalla Bont{\`a}}, {Ferrarese},
  {Corsini}, {Miralda-Escud{\'e}}, {Coccato}, {Sarzi}, {Pizzella}, \&
  {Beifiori}}]{DallaBonta:2009}
{Dalla Bont{\`a}}, E., {Ferrarese}, L., {Corsini}, E.~M., {Miralda-Escud{\'e}},
  J., {Coccato}, L., {Sarzi}, M., {Pizzella}, A., \& {Beifiori}, A. 2009, \apj,
  690, 537

\bibitem[{{Di Matteo} {et~al.}(2005){Di Matteo}, {Springel}, \&
  {Hernquist}}]{DiMatteo:2005}
{Di Matteo}, T., {Springel}, V., \& {Hernquist}, L. 2005, \nat, 433, 604

\bibitem[{{Dierckx}(1993)}]{Dierckx:1993}
{Dierckx}, P. 1993, {Curve and surface fitting with splines}, ed. {Dierckx, P.}

\bibitem[{{Fabbiano} {et~al.}(2004){Fabbiano}, {Baldi}, {Pellegrini},
  {Siemiginowska}, {Elvis}, {Zezas}, \& {McDowell}}]{Fabbiano:2004}
{Fabbiano}, G., {Baldi}, A., {Pellegrini}, S., {Siemiginowska}, A., {Elvis},
  M., {Zezas}, A., \& {McDowell}, J. 2004, \apj, 616, 730

\bibitem[{{Ferrarese} {et~al.}(1996){Ferrarese}, {Ford}, \&
  {Jaffe}}]{Ferrarese:1996}
{Ferrarese}, L., {Ford}, H.~C., \& {Jaffe}, W. 1996, \apj, 470, 444

\bibitem[{{Ferrarese} \& {Merritt}(2000)}]{Ferrarese:2000}
{Ferrarese}, L., \& {Merritt}, D. 2000, \apjl, 539, L9

\bibitem[{{Forestell}(2009)}]{Forestell:2009}
{Forestell}, A.~D. 2009, PhD thesis, The University of Texas at Austin

\bibitem[{{Forestell} \& {Gebhardt}(2010)}]{Forestell:2010}
{Forestell}, A.~D., \& {Gebhardt}, K. 2010, \apj, 716, 370

\bibitem[{{Gebhardt}(2004)}]{Gebhardt:2004}
{Gebhardt}, K. 2004, in Coevolution of Black Holes and Galaxies, ed.
  {L.~C.~Ho}, 248--+

\bibitem[{{Gebhardt} \& {Thomas}(2009)}]{Gebhardt:2009}
{Gebhardt}, K., \& {Thomas}, J. 2009, \apj, 700, 1690

\bibitem[{{Gebhardt} {et~al.}(1996){Gebhardt}, {Richstone}, {Ajhar}, {Lauer},
  {Byun}, {Kormendy}, {Dressler}, {Faber}, {Grillmair}, \&
  {Tremaine}}]{Gebhardt:1996}
{Gebhardt}, K., {et~al.} 1996, \aj, 112, 105

\bibitem[{{Gebhardt} {et~al.}(2000{\natexlab{a}}){Gebhardt}, {Bender}, {Bower},
  {Dressler}, {Faber}, {Filippenko}, {Green}, {Grillmair}, {Ho}, {Kormendy},
  {Lauer}, {Magorrian}, {Pinkney}, {Richstone}, \& {Tremaine}}]{Gebhardt:2000}
---. 2000{\natexlab{a}}, \apjl, 539, L13

\bibitem[{{Gebhardt} {et~al.}(2000{\natexlab{b}}){Gebhardt}, {Richstone},
  {Kormendy}, {Lauer}, {Ajhar}, {Bender}, {Dressler}, {Faber}, {Grillmair},
  {Magorrian}, \& {Tremaine}}]{Gebhardt:2000b}
---. 2000{\natexlab{b}}, \aj, 119, 1157

\bibitem[{{Gebhardt} {et~al.}(2003){Gebhardt}, {Richstone}, {Tremaine},
  {Lauer}, {Bender}, {Bower}, {Dressler}, {Faber}, {Filippenko}, {Green},
  {Grillmair}, {Ho}, {Kormendy}, {Magorrian}, \& {Pinkney}}]{Gebhardt:2003}
---. 2003, \apj, 583, 92

\bibitem[{{Gebhardt} {et~al.}(2007){Gebhardt}, {Lauer}, {Pinkney}, {Bender},
  {Richstone}, {Aller}, {Bower}, {Dressler}, {Faber}, {Filippenko}, {Green},
  {Ho}, {Kormendy}, {Siopis}, \& {Tremaine}}]{Gebhardt:2007}
---. 2007, \apj, 671, 1321

\bibitem[{{Greene} {et~al.}(2010){Greene}, {Peng}, {Kim}, {Kuo}, {Braatz},
  {Violette Impellizzeri}, {Condon}, {Lo}, {Henkel}, \& {Reid}}]{Greene:2010}
{Greene}, J.~E., {et~al.} 2010, \apj, 721, 26

\bibitem[{{Greenhill} {et~al.}(2003){Greenhill}, {Booth}, {Ellingsen},
  {Herrnstein}, {Jauncey}, {McCulloch}, {Moran}, {Norris}, {Reynolds}, \&
  {Tzioumis}}]{Greenhill:2003}
{Greenhill}, L.~J., {et~al.} 2003, \apj, 590, 162

\bibitem[{{G{\"u}ltekin} {et~al.}(2009{\natexlab{a}}){G{\"u}ltekin},
  {Richstone}, {Gebhardt}, {Lauer}, {Pinkney}, {Aller}, {Bender}, {Dressler},
  {Faber}, {Filippenko}, {Green}, {Ho}, {Kormendy}, \&
  {Siopis}}]{Gultekin:2009b}
{G{\"u}ltekin}, K., {et~al.} 2009{\natexlab{a}}, \apj, 695, 1577

\bibitem[{{G{\"u}ltekin} {et~al.}(2009{\natexlab{b}}){G{\"u}ltekin},
  {Richstone}, {Gebhardt}, {Lauer}, {Tremaine}, {Aller}, {Bender}, {Dressler},
  {Faber}, {Filippenko}, {Green}, {Ho}, {Kormendy}, {Magorrian}, {Pinkney}, \&
  {Siopis}}]{Gultekin:2009}
---. 2009{\natexlab{b}}, \apj, 698, 198

\bibitem[{{H{\"a}ring} \& {Rix}(2004)}]{Haering:2004}
{H{\"a}ring}, N., \& {Rix}, H.-W. 2004, \apjl, 604, L89

\bibitem[{{Herrnstein} {et~al.}(2005){Herrnstein}, {Moran}, {Greenhill}, \&
  {Trotter}}]{Herrnstein:2005}
{Herrnstein}, J.~R., {Moran}, J.~M., {Greenhill}, L.~J., \& {Trotter}, A.~S.
  2005, \apj, 629, 719

\bibitem[{{Hirschmann} {et~al.}(2010){Hirschmann}, {Khochfar}, {Burkert},
  {Naab}, {Genel}, \& {Somerville}}]{Hirschmann:2010}
{Hirschmann}, M., {Khochfar}, S., {Burkert}, A., {Naab}, T., {Genel}, S., \&
  {Somerville}, R.~S. 2010, \mnras, 407, 1016

\bibitem[{{Hwang} {et~al.}(2008){Hwang}, {Lee}, {Park}, {Kim}, {Park}, {Sohn},
  {Lee}, {Rey}, {Lee}, \& {Kim}}]{Hwang:2008}
{Hwang}, H.~S., {et~al.} 2008, \apj, 674, 869

\bibitem[{{Jahnke} \& {Maccio}(2010)}]{Jahnke:2010}
{Jahnke}, K., \& {Maccio}, A. 2010, arXiv:1006.0482

\bibitem[{{Kormendy} {et~al.}(1998){Kormendy}, {Bender}, {Evans}, \&
  {Richstone}}]{Kormendy:1998}
{Kormendy}, J., {Bender}, R., {Evans}, A.~S., \& {Richstone}, D. 1998, \aj,
  115, 1823

\bibitem[{{Kormendy} \& {Gebhardt}(2001)}]{Kormendy:2001}
{Kormendy}, J., \& {Gebhardt}, K. 2001, in American Institute of Physics
  Conference Series, Vol. 586, 20th Texas Symposium on relativistic
  astrophysics, ed. {J.~C.~Wheeler \& H.~Martel}, 363--381

\bibitem[{{Kormendy} \& {Richstone}(1995)}]{Kormendy:1995}
{Kormendy}, J., \& {Richstone}, D. 1995, \araa, 33, 581

\bibitem[{{Kronawitter} {et~al.}(2000){Kronawitter}, {Saglia}, {Gerhard}, \&
  {Bender}}]{Kronawitter:2000}
{Kronawitter}, A., {Saglia}, R.~P., {Gerhard}, O., \& {Bender}, R. 2000, \aaps,
  144, 53

\bibitem[{{Kuo} {et~al.}(2010){Kuo}, {Braatz}, {Condon}, {Impellizzeri}, {Lo},
  {Zaw}, {Schenker}, {Henkel}, {Reid}, \& {Greene}}]{Kuo:2010}
{Kuo}, C.~Y., {et~al.} 2010, arXiv:1008.2146

\bibitem[{{Lauer} {et~al.}(1995){Lauer}, {Ajhar}, {Byun}, {Dressler}, {Faber},
  {Grillmair}, {Kormendy}, {Richstone}, \& {Tremaine}}]{Lauer:1995}
{Lauer}, T.~R., {et~al.} 1995, \aj, 110, 2622

\bibitem[{{Lauer} {et~al.}(2005){Lauer}, {Faber}, {Gebhardt}, {Richstone},
  {Tremaine}, {Ajhar}, {Aller}, {Bender}, {Dressler}, {Filippenko}, {Green},
  {Grillmair}, {Ho}, {Kormendy}, {Magorrian}, {Pinkney}, \&
  {Siopis}}]{Lauer:2005}
---. 2005, \aj, 129, 2138

\bibitem[{{Magorrian} {et~al.}(1998){Magorrian}, {Tremaine}, {Richstone},
  {Bender}, {Bower}, {Dressler}, {Faber}, {Gebhardt}, {Green}, {Grillmair},
  {Kormendy}, \& {Lauer}}]{Magorrian:1998}
{Magorrian}, J., {et~al.} 1998, \aj, 115, 2285

\bibitem[{{Marconi} {et~al.}(2001){Marconi}, {Capetti}, {Axon}, {Koekemoer},
  {Macchetto}, \& {Schreier}}]{Marconi:2001}
{Marconi}, A., {Capetti}, A., {Axon}, D.~J., {Koekemoer}, A., {Macchetto}, D.,
  \& {Schreier}, E.~J. 2001, \apj, 549, 915

\bibitem[{{Marconi} \& {Hunt}(2003)}]{Marconi:2003}
{Marconi}, A., \& {Hunt}, L.~K. 2003, \apjl, 589, L21

\bibitem[{{McConnell} {et~al.}(2010){McConnell}, {Ma}, {Graham}, {Gebhardt},
  {Lauer}, {Wright}, \& {Richstone}}]{McConnell:2010}
{McConnell}, N.~J., {Ma}, C., {Graham}, J.~R., {Gebhardt}, K., {Lauer}, T.~R.,
  {Wright}, S.~A., \& {Richstone}, D.~O. 2010, arXiv:1009.0750

\bibitem[{{M{\'e}ndez} {et~al.}(2001){M{\'e}ndez}, {Riffeser}, {Kudritzki},
  {Matthias}, {Freeman}, {Arnaboldi}, {Capaccioli}, \& {Gerhard}}]{Mendez:2001}
{M{\'e}ndez}, R.~H., {Riffeser}, A., {Kudritzki}, R., {Matthias}, M.,
  {Freeman}, K.~C., {Arnaboldi}, M., {Capaccioli}, M., \& {Gerhard}, O.~E.
  2001, \apj, 563, 135

\bibitem[{{M{\'e}ndez} {et~al.}(2008){M{\'e}ndez}, {Teodorescu}, \&
  {Kudritzki}}]{Mendez:2008}
{M{\'e}ndez}, R.~H., {Teodorescu}, A.~M., \& {Kudritzki}, R. 2008, \apjs, 175,
  522

\bibitem[{{M{\'e}ndez} {et~al.}(2009){M{\'e}ndez}, {Teodorescu}, {Kudritzki},
  \& {Burkert}}]{Mendez:2009}
{M{\'e}ndez}, R.~H., {Teodorescu}, A.~M., {Kudritzki}, R., \& {Burkert}, A.
  2009, \apj, 691, 228

\bibitem[{{Navarro} {et~al.}(1996){Navarro}, {Frenk}, \&
  {White}}]{Navarro:1996}
{Navarro}, J.~F., {Frenk}, C.~S., \& {White}, S.~D.~M. 1996, \apj, 462, 563

\bibitem[{{Pellegrini} {et~al.}(2007){Pellegrini}, {Baldi}, {Kim}, {Fabbiano},
  {Soria}, {Siemiginowska}, \& {Elvis}}]{Pellegrini:2007}
{Pellegrini}, S., {Baldi}, A., {Kim}, D.~W., {Fabbiano}, G., {Soria}, R.,
  {Siemiginowska}, A., \& {Elvis}, M. 2007, \apj, 667, 731

\bibitem[{{Peng}(2007)}]{Peng:2007}
{Peng}, C.~Y. 2007, \apj, 671, 1098

\bibitem[{{Pierce} {et~al.}(2006){Pierce}, {Beasley}, {Forbes}, {Bridges},
  {Gebhardt}, {Faifer}, {Forte}, {Zepf}, {Sharples}, {Hanes}, \&
  {Proctor}}]{Pierce:2006}
{Pierce}, M., {et~al.} 2006, \mnras, 366, 1253

\bibitem[{{Pinkney} {et~al.}(2003){Pinkney}, {Gebhardt}, {Bender}, {Bower},
  {Dressler}, {Faber}, {Filippenko}, {Green}, {Ho}, {Kormendy}, {Lauer},
  {Magorrian}, {Richstone}, \& {Tremaine}}]{Pinkney:2003}
{Pinkney}, J., {et~al.} 2003, \apj, 596, 903

\bibitem[{{Press} {et~al.}(1992){Press}, {Teukolsky}, {Vetterling}, \&
  {Flannery}}]{Press:1992}
{Press}, W.~H., {Teukolsky}, S.~A., {Vetterling}, W.~T., \& {Flannery}, B.~P.
  1992, {Numerical recipes in C. The art of scientific computing}, ed. {Press,
  W.~H., Teukolsky, S.~A., Vetterling, W.~T., \& Flannery, B.~P. }

\bibitem[{{Richstone} {et~al.}(2004){Richstone}, {Gebhardt}, {Aller}, {Bender},
  {Bower}, {Dressler}, {Faber}, {Filippenko}, {Green}, {Ho}, {Kormendy},
  {Lauer}, {Magorrian}, {Pinkney}, {Siopis}, \& {Tremaine}}]{Richstone:2004}
{Richstone}, D., {et~al.} 2004, arXiv:0403257

\bibitem[{{Richstone} \& {Tremaine}(1988)}]{Richstone:1988}
{Richstone}, D.~O., \& {Tremaine}, S. 1988, \apj, 327, 82

\bibitem[{{Rix} {et~al.}(1997){Rix}, {de Zeeuw}, {Cretton}, {van der Marel}, \&
  {Carollo}}]{Rix:1997}
{Rix}, H., {de Zeeuw}, P.~T., {Cretton}, N., {van der Marel}, R.~P., \&
  {Carollo}, C.~M. 1997, \apj, 488, 702

\bibitem[{{Romanowsky} {et~al.}(2003){Romanowsky}, {Douglas}, {Arnaboldi},
  {Kuijken}, {Merrifield}, {Napolitano}, {Capaccioli}, \&
  {Freeman}}]{Romanowsky:2003}
{Romanowsky}, A.~J., {Douglas}, N.~G., {Arnaboldi}, M., {Kuijken}, K.,
  {Merrifield}, M.~R., {Napolitano}, N.~R., {Capaccioli}, M., \& {Freeman},
  K.~C. 2003, Science, 301, 1696

\bibitem[{{Schwarzschild}(1979)}]{Schwarzschild:1979}
{Schwarzschild}, M. 1979, \apj, 232, 236

\bibitem[{{Shapiro} {et~al.}(2006){Shapiro}, {Cappellari}, {de Zeeuw},
  {McDermid}, {Gebhardt}, {van den Bosch}, \& {Statler}}]{Shapiro:2006}
{Shapiro}, K.~L., {Cappellari}, M., {de Zeeuw}, T., {McDermid}, R.~M.,
  {Gebhardt}, K., {van den Bosch}, R.~C.~E., \& {Statler}, T.~S. 2006, \mnras,
  370, 559

\bibitem[{{Shen} \& {Gebhardt}(2010)}]{Shen:2010}
{Shen}, J., \& {Gebhardt}, K. 2010, \apj, 711, 484

\bibitem[{{Silk} \& {Rees}(1998)}]{Silk:1998}
{Silk}, J., \& {Rees}, M.~J. 1998, \aap, 331, L1

\bibitem[{{Siopis} {et~al.}(2009){Siopis}, {Gebhardt}, {Lauer}, {Kormendy},
  {Pinkney}, {Richstone}, {Faber}, {Tremaine}, {Aller}, {Bender}, {Bower},
  {Dressler}, {Filippenko}, {Green}, {Ho}, \& {Magorrian}}]{Siopis:2009}
{Siopis}, C., {et~al.} 2009, \apj, 693, 946

\bibitem[{{Soria} {et~al.}(2006){Soria}, {Fabbiano}, {Graham}, {Baldi},
  {Elvis}, {Jerjen}, {Pellegrini}, \& {Siemiginowska}}]{Soria:2006}
{Soria}, R., {Fabbiano}, G., {Graham}, A.~W., {Baldi}, A., {Elvis}, M.,
  {Jerjen}, H., {Pellegrini}, S., \& {Siemiginowska}, A. 2006, \apj, 640, 126

\bibitem[{{Springel} {et~al.}(2005){Springel}, {Di Matteo}, \&
  {Hernquist}}]{Springel:2005}
{Springel}, V., {Di Matteo}, T., \& {Hernquist}, L. 2005, \apjl, 620, L79

\bibitem[{{Thomas} {et~al.}(2005){Thomas}, {Saglia}, {Bender}, {Thomas},
  {Gebhardt}, {Magorrian}, {Corsini}, \& {Wegner}}]{Thomas:2005}
{Thomas}, J., {Saglia}, R.~P., {Bender}, R., {Thomas}, D., {Gebhardt}, K.,
  {Magorrian}, J., {Corsini}, E.~M., \& {Wegner}, G. 2005, \mnras, 360, 1355

\bibitem[{{Thomas} {et~al.}(2007){Thomas}, {Saglia}, {Bender}, {Thomas},
  {Gebhardt}, {Magorrian}, {Corsini}, \& {Wegner}}]{Thomas:2007}
---. 2007, \mnras, 382, 657

\bibitem[{{Thomas} {et~al.}(2009){Thomas}, {Saglia}, {Bender}, {Thomas},
  {Gebhardt}, {Magorrian}, {Corsini}, \& {Wegner}}]{Thomas:2009}
---. 2009, \apj, 691, 770

\bibitem[{{Thomas} {et~al.}(2004){Thomas}, {Saglia}, {Bender}, {Thomas},
  {Gebhardt}, {Magorrian}, \& {Richstone}}]{Thomas:2004}
{Thomas}, J., {Saglia}, R.~P., {Bender}, R., {Thomas}, D., {Gebhardt}, K.,
  {Magorrian}, J., \& {Richstone}, D. 2004, \mnras, 353, 391

\bibitem[{{Tremaine} {et~al.}(2002){Tremaine}, {Gebhardt}, {Bender}, {Bower},
  {Dressler}, {Faber}, {Filippenko}, {Green}, {Grillmair}, {Ho}, {Kormendy},
  {Lauer}, {Magorrian}, {Pinkney}, \& {Richstone}}]{Tremaine:2002}
{Tremaine}, S., {et~al.} 2002, \apj, 574, 740

\bibitem[{{Valluri} {et~al.}(2004){Valluri}, {Merritt}, \&
  {Emsellem}}]{Valluri:2004}
{Valluri}, M., {Merritt}, D., \& {Emsellem}, E. 2004, \apj, 602, 66

\bibitem[{{van den Bosch} \& {de Zeeuw}(2010)}]{vandenBosch:2010}
{van den Bosch}, R.~C.~E., \& {de Zeeuw}, P.~T. 2010, \mnras, 401, 1770

\bibitem[{{van der Marel} {et~al.}(1998){van der Marel}, {Cretton}, {de Zeeuw},
  \& {Rix}}]{vanderMarel:1998}
{van der Marel}, R.~P., {Cretton}, N., {de Zeeuw}, P.~T., \& {Rix}, H. 1998,
  \apj, 493, 613

\bibitem[{{Weijmans} {et~al.}(2009){Weijmans}, {Cappellari}, {Bacon}, {de
  Zeeuw}, {Emsellem}, {Falc{\'o}n-Barroso}, {Kuntschner}, {McDermid}, {van den
  Bosch}, \& {van de Ven}}]{Weijmans:2009}
{Weijmans}, A., {et~al.} 2009, \mnras, 398, 561

\bibitem[{{Woo} {et~al.}(2010){Woo}, {Treu}, {Barth}, {Wright}, {Walsh},
  {Bentz}, {Martini}, {Bennert}, {Canalizo}, {Filippenko}, {Gates}, {Greene},
  {Li}, {Malkan}, {Stern}, \& {Minezaki}}]{Woo:2010}
{Woo}, J., {et~al.} 2010, \apj, 716, 269

\bibitem[{{Zepf} {et~al.}(2000){Zepf}, {Beasley}, {Bridges}, {Hanes},
  {Sharples}, {Ashman}, \& {Geisler}}]{Zepf:2000}
{Zepf}, S.~E., {Beasley}, M.~A., {Bridges}, T.~J., {Hanes}, D.~A., {Sharples},
  R.~M., {Ashman}, K.~M., \& {Geisler}, D. 2000, \aj, 120, 2928

\end{thebibliography}
\end{document}